\documentclass[article]{IEEEtran}
\usepackage{graphicx}
\usepackage{setspace}
\usepackage{longtable}
\usepackage{url}
\usepackage{cite}
\usepackage{ulem}
\usepackage{epstopdf}
\DeclareGraphicsExtensions{.png}
\usepackage[usenames, dvipsnames]{color}

\title{Software Defined Optical Networks (SDONs): \\
     A Comprehensive Survey
\thanks{Please direct correspondence to M.~Reisslein}}

\author{Akhilesh Thyagaturu, Anu Mercian, Michael P.~McGarry,
 Martin Reisslein, and Wolfgang Kellerer
\thanks{A.~Thyagaturu, A. Mercian, and M.~Reisslein are with the
 School of Electrical, Computer, and Energy Eng., Arizona State University,
  Tempe, AZ 85287-5706, USA,
Phone: 480-965-8593, Fax: 480-965-8325,
(e-mail: \{athyagat, amercian, reisslein\}@asu.edu).}
\thanks{M.~McGarry is with the Dept. of Electr. and Comp.
Eng., University of Texas at El Paso, El Paso, TX 79968, USA
(email: mpmcgarry@utep.edu).}
\thanks{W.~Kellerer is with the
Lehrstuhl f\"ur Kommunikationsnetze, Technische Universit\"at M\"unchen,
Munich, 80290, Germany, (email: {wolfgang.kellerer}@tum.de).}  }

\begin{document}
\maketitle

\begin{abstract}
The emerging Software Defined Networking (SDN) paradigm separates the
data plane from the control plane and centralizes network control
in an SDN controller.
Applications interact with controllers to implement
network services, such as network transport with Quality of Service (QoS).
SDN facilitates the virtualization of network functions so that
multiple virtual networks can operate over a given installed
physical network infrastructure.
Due to the specific characteristics of optical (photonic)
communication components and the high optical transmission capacities,
SDN based optical networking poses particular challenges, but holds
also great potential.
In this article, we comprehensively survey studies that
examine the SDN paradigm in optical networks; in brief, we survey the area of
Software Defined Optical Networks (SDONs).
We mainly organize the SDON studies into studies focused on
the infrastructure layer, the control layer, and the application layer.
Moreover, we cover SDON studies focused on network virtualization, as well
as SDON studies focused on the orchestration of multilayer and multidomain
networking. Based on the survey, we identify open challenges for
SDONs and outline future directions.
\end{abstract}


\bstctlcite{IEEEexample:BSTcontrol}

\section{Introduction} \label{Sec1_Intro}
At least a decade ago \cite{ComerNetMgmtBook} it was recognized that new
network abstraction layers for network control functions needed to be
developed to both simplify and automate network management. Software
Defined Networking (SDN)~\cite{HuHB14,jai2013b4,Vau11} is the design
principle that emerged to structure the development of those new
abstraction layers.
Fundamentally, SDN is defined by three architectural
principles~\cite{SDNarch11,KreRVR15}:
$(i)$ the separation of control plane functions
and data plane functions, $(ii)$ the logical centralization of control, and
$(iii)$ programmability of network functions. The first two
architectural principles are related in that they combine to allow for network
control functions to have a wider perspective on the network. The idea is that
networks can be made easier to manage (i.e., control and monitor) with a
move away from significantly distributed control. A tradeoff is then
considered that balances ease of management arising from control
centralization and scalability issues that naturally arise from that
centralization.

\begin{figure*}[t!]	\centering
	\includegraphics[width=6in]{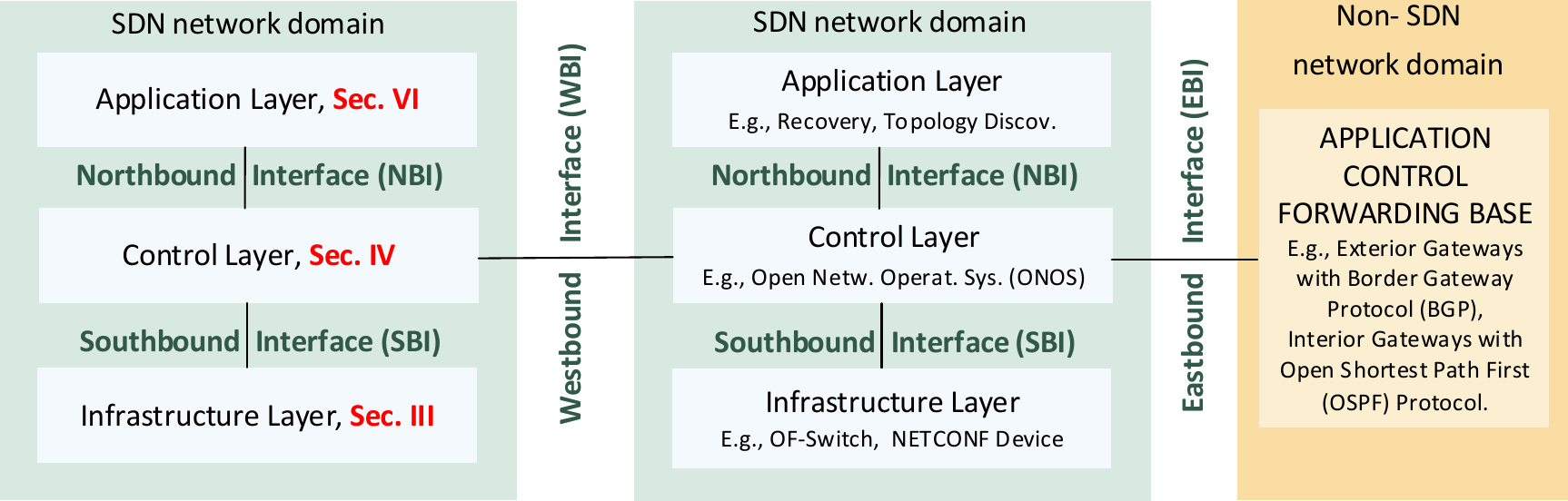}
\caption{Illustration of Software Defined Networking (SDN) abstraction
  layers: The infrastructure layer
  implements the data plane, e.g., with OpenFlow (OF) switches~\cite{LaKR14}
  or network elements (devices) controlled with the NETCONF
  protocol~\cite{rfc6241}. A controller at the control layer,
  e.g., the ONOS controller~\cite{onos}, controls the infrastructure layer based
  on the application layer requirements.
  The interface
  between the application and control layers is commonly referred to
  as the NorthBound Interface (NBI), while the interface between the
  control and infrastructure layers is commonly referred to as the
  SouthBound Interface (SBI).
  The WestBound Interface (WBI) interconnects multiple SDN domains, while
  the EastBound Interface (EBI) interconnects with non-SDN domains.}
	\label{fig:sdnlayers}
\end{figure*}

The SDN abstraction layering consists of three generally accepted
layers~\cite{SDNarch11} inspired by computing systems,
from the bottom layer to the top layer:
$(i)$ the \textit{infrastructure} layer, $(ii)$ the \textit{control} layer, and
($iii$) the \textit{application} layer,
as illustrated in Fig.~\ref{fig:sdnlayers}.
The interface between the
application layer and the control layer is referred to as the
NorthBound Interface (NBI), while the interface between the control layer
and the infrastructure layer is referred to as the SouthBound
Interface (SBI). There are a variety of standards emerging for these
interfaces, e.g., the OpenFlow protocol~\cite{LaKR14} for the SBI.

The \textit{application} layer is modeled after software applications
that utilize computing resources to complete tasks.
The \textit{control} layer is modeled after a computer's Operating System
(OS) that manages computer resources (e.g., processors and memory),
provides an abstraction layer to simplify interfacing with the
computer's devices, and provides a common set of services that all
applications can leverage. Device drivers in a computer's OS hide the
details of interfacing with many different devices from the
applications by offering a simple and unified interface for various
device types. In the SDN model both the unified SBI
as well as the control layer functionality provide
the equivalent of a device driver for interfacing with devices in the
\textit{infrastructure} layer, e.g., packet switches.

Optical networks play an important role in our modern information
technology due to their high transmission capacities.
At the same time, the specific optical (photonic) transmission
and switching characteristics, such as circuit, burst, and packet
switching on wavelength channels, pose challenges for controlling
optical networks.
This article presents a comprehensive survey of Software
Defined Optical Networks (SDONs).
SDONs seek to leverage the flexibility of
SDN control for supporting networking applications with an
underlying optical network infrastructure.
This survey comprehensively covers SDN related mechanisms that have
been studied to date for optical networks.

\subsection{Related Work}
The general principles of SDN have been extensively covered in several
surveys, see for instance,~\cite{aky2014roa,Chen2015a,cui2016big,far2015soft, FeRZ14, HuHB14, Jain2013,
  jar2014sur, JaZH14, kha2015jons, KreRVR15, LaKR14, li2014sof, Lopes2015,MiCK13,
  NuMNO14,Racherla,Trois2016,van2014sca, wic2015sof, XiWF14}.
SDN security has been surveyed in~\cite{ahm2015sec,sco2015sur}, while
management of SDN networks has been surveyed in~\cite{wic2015sof} and
SDN-based satellite networking is considered in~\cite{ber2015sof}.

To date, there have been relatively few overview and survey articles
on SDONs.
Zhang et al.~\cite{zha2013surflexi} have presented a thorough survey
on flexible optical networking based on Orthogonal Frequency
Division Multiplexing (OFDM) in core (backbone) networks. The survey
briefly notes how OFDM-based elastic networking can facilitate
network virtualization and surveys a few studies on OFDM-based
network virtualization in core networks.

Bhaumik et al.~\cite{BhZCS14} have presented an overview of
SDN and network virtualization concepts and outlined principles
for extending SDN and network virtualization concepts to
the field of optical networking.
Their focus has been mainly on industry efforts, reviewing white papers
on SDN strategies from leading networking companies, such as
Cisco, Juniper, Hewlett-Packard, Alcatel-Lucent, and Huawei.
A few selected academic research projects on general SDN optical
networks, namely projects reported in the journal
articles~\cite{ChNS13,liu2013field} and a few related conference papers,
have also been reviewed by Bhaumik et al.~\cite{BhZCS14}.
In contrast to Bhaumik et al.~\cite{BhZCS14}, we provide a comprehensive
up-to-date review of academic research on
SDONs. Whereas Bhaumik et al.~\cite{BhZCS14}
presented a small sampling of SDON research organized by research projects,
we present a comprehensive SDON survey that is organized according
to the SDN infrastructure, control, and application layer architecture.

For the SDON sub-domain of access networks,
Cvijetic~\cite{Cvi14} has given an overview of access network challenges
that can be addressed with SDN.
These challenges include lack of support for on-demand modifications of
traffic transmission policies and rules and limitations to
vendor-proprietary policies, rules, and software.
Cvijetic~\cite{Cvi14} also offers a very brief overview of
research progress for SDN-based optical access networks, mainly focusing on
studies on the physical (photonics) infrastructure layer.
Cvijetic~\cite{CviSept14} has further expanded the overview of
SDON challenges by considering the incorporation of 5G wireless
systems. Cvijetic~\cite{CviSept14} has noted that
SDN access networks are highly promising for low-latency and high-bandwidth
back-hauling from 5G cell base stations and briefly surveyed
the requirements and areas of future research required for
integrating 5G with SDON access networks.
A related overview of general software defined access networks
based on a variety of physical transmission media, including copper
Digital Subscriber Line (DSL)~\cite{dslbook}
and Passive Optical Networks (PONs), has been
presented by Kerpez et al.~\cite{KeCGG14}.

Bitar~\cite{Bi14} has surveyed use cases for SDN controlled broadband access,
such as on-demand bandwidth boost, dynamic service
re-provisioning, as well as value-added services and service protection.
Bitar~\cite{Bi14} has discussed the commercial perspective of the
access networks that are enhanced with SDN to add
cost-value to the network operation.
Almeida Amazonas et al.~\cite{AmSS14} have surveyed the key issues
of incorporating SDN in optical and wireless access networks.
They briefly outlined the obstacles posed by
 the different specific physical characteristics of
optical and wireless access networks.

Although our focus is on optical networks, for completeness we note that
for the field of wireless and mobile networks, SDN based networking mechanisms
have been surveyed
in~\cite{ArSR15,BeDSB14,haq2016wir,jag2014sof,SaCKA15,soo2015sof,yan2014sof}
while network virtualization has been surveyed in~\cite{LiY15} for
general wireless networks and in~\cite{kha2015wir} for
wireless sensor networks.
SDN and virtualization strategies for LTE wireless cellular networks
have been surveyed in~\cite{ngu2015sdn}. SDN-based 5G wireless
network developments for mobile networks have been outlined
in~\cite{Huawei14,PeLZW14,TrGVS15,YaKO14}.

\subsection{Survey Organization}
We have mainly organized
our survey according to the three-layer SDN architecture
illustrated in Fig.~\ref{fig:sdnlayers}.
In particular, we have organized the survey in a bottom-up manner,
surveying first SDON studies focused on the
infrastructure layer in Section~\ref{sdninfra:sec}.
Subsequently, we survey SDON studies focused on the control layer
in Section~\ref{sdnctl:sec}.
The virtualization of optical networks is commonly closely related to
the SDN control layer. Therefore, we survey SDON studies
focused on virtualization in Section~\ref{virt:sec},
right after the SDON control layer section.
Resuming the journey up the layers in Fig.~\ref{fig:sdnlayers},
we survey SDON studies focused on the application layer in
Section~\ref{sdnapp:sec}.
We survey mechanisms for the overarching orchestration of the application
layer and lower layers, possibly across multiple network domains
(see Fig.~\ref{fig_control_orch}),
in Section~\ref{orch:sec}.
Finally, we outline open challenges and future research directions in
Section~\ref{sec:open} and conclude the survey in Section~\ref{sec:conclusion}.

\section{Background}
\label{bg:sec}
This section first provides background on
Software Defined Networking (SDN), followed by background on virtualization
and optical networking.
SDN, as defined by the Internet Engineering
Task Force (IETF)~\cite{rfc7426},
is a networking paradigm enabling the programmability of networks.
SDN abstracts and separates the data forwarding
plane from the control plane, allowing faster
technological development both in data and control planes.
We provide background on the SDN architecture, including its architectural
layers in Subsection~\ref{sdnarch:sec}.
The network programmability provides the flexibility to dynamically initialize,
control, manipulate, and manage the end-to-end network behavior via open
interfaces, which are reviewed in Subsection~\ref{sdnint:sec}.
Subsequently, we provide background on network virtualization in
Subsection~\ref{bgnetvit:sec}
and on optical networking in Subsection~\ref{bg_access:sec}.

\subsection{Software Defined Networking (SDN) Architectural Layers}
\label{sdnarch:sec}
SDN offers a simplified view of the underlying network infrastructure
for the network control and monitoring applications
through the abstraction of each independent network layer.
Fig.~\ref{fig:sdnlayers} illustrates the
three-layer SDN architecture model consisting of application,
control, and infrastructure layers as defined by the
Open Networking Foundation (ONF)~\cite{SDNarch11}. The ONF is
the organization that is responsible for
the publication of specifications for the OpenFlow protocol.
The OpenFlow protocol~\cite{HuHB14,keo2008of,LaKR14}
has been the first protocol for the SouthBound Interface (SBI,
also referred to as Data-Controller Plane Interface (D-CPI))
between the control and infrastructure layers.
Each layer operates independently, allowing multiple solutions
to coexist within each layer, e.g., the
infrastructure layer can be built from any
programmable devices, which are commonly referred to as
network elements~\cite{SDNarch11_521} or network devices~\cite{rfc7426}
(or sometimes as forwarding elements~\cite{rfc3746}).
We will use the terminology network element throughout this survey.
The SouthBound Interface (SBI) and the NorthBound
Interface (NBI, also referred to as Application-Controller Plane
Interface (A-CPI)) are defined as the primary interfaces
interconnecting the SDN layers through abstractions.
An SDN network architecture can coexist with both
concurrent SDN architectures and non-SDN legacy network architectures.
Additional interfaces are defined namely the
EastBound Interface (EBI) and the WestBound
Interface (WBI)~\cite{JaZH14} to interconnect the SDN architecture
with external network architectures
(the EBI and WBI are also collectively
referred to as Intermediate-Controller Plane Interfaces (I-CPIs)).
Generally, EBIs establish communication links to
legacy network architectures (i.e., non-SDN networks); whereas,
links to concurrent (side-by-side) SDN architectures are facilitated
by the WBIs.

\subsubsection{Infrastructure Layer}
The infrastructure layer includes an environment for
(payload) data traffic forwarding (data plane)
either in virtual or actual hardware.
The data plane comprises a network of network elements,
which expose their capabilities through the SBI
to the control plane. In traditional networking, control mechanisms are
embedded within an infrastructure,
i.e., decision making capabilities are embedded within the
infrastructure to perform network actions, such as switching or routing.
Additionally, these forwarding actions in the traditional network elements
are autonomously
established based on self-evaluated topology information that is often
obtained through proprietary vendor-specific algorithms.
Therefore, the configuration setups of traditional network elements
are generally not reconfigurable without a service disruption,
limiting the network flexibility.
In contrast, SDN decouples the autonomous control functions, such as
forwarding algorithms and neighbor discovery of
the network nodes, and moves these control functions out of the infrastructure
to a centrally controlled logical node, the controller.
In doing so, the network elements act only as dumb switches which
act upon the instructions of the controller. This decoupling reduces the
network element complexity and improves reconfigurability.

In addition to decoupling the
control and data planes, packet modification capabilities at the line-rates
of network elements have been significantly improved with SDN.
P4~\cite{bosshart2014p4} is a programmable
protocol-independent packet processor, that can arbitrarily
match the fields within any formatted packet
and is capable of applying any arbitrary actions (as programmed)
on the packet before forwarding. A similar forwarding mechanism,
Protocol-oblivious Forwarding (PoF) has been
proposed by Huawei Technologies~\cite{Song2013}.

\subsubsection{Control Layer}
The control layer is responsible for programming (configuring)
the network elements (switches) via the SBIs.
The SDN controller is a logical entity that identifies the south bound
instructions to configure the network infrastructure
based on application layer requirements.
To efficiently manage the network, SDN controllers can
request information from the SDN infrastructures,
such as flow statistics, topology information, neighbor relations,
and link status from the network elements (nodes).
The software entity that implements the SDN
controller is often referred to as \textit{Network Operating System (NOS)}.
Generally, a NOS can be implemented independently of
SDN, i.e., without supporting SDN.
On the other hand, in addition to supporting SDN operations,
a NOS can provide advanced capabilities, such as virtualization,
application scheduling, and database management.
The Open Network Operating System (ONOS)~\cite{onos}
is an example of an SDN based NOS
with a distributed control architecture designed to operate over
Wide Area Networks (WANs).
Furthermore, Cisco has recently developed the one Platform Kit
(onePK)~\cite{onepk}, which consists of a set of
Application Program Interfaces (APIs)
that allow the network applications to control Cisco network devices
without a command line interface.
The onePK libraries act as an SBI for Cisco ONE controllers and
are based on C and Java compilers.

\subsubsection{Application Layer}
The application layer comprises network applications and services
that utilize the control plane to realize
network functions over the physical or virtual infrastructure.
Examples of network applications include
network topology discovery, provisioning, and fault restoration.
The SDN controller presents an abstracted view of the
network to the SDN applications to facilitate the realization of
application functionalities.
The applications can also include higher levels of network management,
such as network data analytics, or specialized functions requiring processing
in large data centers. For instance, the Central Office Re-architected as a
Data center (CORD)~\cite{cord} is an SDN application based
on ONOS~\cite{onos},
that implements the typical central office network functions, such as
optical line termination, as well as BaseBand Unit (BBU) and
Data Over Cable Interface (DOCSIS)~\cite{fel2001doc} processing as
virtualized software entities, i.e., as SDN applications.

\begin{figure}[t!]
	\centering 	
	\includegraphics[width=3.4in]{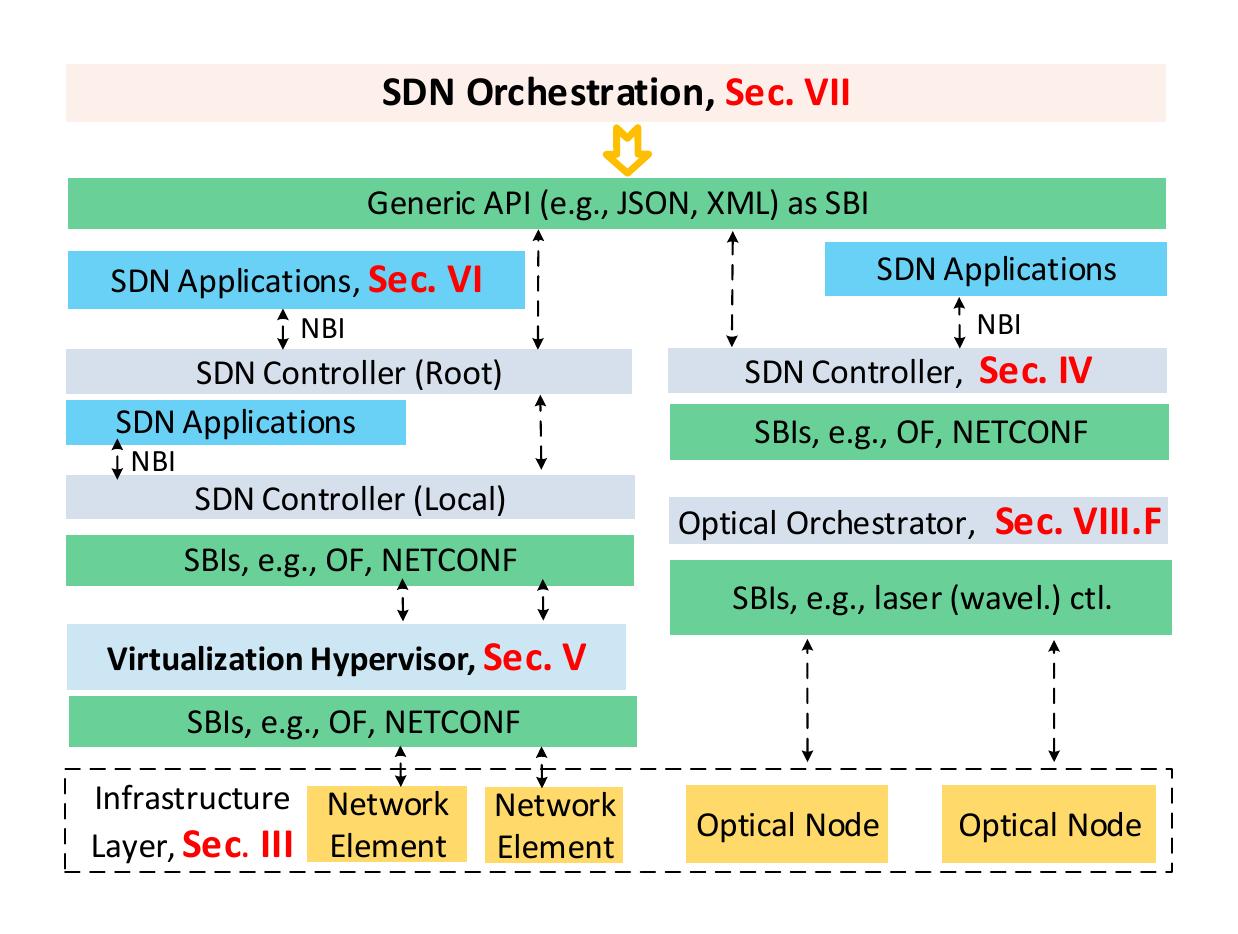}
\caption{Overview of SDN orchestrator and SDN controllers: The SDN
  orchestration coordinates and manages at a higher abstracted layer,
  above the SDN applications and SDN controllers.  SDN controllers,
  which may be in a hierarchy (see left part),
  implement the orchestrator decisions. A virtualization hypervisor may
  intercept the SouthBound Interfaces (SBIs) to create multiple virtual
  networks from a given physical network infrastructure.
  (The optical orchestrator on the right can be ignored
  for now and will be addressed in Section~\ref{multilayer:fut}.)}
	\label{fig_control_orch}
\end{figure}
\subsubsection{Orchestration Layer}  \label{intro:orch:sec}
Although the orchestration layer is commonly not considered one of the
main SDN architectural layers illustrated in Fig.~\ref{fig:sdnlayers},
as SDN systems become more complex, orchestration becomes increasingly
important. We introduce therefore the orchestration layer as
an important SDN architectural layer in this background section.
Typically, an SDN orchestrator is the entity that coordinates
software modules within a single SDN controller,
a hierarchical structure of multiple SDN controllers, or a set of
multiple SDN controllers in a ``flat'' arrangement (i.e., without
a hierarchy) as illustrated in Fig.~\ref{fig_control_orch}.
An SDN controller in contrast
can be viewed as a logically centralized single
control entity. This logically centralized single control entity appears as the
directly controlling entity to the network elements.
The SDN controller is responsible for signaling the control actions or rules
that are typically predefined (e.g., through OpenFlow) to the network elements.
In contrast, the SDN orchestrator makes control
decisions that are generally not predefined.
More specifically, the SDN orchestrator could make an automated decision
with the help of SDN applications or seek a manual
recommendation from user inputs; therefore,
results are generally not predefined. These
orchestrator decisions (actions/configurations)
are then delegated via the SDN controllers
and the SBIs to the network elements.

Intuitively speaking, SDN orchestration can be viewed as a distinct
abstracted (higher) layer for coordination and management that is
positioned above the SDN control and application layers.  Therefore,
we generalize the term SDN orchestrator as an entity that realizes a
wider, more general (more encompassing) network functionality as
compared to the SDN controllers. For instance, a cloud SDN
orchestrator can instantiate and tear down Virtual Machines (VMs)
according to the cloud
workload, i.e., make decisions that span across multiple network
domains and layers.  In contrast, SDN controllers realize more
specific network functions, such as routing and path computation.

\subsection{SDN Interfaces}  \label{sdnint:sec}
\subsubsection{Northbound Interfaces (NBIs)}
A logical interface that interconnects the SDN controller and a
software entity operating at the application layer is commonly
referred to as a NorthBound Interface (NBI), or as Application-Controller Plane
Interface (A-CPI).

\paragraph{REST}
REpresentational State Transfer (REST)~\cite{REST15} is generally defined as
a software architectural style that supports flexibility, interoperability,
and scalability.
In the context of the SDN NBI, REST is commonly defined as
an API that meets the REST architectural style~\cite{LiChou2016},
i.e., is a so-called RESTful API:
\begin{itemize}
\item Client-Sever: Two software entities should follow the
  client-server model.  In SDN, a controller can be a server and the
  application can be the client.  This allows multiple
  heterogeneous SDN applications to coexist and operate over a common
  SDN controller.
\item Stateless: The client is responsible for managing all the states and
the server acts upon the client's request.
In SDN, the applications collect and maintain the states of the network, while
the controller follows the instructions from the applications.
\item Caching: The client has to support the temporary local storage
of information such that interactions between the
client and server are reduced so as to improve performance and scalability.
\item Uniform/Interface Contract:  An overarching technical
interface must be followed across all services using the REST API.
For example, the same data format, such as Java Script Object Notation (JSON)
or eXtended Markup Language (XML), has to be followed for all
interactions sharing the common interface.
\item Layered System: In a multilayered architectural solution,
  the interface should only be concerned with the next immediate node
  and not beyond. Thus, allowing more layers to be inserted, modified,
  or removed without affecting the rest of the system.
\end{itemize}

\subsubsection{Southbound Interfaces (SBIs)}
A logical interface that interconnects the SDN controller and the network
element operating on the infrastructure layer (data plane) is commonly
referred to as a SouthBound Interface (SBI), or as the
Data-Controller Plane Interface (D-CPI).
Although a higher level connection, such as a UDP or TCP connection, is
sufficient for enabling the communication between two entities of the SDN
architecture, e.g., the controller and the network elements,
specific SBI protocols have been proposed.
These SBI protocols are typically not interoperable
and thus are limited to work with SBI protocol-specific network elements
(e.g., an OpenFlow switch does not work with the NETCONF protocol).

\paragraph{OpenFlow Protocol}
The interaction between an OpenFlow switching element
(data plane) and an OpenFlow controller (control plane) is carried
out through the OpenFlow protocol~\cite{keo2008of,LaKR14}. This SBI (or D-CPI)
is therefore also sometimes referred to as the OpenFlow control channel.
SDN mainly operates through packet flows that are identified through
matches on prescribed packet fields that are specified in the
OpenFlow protocol specification. For matched packets, SDN switches
then take prescribed actions, e.g., process the flow's packets in a
particular way, such as dropping the packet, duplicating it on a different
port or modifying the header information.

\paragraph{Path Computation Element Protocol (PCEP)}
The PCEP enables communication between the Path Computation Client (PCC) of
the network elements and the Path Computation Element (PCE) residing within
the controller. The PCE centrally computes the paths based on constraints
received from the network elements. Computed paths are then forwarded to the
individual network elements through the PCEP protocol~\cite{rfc4655,rfc5440}.

\paragraph{Network Configuration (NETCONF) Protocol}
The NETCONF protocol~\cite{rfc6241}
provides mechanisms to configure, modify, and delete configurations
on a network device.
Configuration of the data and protocol messages are encoded in the NETCONF
protocol using an eXtensible Markup Language (XML).
Remote procedure calls are used to realize the NETCONF protocol
operations. Therefore, only devices that are enabled
with required remote procedure calls allow the NETCONF protocol to remotely
modify device configurations.

\paragraph{Border Gateway Protocol Link State Distribution (BGP-LS) Protocol}
The central controller needs a topology information database,
also known as Traffic Engineering Database (TED), for
optimized end-to-end path computation.
The controller has to request the information for building the TED,
such as topology and bandwidth utilization, via the SBIs
from the network elements.
This information can be gathered by a BGP extension,
which is referred to as BGP-LS.

\subsection{Network Virtualization} \label{bgnetvit:sec}
Analogously to the virtualization of computing
resources~\cite{gold1974sur,Douglis2013a},
network virtualization abstracts the underlying physical network
infrastructure so that one or multiple virtual networks
can operate on a given physical
network~\cite{bel2012res,duan2012sur,fis2013vir,han2015net,leo2003vir,mij2015net,pen2015gue,Jain2013,wan2013net}.
Virtual networks can span over a
single or multiple physical infrastructures (e.g., geographically
separated WAN segments).
Network Virtualization (NV) can flexibly create independent virtual
networks (slices) for distinct
users over a given physical infrastructure.  Each network slice
can be created with prescribed resource allocations.
When no longer required, a
slice can be deleted, freeing up the reserved physical resources.

Network hypervisors~\cite{Khan2012,she2009flo}
are the network elements that abstract the
physical network infrastructure (including network elements,
communication links, and control functions) into logically isolated
virtual network slices.
In particular, in the case of an underlying physical SDN network,
an SDN hypervisor can create multiple isolated virtual SDN
networks~\cite{ble2015sur,dru2013sca}.
Through hypervisors, NV supports the implementation of a wide
range of network services belonging to the link and network protocol
layers (L2 and L3), such as switching and routing.
Additionally, virtualized infrastructures can also support
higher layer services, such as load-balancing of servers and firewalls.
The implementation of such higher layer services in a virtualized environment
is commonly referred to as Network Function Virtualization
(NFV)~\cite{haw2014nfv,LiChen2015,lin2016dem,mat2015tow,ye2016joi}.
NFV can be viewed as a special case of NV in which network
functions, such as address translation and intrusion detection functions,
are implemented in a virtualized environment. That is,
the virtualized functions are implemented in the form of software
entities (modules) running on a data center (DC) or
the cloud~\cite{mij2015net}.
In contrast, the term NV emphasizes the virtualization of
the network resources, such as communication links and network nodes.

\subsection{Optical Networking Background}   \label{bg_access:sec}

\subsubsection{Optical Switching Paradigms}
Optical networks are networks that either maintain signals in the
optical domain or at least utilize transmission channels that carry
signals in the optical domain. In optical networks that
maintain signals in the optical domain, switching can be
performed at the \textit{circuit, packet, or burst} granularities.

\paragraph{Circuit Switching}
Optical \textit{circuit} switching can be performed in space,
waveband, wavelength, or time. The optical spectrum is divided into
wavelengths either on a fixed wavelength grid or on a flexible
wavelength grid. Spectrally adjacent wavelengths can be coalesced into
wavebands. The fixed wavelength grid standard (ITU-T G.694.1)
specifies specific center frequencies that are either 12.5~GHz, 25~GHz,
50~GHz, or 100~GHz apart. The flexible DWDM grid (flexi-grid) standard (ITU-T
G.694.1)~\cite{gon2015opt,jue2014sof,tom2014tut,zha2013surflexi}
allows the center frequency to be any multiple of 6.25~GHz
away from 193.1~THz and the spectral width to be any multiple of
12.5~GHz. Elastic Optical Networks (EONs)~\cite{cha2015rou,tal2014spe,yu2014spe}
that take advantage of the flexible
grid can make more efficient use of the optical spectrum but can cause
spectral fragmentation, as lightpaths are set up and torn down, the spectral
fragmentation counteracts the more efficient spectrum
utilization~\cite{Ger2012}.

\paragraph{Packet Switching}
Optical \textit{packet} switching performs packet-by-packet switching
using header fields in the optical domain as much as possible. An
all-optical packet switch requires~\cite{ram2009opt}:
\begin{itemize}
  \item Optical synchronization, demultiplexing, and multiplexing
  \item Optical packet forwarding table computation
  \item Optical packet forwarding table lookup
  \item Optical switch fabric
  \item Optical buffering
\end{itemize}
Optical packet switches typically relegate some of these design
elements to the electrical domain. Most commonly the packet forwarding
table computation and lookup is performed electrically. When there is
contention for a destination port, a packet needs to be buffered
optically, this buffering can be accomplished with rather impractical
fiber delay lines. Fiber delay lines are fiber optic cables whose lengths are
configured to provide a certain time delay of the optical signal; e.g.,
100 meters of fiber provides 500~ns of delay. An alternative to
buffering is to either drop the packet or to use deflection routing,
whereby a packet is routed to a different output that may or may not
lead to the desired destination.

\paragraph{Burst Switching}
Optical \textit{burst} switching alleviates the requirements of
optical packet forwarding table computation, forwarding table lookup,
as well as buffering while accommodating bursty traffic that would
lead to poor utilization of optical circuits. In essence, it permits
the rapid establishment of short-lived optical circuits to support the
transfer of one or more packets coalesced into a burst. A control
packet is sent through the network that establishes the lightpath for
the burst and then the burst is transmitted on the short-lived circuit
with no packet lookup or buffering required along the path~\cite{ram2009opt}.
Since the circuit is only established for the length of the burst, network
resources are not wasted during idle periods. To avoid any buffering
of the burst in the optical network, the burst transmission can begin
once the lightpath establishment has been confirmed (tell-and-wait) or
a short time period after the control packet is sent
(just-enough-time). \textit{Note}: Sending the burst immediately after
the control packet (tell-and-go) would require some buffering of the
optical burst at the switching nodes.

\subsubsection{Optical Network Structure}  \label{optnetstruct:sec}
Optical networks are typically structured into three main tiers, namely
access networks, metropolitan (metro) area networks,
and backbone (core) networks~\cite{sim2014opt}.

\paragraph{Access Networks}
In the area of optical access networks~\cite{for2015nex}, so-called Passive
Optical Networks (PONs), in particular, Ethernet PONs (EPONs) and
Gigabit PONs (GPONs)~\cite{haj2006epo,sku2009com},
have been widely studied.
A PON has typically an inverse tree structure with a central
Optical Line Terminal (OLT) connecting multiple distributed Optical
Network Units (ONUs; also referred to as Optical Network Terminals, ONTs)
to metro networks.
In the downstream (OLT to ONUs) direction, the OLT broadcasts transmissions.
However, in the upstream (ONUs to OLT) direction, the transmissions of the
distributed ONUs need to be coordinated to avoid collisions on the
shared upstream wavelength channel.
Typically, a cyclic polling based Medium Access Control (MAC) protocol,
e.g., based on the MultiPoint Control Protocol (MPCP, IEEE 802.3ah),
is employed.
The ONUs report their bandwidth demands to the OLT and the OLT then
assigns upstream transmission windows according to a Dynamic
Bandwidth Allocation (DBA)
algorithm~\cite{kan2012ban,mcg2010sho,mcg2012inv,zhe2009sur}.
Conventional PONs cover distances up to 20~km, while so-called
Long-Reach (LR) PONs cover distances up to
around 100~km~\cite{mer2013off,nag2016n,son2010lon}.

Recently, hybrid access networks that combine multiple transmission
media, such as Fiber-Wireless (FiWi)
networks~\cite{aur2014fiw,gha2011fib,liu2016new,sar2015arc,tsa2011sur} and
PON-DSL networks~\cite{gur2014pon}, have been explored to take
advantage of the respective strengths of the different transmission
media.

\paragraph{Networks Connected to Access Networks}
Optical access networks provide Internet connectivity for a wide range
of peripheral networks. Residential (home) wired or wireless
local area networks~\cite{che2014sur}
typically interconnect individual end devices (hosts) in a home or small
business and may connect directly with an optical access network.
Cellular wireless networks provide Internet access to a wide range of
mobile devices~\cite{cap2013dow,dam2011sur,sch2013pus}.
Specialized cellular backhaul
networks~\cite{LiPCYW14,LiZZW13,park2014fro,PeWLP15,raz2013bri,tip2011evo,YaLJS13} relay the traffic
to/from base stations of wireless cellular
networks to either wireless access
networks~\cite{aky2005wir,alo2012sur,ben2012wir,kur2007sur,pat2011sur,vij2013dis} or optical access networks.
Moreover, optical access networks are often employed to connect
Data Center (DC) networks to the Internet. DC networks interconnect highly
specialized server units
  that process and store large data amounts with specialized
networking technologies~\cite{cai2013sur,kac2012sur,sam2016sof,yan2016sud,zha2013sur}.
Data centers are
typically employed to provide the so-called  ``cloud'' services for
 commercial and social media applications.

\paragraph{Metropolitan Area Networks}
Optical Metropolitan (metro) Area Networks (MANs) interconnect the optical
access networks in a metropolitan area with each other and with
wide-area (backbone, core) networks.
MANs have typically a ring or star
topology~\cite{bia2013cos,bia2015sho,cha2013tow,mai2003hyb,rot2013rou,sch2003wav}
and commonly employ optical networking technologies.

\paragraph{Backbone Networks}
Optical backbone (wide area) networks interconnect the individual MANs
on a national
or international scale. Backbone networks have typically a mesh structure
and employ very high speed optical transmission links.
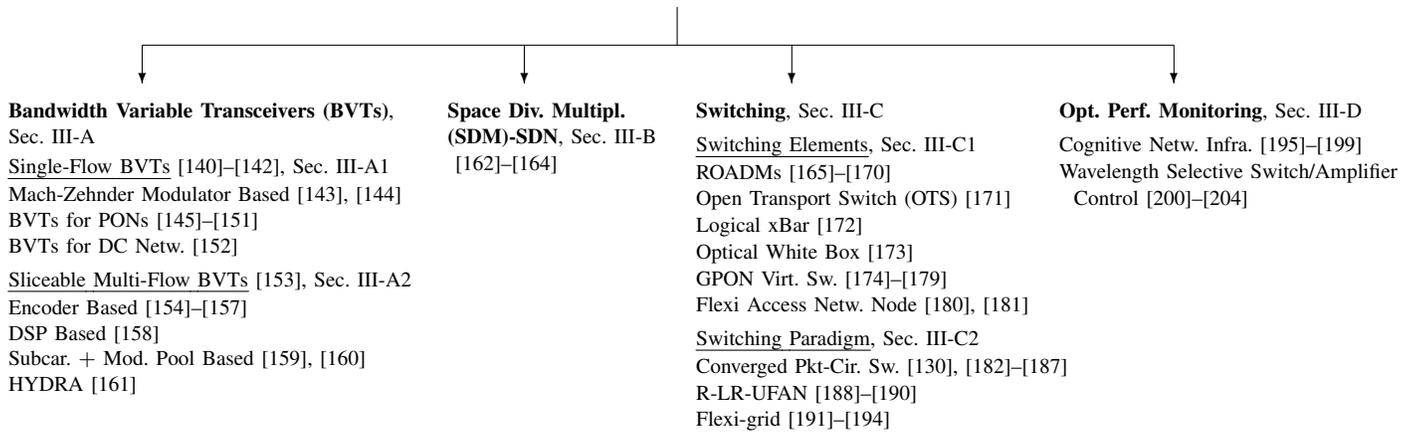
\begin{figure*}[t!]
\footnotesize
\setlength{\unitlength}{0.10in} 
\centering
\begin{picture}(40,33)
\put(4,33){\textbf{SDN Controlled Photonic Communication Infrastructure Layer, Sec.~\ref{sdninfra:sec}}}
\put(-9,30){\line(1,0){54}}
\put(19,30){\line(0,1){2}}
\put(-9,30){\vector(0,-1){2}}

\put(-16,27){\makebox(0,0)[lt]{\shortstack[l]{			
\textbf{Bandwidth Variable Transceivers (BVTs)},\\ Sec.~\ref{transc:sec}	\\	\\ 		
\uline{Single-Flow BVTs}~\cite{aut2013eva,ElA12,gri2010fle}, Sec.~\ref{SF-BVT:sec}\\
Mach-Zehnder Modulator Based~\cite{Choi2013,Liu2013c}\\
BVTs for PONs~\cite{LaAVKS14,IiSSK12,yeh2010usi,yu2008cen,VaBPF13,bol2014dig,Bolea2015a}\\
BVTs for DC Netw.~\cite{Malacarne2014} \\ \\
\uline{Sliceable Multi-Flow BVTs}~\cite{jin2012mul}, Sec.~\ref{MF-BVT:sec}	\\				
Encoder Based~\cite{Sambo2014a,sam2015nex,sam2014sli,cug2016tow} \\					
DSP Based~\cite{Moreolo2016} \\					
Subcar. $+$ Mod. Pool Based~\cite{Ou2016,ou2015onl} \\					
HYDRA~\cite{mat2015hyd}  
}}}

\put(11,30){\vector(0,-1){2}}

\put(7,27){\makebox(0,0)[lt]{\shortstack[l]{			
\textbf{Space Div. Multipl.} \\
\textbf{(SDM)-SDN}, Sec.~\ref{SDM_SDN:sec} \\
\cite{ama2013ful,ama2014sof,Galve2016}
}}}

\put(25,30){\vector(0,-1){2}}
\put(20,27){\makebox(0,0)[lt]{\shortstack[l]{						
\textbf{Switching}, Sec.~\ref{infra_sw:sec} \\ \\ 				
\uline{Switching Elements}, Sec.~\ref{sw_elem:sec}	\\
ROADMs~\cite{Co13,ama2013int,rof2013all,you2013eng,Way2013wav,Garrich2015}	\\			
Open Transport Switch (OTS)~\cite{SaSPL13}	\\					
Logical xBar~\cite{PaSMK13}	\\					
Optical White Box~\cite{Nejabati2015}	\\					
GPON Virt. Sw.~\cite{Lee2016,gu2014sof,Amokrane2014,amo2015dyn,Yeh2015,gu2016eff}	\\				
Flexi Access Netw. Node~\cite{FoG13,kon2015sdn} \\ \\
\uline{Switching Paradigm}, Sec.~\ref{ws_para:sec}	\\			
Converged Pkt-Cir. Sw.~\cite{DaPM09,DaPMSG10,VeBB13,AzNEJ11,ShJKG12,kac2012sur,CeLR13}	\\					
R-LR-UFAN~\cite{yin2013ult,ShYD14,Yin2015}	\\
Flexi-grid~\cite{Cv13,CvTJSM14,OlSCH13,ZhZYY13}	
}}}	
\put(45,30){\vector(0,-1){2}}
\put(39, 27){\makebox(0,0)[lt]{\shortstack[l]{					
\textbf{Opt. Perf. Monitoring}, Sec.~\ref{opm:sec} \\ \\
Cognitive Netw. Infra.~\cite{MiDJF13,cab2014cog,dur2016exp,Oliveira2015,Giglio2015} \\
Wavelength Selective Switch/Amplifier \\ \ \ Control~\cite{Moura2015,mou2016cog,pao2015sup,Carvalho2015,Wang2015f}
}}}	
\end{picture}
\vspace{-2.3cm}
\caption{Classification of physical infrastructure layer SDON studies. }
\label{infra_class:fig}
\end{figure*}
\section{SDN Controlled Photonic Communication Infrastructure Layer}
\label{sdninfra:sec}
This section surveys mechanisms for controlling physical layer
aspects of the optical (photonic) communication infrastructure through SDN.
Enabling the SDN control down to the photonic level operation of
optical communications allows for flexible adaptation of the
photonic components supporting optical networking
functionalities~\cite{ChNS13,gri2013ext,jin2013vir,RoD08}.
As illustrated in Fig.~\ref{infra_class:fig},
this section first surveys transmitters and receivers (collectively
referred to as transceivers or transponders) that permit SDN control of the
optical signal transmission characteristics, such as modulation format.
We also survey SDN controlled space division
multiplexing (SDM), which provides an emerging avenue for highly
efficient optical transmissions.
Then, we survey SDN controlled optical switching, covering first
switching elements and then overall switching paradigms, such as converged
packet and circuit switching.
Finally, we survey cognitive photonic communication infrastructures that
monitor the optical signal quality. The optical signal quality
information can be used to dynamically control the transceivers
as well as the filters in switching elements.

\subsection{Transceivers}  \label{transc:sec}
Software defined optical transceivers are optical transmitters and
receivers that can be flexibly configured by SDN to transmit or
receive a wide range of optical signals~\cite{Hillerkuss2016}.
Generally, software defined optical transceivers vary the modulation
format~\cite{win2006adv} of the transmitted optical signal by
adjusting the transmitter and receiver operation through Digital
Signal Processing (DSP)
techniques~\cite{cha2014adv,sch2010rea,yos2013dsp}. These
transceivers have evolved in recent years from Bandwidth Variable
Transceivers (BVTs) generating a single signal flow to sliceable
multi-flow BVTs. Single-flow BVTs permit SDN control to adjust the
transmission bandwidth of the single generated signal flow. In
contrast, sliceable multi-flow BVTs allow for the independent SDN
control of multiple communication traffic flows generated by a
single BVT.

\subsubsection{Single-Flow Bandwidth Variable Transceivers (BVTs)}
\label{SF-BVT:sec} Software defined optical transceivers have
initially been examined in the context of adjusting a single optical
signal flow for flexible WDM
networking~\cite{aut2013eva,ElA12,gri2010fle}. The goal has been to
make the photonic transmission characteristics of a given
transmitter fully programmable. We proceed to review a
representative single-flow BVT design for general optical mesh
networks in detail and then summarize related single-flow BVTs for
PONs and data center networks.

\paragraph{Mach-Zehnder Modulator Based Flexible Transmitter}
Choi and Liu et al.~\cite{Choi2013,Liu2013c} have demonstrated a
flexible transmitter based on Mach-Zehnder Modulators
(MZMs)~\cite{bar2003wid} and a corresponding flexible receiver for
SDN control in a general mesh network. The flexible transceiver
employs a single dual-drive MZM that is fed by two binary electric
signals as well as a parallel arrangement of two MZMs which are fed
by two additional electrical signals. Through adjusting the direct
current bias voltages and amplitudes of drive signals the
combination of MZMs can vary the amplitude and phase of the
generated optical signal~\cite{cho2012ber}. Thus, modulation formats
ranging from Binary Phase Shift Keying (BPSK) to Quadrature Phase
Shift Keying (QPSK) as well as 8 and 16 quadrature amplitude
modulation~\cite{win2006adv} can be generated. The amplitudes and
bias voltages of the drive signals can be signaled through an SDN OpenFlow
control plane to achieve the different modulation formats. The
corresponding flexible receiver consists of a polarization filter
that feeds four parallel photodetectors, each followed by an
Analog-to-Digital Converter (ADC). The outputs of the four parallel
ADCs are then processed with DSP techniques to automatically
(without SDN control) detect the modulation format. Experiments
in~\cite{Choi2013,Liu2013c} have evaluated the bit error rates and
transmission capacities of the different modulation formats and have
demonstrated the SDN control.

\paragraph{Single-Flow BVTs for PONs}
Flexible optical networking with real-time bandwidth adjustments
is also highly desirable for PON access and metro networks,
albeit the BVT technologies for access and metro networks should
have low cost and complexity~\cite{LaAVKS14}.
Iiyama et al.~\cite{IiSSK12} have developed a DSP based approach
that employs SDN to coordinate the downstream PON transmission of
On-Off Keying (OOK) modulation~\cite{yeh2010usi} and
Quadrature Amplitude Modulation (QAM)~\cite{yu2008cen} signals.
The OOK-QAM-SDN scheme involves a
novel multiplexing method, wherein all the data
are simultaneously sent from the OLT to the ONUs and the ONUs filter
the data they need.
The experimental setup in~\cite{IiSSK12} also demonstrated
digital software ONUs that concurrently transmit data by exploiting
the coexistence of OOK and QAM.
The OOK-QAM-SDN evaluations demonstrated the control of the receiving
sensitivity which is very useful for a wide range of transmission environments.

In a related study, Vacondio et al.~\cite{VaBPF13} have examined
Software-Defined Coherent Transponders (SDCT)
for TDMA PON access networks.
The proposed SDCT digitally processes the burst transmissions to
achieve improved burst mode transmissions according to the distance of a
user from the OLT.
The performance results indicate that the proposed flexible
approach more than doubles the average transmission capacity
per user compared to  a static approach.

\begin{figure*}[t]
    \centering
    \vspace{-.1cm}
    \includegraphics[width=6in]{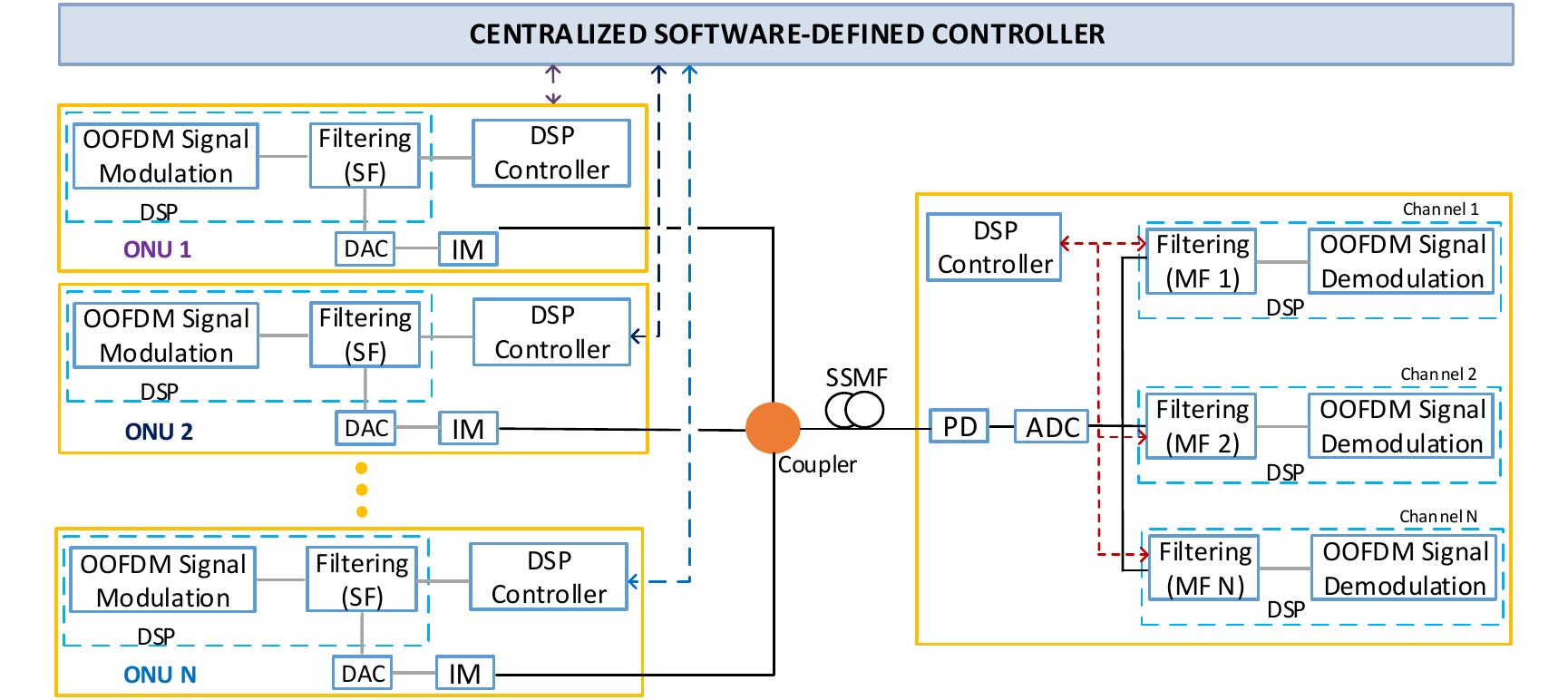}
    \vspace{-.1cm}
    \caption{Illustration of DSP reconfigurable ONU and OLT
  designs~\cite{Bolea2015a}:
Each ONU passes the electrical
Optical OFDM signal~\cite{bol2014dig} through a Shaping Filter (SF)
that is SDN-configured by the DSP controller, followed by
a Digital-to-Analog Converter (DAC) and Intensity Modulator (IM) to generate
the optical signal. The centralized SDN controller configures the
corresponding OLT Matching Filter (MF) and ensures that all ONU filters
are orthogonal.}
    \vspace{-.1cm}
    \label{fig:bol2015}
\end{figure*}
Bolea et al.~\cite{bol2014dig,Bolea2015a} have recently
developed low-complexity DSP reconfigurable ONU and OLT designs for
SDN-controlled PON communication.
The proposed communication is based on carrierless amplitude and phase
modulation~\cite{rod2011car} enhanced with optical Orthogonal
frequency Division Multiplexing (OFDM)~\cite{bol2014dig}.
The different OFDM channels are manipulated through DSP filtering.
As illustrated in Fig.~\ref{fig:bol2015}, the ONU consists of a DSP controller
that controls the filter coefficients of the shaping filter.
The filter output is then passed through a Digital-to-Analog Converter (DAC)
and intensity modulator for electric-optical conversion.
At the OLT, a photo diode converts the optical signal to an electrical signal,
which then passes through an Analog-to-Digital Converter (ADC).
The SDN controlled OLT DSP controller
sets the filter coefficients in the matching filter to
correspond to the filtering in the sending ONU.
The OLT DSP controller is also responsible for ensuring the
orthogonality of all the ONU filters in the PON.
The performance evaluations in~\cite{Bolea2015a} indicate that
the proposed DSP reconfigurable ONU and OLT system
achieves ONU signal bitrates around 3.7~Gb/s for eight ONUs transmitting
upstream over a 25~km PON.
The performance evaluations also illustrate that long DSP filter lengths,
which increase the filter complexity, improve performance.

\paragraph{Single-Flow BVTs for Data Center Networks}
Malacarne et al.~\cite{Malacarne2014} have developed a low-complexity
and low-cost bandwidth adaptable transmitter for data center
networking.
The transmitter can multiplex Amplitude Shift Keying (ASK),
specifically On-Off Keying (OOK), and Phase Shift Keying (PSK)
on the same optical carrier signal without any
special synchronization or temporal alignment mechanism.
In particular, the transmitter design~\cite{Malacarne2014}
uses the OOK electronic signal to drive a Mach-Zehnder Modulator (MZM)
that is fed by the optical pulse modulated signal.
SDN control can activate (or de-activate) the OOK signal stream, i.e.,
adapt from transmitting only the PSK signal to transmitting
both the PSK and OOK signal and thus providing a higher transmission bit rate.

\subsubsection{Sliceable Multi-Flow Bandwidth Variable Transceivers}
\label{MF-BVT:sec} Whereas the single-flow transceivers surveyed in
Section~\ref{SF-BVT:sec} generate a single optical signal flow,
parallelization efforts have resulted in multi-flow transceivers
(transponders)~\cite{jin2012mul}. Multi-flow transceivers can
generate multiple parallel optical signal flows and thus form the
infrastructure basis for network virtualization.

\paragraph{Encoder Based Programmable Transponder}
Sambo et al.~\cite{Sambo2014a,sam2015nex} have developed an
SDN-programmable bandwidth-variable multi-flow transmitter and
corresponding SDN-programmable multi-flow bandwidth variable
receiver, referred to jointly as programmable bandwidth-variable
transponder. The transmitter mainly consists of a programmable
encoder and multiple parallel Polarization-Multiplexing Quadrature
Phase Shift Keying (PM-QPSK~\cite{win2006adv}) laser transmitters,
whose signals are multiplexed by a coupler. The encoder is
SDN-controlled to implement Low-Density Parity-Check (LDPC)
coding~\cite{bon2011low} with different code rates. At the receiver,
the SDN control sets the local oscillators and LDPC decoder. The
developed transponder allows the setting of the number of
subcarriers, the subcarrier bitrate, and the LDPC coding rate
through SDN. Related frequency conversion and defragmentation issues
have been examined in \cite{Sambo2015}. In~\cite{sam2014sli}, a
low-cost version of the SDN programmable transponder with a
multiwavelength source has been developed. The multiwavelength
source is based on a micro-ring resonator~\cite{ras2009dem} that
generates multiple signal carriers with only a single laser.
Automated configuration procedures for the comprehensive set of
transmission parameters, including modulation format, coding configuration,
and carriers have been explored in~\cite{cug2016tow}.

\paragraph{DSP Based Sliceable BVT}
Moreolo et al.~\cite{Moreolo2016} have developed an SDN controlled
sliceable BVT based on adaptive Digital Signal Processing (DSP) of
multiple parallel signal subcarriers. Each subcarrier is fed by a
DSP module that configures the modulation format, including the bit
rate setting, and the power level of the carrier by adapting a gain
coefficient. The output of the DSP module is then passed through
digital to analog conversion that drives laser sources. The parallel
flows can be combined with a wavelength selective switch; the
combined flow can be sliced into multiple distinct sub-flows for
distinct destinations. The functionality of the developed DSP based
BVT has been verified for a metropolitan area network with links
reaching up to 150~km.

\begin{figure}[t]
    \centering
    \vspace{-.1cm}
    \includegraphics[width=3.2in]{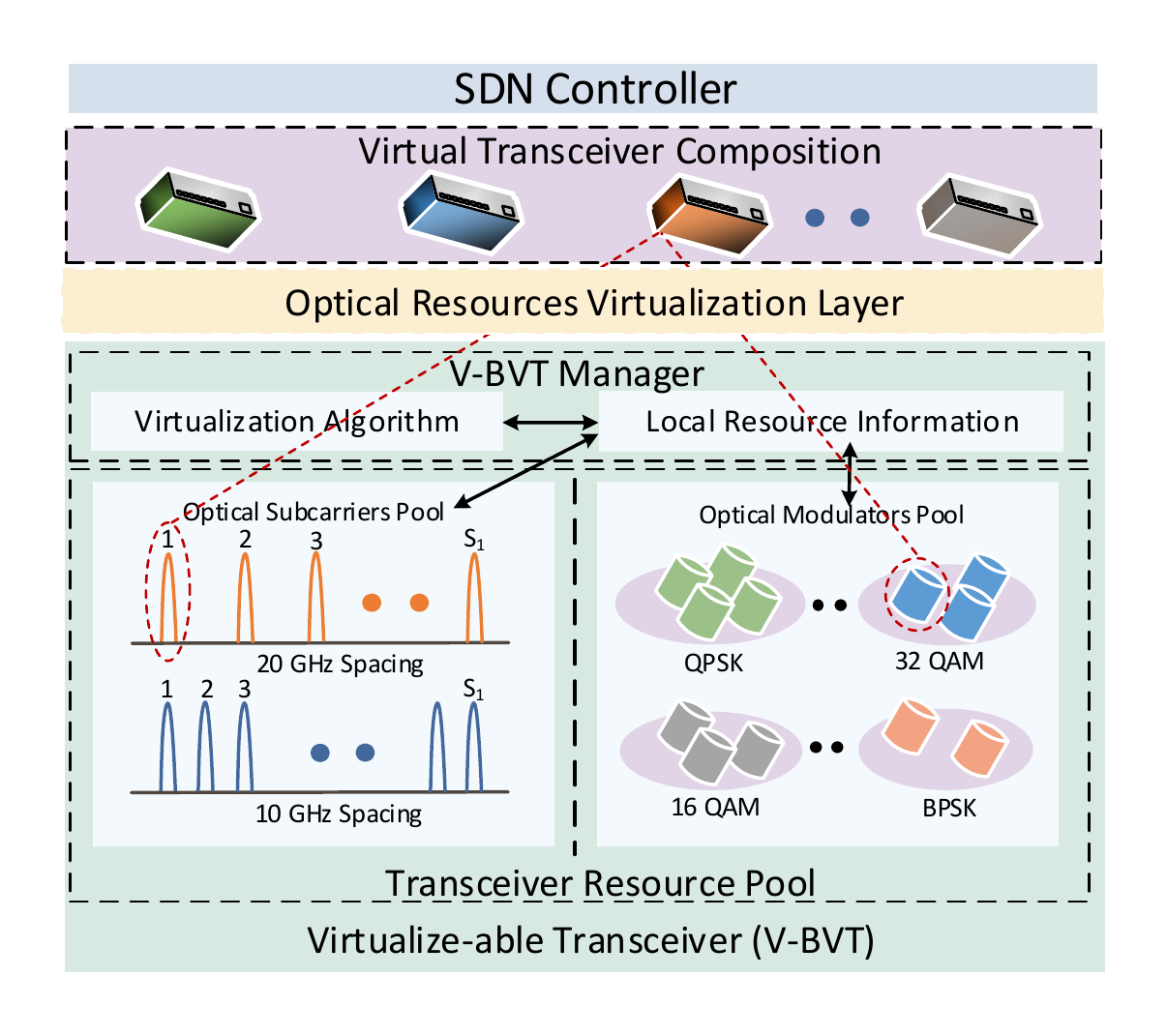}
    \vspace{-.1cm}
    \caption{Illustration of Subcarrier and Modulator Pool Based
Virtualizable Bandwidth Variable Transceiver (V-BVT)~\cite{Ou2016}:
Through SDN control, the V-BVT Manager composes virtual transceivers
by combining subcarriers from the optical subcarriers pool with
modulators from the optical modulators pool.}
    \vspace{-.1cm}
    \label{fig:ou2016}
\end{figure}
\paragraph{Subcarrier and Modulator Pool Based Virtualizable BVT}
Ou et al.~\cite{Ou2016,ou2015onl} have developed a Virtualizable BVT
(V-BVT) based on a combination of an optical subcarriers pool
with an independent optical modulators pool, as illustrated in
Fig.~\ref{fig:ou2016}.
The emphasis of the design is on implementing
Virtual Optical Networks (VONs) at the transceiver level.
The optical subcarriers pool contains multiple optical carriers,
whereby channel spacing and central frequency (wavelength channel)
can be selected.
The optical modulators pool contains optical modulators that can
generate a wide variety of modulation formats.
The SDN control interacts with a V-BVT Manager that implements a
virtualization algorithm. The virtualization algorithm
generates a transceiver slice by combining a particular
set of subcarriers (with specific number of subcarriers, channel
spacing, and central frequencies) from the optical subcarriers pool
with a particular modulation (with specific number of modulators and
modulation formats) from the optical modulators pool.
The evaluations in~\cite{Ou2016} have evaluated the proposed V-BVT in
a network testbed with path lengths up to 200~km with 20~GHz channel
spacing and a variety of modulation formats, including
BPSK as well as 16QAM and 32QAM.

\paragraph{S-BVT Based Hybrid Long-Reach Fiber Access Network (HYDRA)}
HYDRA~\cite{mat2015hyd} is a novel hybrid long-reach fiber access
network architecture based on sliceable BVTs. HYDRA supports
low-cost end-user ONUs through an Active Remote Node (ARN) that
directly connects via a distribution fiber segment, a passive remote
node, and a trunk fiber segment to the core (backbone) network,
bypassing the conventional metro network. The ARN is based on an SDN
controlled S-BVT to optimize the modulation format. With the
modulation format optimization, the ARN can optimize the
transmission capacity for the given distance (via the distribution
and trunk fiber segments) to the core network. The evaluations
in~\cite{mat2015hyd} demonstrate good bit error rate performance of
representative HYDRA scenarios with a 200~km trunk fiber segment and
distribution fiber lengths up to 100~km. In particular, distribution
fiber lengths up to around 70~km can be supported without Forward
Error Correction (FEC), whereas distribution fiber lengths above
70~km would require standard FEC. The consolidation of the access
and metro network infrastructure~\cite{wan2016mig} achieved through
the optimized S-BVT transmissions can significantly reduce the
network cost and power consumption.

\subsection{Space Division Multiplexing (SDM)-SDN}  \label{SDM_SDN:sec}
Amaya et al.~\cite{ama2013ful,ama2014sof} have
demonstrated SDN control of
Space Division Multiplexing (SDM)~\cite{ric2013spa} in optical networks.
More specifically, Amaya et al.~employ SDN to control the
physical layer so as to achieve a bandwidth-flexible and programmable
SDM optical network. The SDN control can perform network slicing,
resulting in sliceable superchannels.
A superchannel consists of multiple spatial carriers to support
dynamic bandwidth and QoS provisioning.

Galve et al.~\cite{Galve2016} have built on the flexible SDN controlled
SDM communication principles to develop a reconfigurable Radio
Access Network (RAN). The RAN connects the BaseBand processing Units (BBUs)
in a shared central office with the corresponding distributed
Remote Radio Heads (RRHs) located at Base Stations (BSs).
A multicore fiber operated with SDM~\cite{ric2013spa} connects the
RRHs to the BBUs in the central office.
Galve et al.~introduce a radio over fiber operation mode where
SDN controlled switching maps the subcarriers dynamically to
spatial output ports.
A complementary digitized radio over fiber operating mode maintains a
BBU pool. Virtual BBUs are dynamically allocated to the cores of
the SDM operated multicore fiber.

\subsection{SDN-Controlled Switching} \label{infra_sw:sec}
\subsubsection{Switching Elements}   \label{sw_elem:sec}
\paragraph{ROADM} \label{ROADMs:par}
The Reconfigurable Optical Add-Drop Multiplexer (ROADM) is an
important photonic switching device for optical networks. Through
wavelength selective optical switches, a ROADM can drop (or add) one
or multiple wavelength channels carrying optical data signals from
(to) a fiber without requiring the conversion of the optical signal
to electric signals~\cite{he2014sur}. The ROADM thus provides an
elementary switching functionality in the optical wavelength domain.
Initial ROADM based node architectures for cost-effectively
supporting flexible SDN networks have been presented
in~\cite{Co13}. Conventional ROADM networks have
typically statically configured wavelength channels that transport
traffic along a pre-configured route. Changes of wavelength channels
or routes in the statically configured networks incur presently high
operational costs due to required physical interventions and are
therefore typically avoided. New ROADM node designs allow changes of
wavelength channels and routes through a management control plane.
Due to these two flexibility dimensions (wavelength and route),
these new ROADM nodes are referred to as ``colorless'' and
``directionless''. First designs for such colorless and
directionless ROADM nodes have been outlined in~\cite{Co13} and
further elaborated in~\cite{ama2013int,rof2013all}.
In addition to the colorless and directionless properties, the
contentionless property has emerged for ROADMs~\cite{gri2010fle}.
Contentionless ROADM operation means that any port can be
routed on any wavelength (color) in any direction without causing
resource contention.
Designs for such Colorless-Directionless-Contentionless (CDC)
ROADMs have been proposed in~\cite{you2013eng,Way2013wav}.
In general, the ROADM designs consist of an express bank that
interconnects the input and output ports coming from/leading to other ROADMs,
and an add-drop bank that connects the express bank with the local
receivers for dropped wavelength channels or transmitters for added
wavelength channels. The recent designs have focused on
the add-drop bank and explored different arrangements
of wavelength selective switches and multicast switches to
provide add-drop bank functionality with the CDC
property~\cite{you2013eng,Way2013wav}.

Garrich et al.~\cite{Garrich2015} have recently designed
and demonstrated a CDC ROADM with an add-drop bank based on
an Optical Cross-Connect (OXC) backplane~\cite{wan2012mul}.
The OXC backplane allows for highly flexible add/drop configurations
implemented through SDN control.
The backplane based ROADM has been analytically
compared with prior designs based on
wavelength selective and multicast switches and has been
shown to achieve higher flexibility and lower losses.
An experimental evaluation has tested the backplane based ROADM for
a metropolitan area mesh network extending over 100~km
with an aggregate traffic load of  close to 9~Tb/s.

\paragraph{Open Transport Switch (OTS)}
The Open Transport Switch (OTS)~\cite{SaSPL13} is an OpenFlow-enabled
optical virtual switch design. The OTS design abstracts the details
of the underlying physical switching layer (which could be packet
switching or circuit switching) to a virtual switch element. The OTS
design introduces three agent modules (discovery, control, and data
plane) to interface with the physical switching hardware. These
agent modules are controlled from an SDN controller through extended
OpenFlow messages. Performance measurements for an example testbed
network setup indicate that the circuit path computation latencies
on the order of 2--3 s that can be reduced through faster processing
in the controller.

\paragraph{Logical xBar}
The logical xBar~\cite{PaSMK13} has been defined to represent a
programmable switch. An elementary (small) xBar could consist of a
single OpenFlow switch. Multiple small xBars can be recursively
merged to form a single large xBar with a single forwarding table.
The xBar concept envisions that xBars are the building blocks for
forming large networks. Moreover, labels based on SDN and MPLS are
envisioned for managing the xBar data plane forwarding.
The xBar concepts have been further advanced in the Orion
study~\cite{fu2014ori} to achieve low computational complexity of
the SDN control plane.

\paragraph{Optical White Box}
Nejabati et al.~\cite{Nejabati2015} have proposed an optical white
box switch design as a building block for a completely softwarized
optical network. The optical white box design combines a
programmable backplane with programmable switching node elements.
More specifically, the backplane consists of two slivers, namely an
optical backplane sliver and an electronic backplane sliver. These
slivers are set up to allow for flexible arbitrary connections
between the switch node elements. The switch node elements include
programmable interfaces that build on SDN-controlled BVTs (see
Section~\ref{transc:sec}), protocol agnostic switching, and DSP elements.
The protocol agnostic switching
element is envisioned to support both wavelength channel and time
slot switching in the optical backplane as well as programmable
switching with a high-speed packet processor in the electronic
backplane. The DSP elements support both the network processing and
the signal processing for executing a wide range of network
functions. A prototype of the optical white box has been built with
only a optical backplane sliver consisting of a $192 \times 192$
optical space switch. Experiments have indicated that the creation of
a virtual switching node with the OpenDayLight SDN controller takes
roughly 400~ms.

\paragraph{GPON Virtual Switch}
Lee et al.~\cite{Lee2016} have developed a GPON virtual switch
design that makes the GPON fully programmable similar to a
conventional OpenFlow switch. Preliminary steps towards the GPON
virtual switch design have been taken by Gu et al.~\cite{gu2014sof}
who developed components for SDN control of a PON in a data center
and Amokrane et al.~\cite{Amokrane2014,amo2015dyn} who developed a
module for mapping OpenFlow flow control requests into PON configuration
commands. Lee et al.~\cite{Lee2016} have expanded on this groundwork
to abstract the entire GPON into a virtual OpenFlow switch. More
specifically, Lee et al. have comprehensively designed a hardware
architecture and a software architecture to allow SDN control to
interface with the virtual GPON as if it were a standard OpenFlow switch.
The experimental performance evaluation of the designed GPON virtual
switch measured response times for flow entry modifications from an
ONU port (where a subscriber connects to the virtual GPON switch) to
an SDN external port around 0.6~ms, which compares to 0.2~ms for a
corresponding flow entry modification in a conventional OFsoftswitch
and 1.7~ms in a EdgeCore AS4600 switch. In a related study on SDN
controlled switching in a PON, Yeh et al.~\cite{Yeh2015} have
designed an ONU with an optical switch that selects OFDM subchannels
in a TWDM-PON. The switch in the ONU allows for flexible dynamic
adaption of the downstream bandwidth through SDN.
Gu et al.~\cite{gu2016eff} have examined the flexible SDN controlled
re-arrangement of ONUs to OLTs so as to efficiently support PON
service with network coding~\cite{bas2013net}.

\paragraph{Flexi Access Network Node}
A flexi-node for an access network that flexibly aggregates
traffic flows from a wide range of networks,
such as local area networks and base stations of wireless
networks has been proposed in~\cite{FoG13}.
The flexi-node design is motivated by the shortcomings of the
currently deployed core/metro
network architectures that attempt to consolidate the access and
metro networks.
This consolidation forces all traffic in the access network to
traverse the metro network, even if the traffic is destined to
destination nodes in the coverage area of an access network.
In contrast, the proposed flexi-node encompasses
electrical and optical forwarding capabilities that can be
controlled through SDN. The flexi-node can thus serve as an
effective aggregation node in access-metro networks.
Traffic that is destined to other nodes in the coverage area of
an access network can be sent directly to the access network.

Kondepu et al. have similarly presented an SDN based PON aggregation
node~\cite{kon2015sdn}.
In their architecture, multiple ONUs
communicate with the SDN controller within the aggregation node
to request the scheduling of upstream transmission resources.
ONUs are then serviced by multiple Optical Service Units (OSUs)
which exist
within the aggregation node alongside with the SDN controller.
OSUs are then configured by the controller based on Time and Wavelength
Division Multiplexed (TWDM) PON. The OSUs step between normal and sleep-mode
depending on the traffic loads, thus saving power.

\subsubsection{Switching Paradigms}  \label{ws_para:sec}
\paragraph{Converged Packet-Circuit Switching}
Hybrid packet-circuit optical network infrastructures controlled by SDN
have been explored in a few studies.
Das et al.~\cite{DaPM09} have described how to unify the control and
management of circuit- and packet-switched networks using OpenFlow.
Since packet- and circuit-switched networking are extensively employed
in optical networks, examining their integration is an important research
direction.
Das et al.~have given a high-level overview of a flow abstraction for
each type of switched network and a common control paradigm.
In their follow-up work, Das et al.~\cite{DaPMSG10}
have described how a packet and circuit switching
network can be implemented in the context of an OpenFlow-protocol based testbed. The testbed is a standard Ethernet network that could generally
be employed in any access network with Time Division Multiplexing (TDM).
Veisllari et al.~\cite{VeBB13} studied packet/circuit hybrid optical
long-haul metro access networks.
Although Veisllari et al.~indicated that SDN can be used for load balancing in
the proposed packet/circuit network, no detailed study of
such an SDN-based load balancing has been conducted in~\cite{VeBB13}.
Related switching paradigms that integrate SDN with Generalized
Multiple Protocol Label Switching (GMPLS)
have been examined in~\cite{AzNEJ11,ShJKG12},
while data center specific aspects have been surveyed in~\cite{kac2012sur}.

Cerroni et al.~\cite{CeLR13} have further developed the concept of
unifying circuit- and packet-switching networks with OpenFlow,
which was initiated by Das et al.~\cite{DaPM09,DaPMSG10}.
The unification is accomplished with SDN on the network layer and
can be used in core networks. Specifically, Cerroni et
al.~\cite{CeLR13} have described an extension of the OpenFlow flow
concept to support hybrid networks. OpenFlow message format
extensions to include matching rules and flow entries have also been
provided. The matching rules can represent different transport
functions, such as a channel on which a packet is received in
optical circuit-switched WDM networks, time slots in TDM networks,
or transport class services (such as guaranteed circuit service or
best effort packet service). Cerroni et al.~\cite{CeLR13} have
presented a testbed setup and reported performance results for
throughput (in bit/s and packets/s) to demonstrate the feasibility
of the proposed unified OpenFlow switching network.

\paragraph{R-LR-UFAN}  \label{R_LR_UFAN:sec}
The Reconfigurable Long-Reach UltraFlow Access Network
(R-LR-UFAN)~\cite{yin2013ult,ShYD14} provides flexible dual-mode transport
service based on either the Internet Protocol (IP) or
Optical Flow Switching (OFS).
OFS~\cite{chan2012opt} provides dedicated end-to-end network paths
through purely optical switching, i.e., there is no electronic
processing or buffering at intermediate network nodes. The R-LR-UFAN
architecture employs multiple feeder fibers to form subnets within
the network.
UltraFlow coexists alongside the conventional PON OLT and ONUs. The
R-LR-UFAN introduces new entities, namely the Optical Flow Network
Unit (OFNU) and the SDN-controlled Optical Flow Line Terminal
(OFLT). A Quasi-PAssive Reconfigurable (QPAR) node~\cite{Yin2015} is
introduced between the OFNU and OFLT. The QPAR node can re-route
intra PON traffic between OFNUs without having to pass through the
OLFTs. The optically rerouted intra-PON channels can be used for
communication between wireless base stations supporting inter cell
device-to-device communication. The testbed evaluations indicate
that for an intra-PON traffic ratio of 0.3, the QPAR strategy
achieves power savings up to 24\%.

\paragraph{Flexi-grid}
The principle of flexi-grid (elastic) optical
networking~\cite{cha2015rou,gon2015opt,jue2014sof,she2016sur,tal2014spe,tom2014tut,yu2014spe,zha2013surflexi}
has been explored in several SDN infrastructure studies.
Generally, flexi-grid networking strives to enhance the efficiency
of the optical transmissions by adapting physical (photonic)
transmission parameters, such as modulation format, symbol rate,
number and spacing of subcarrier wavelength channels, as well as the
ratio of forward error correction to payload.
Flexi-grid transmissions have become feasible with
high-capacity flexible transceivers.
Flexi-grid transmissions use narrower
frequency slots (e.g., 12.5~GHz) than classical Wavelength Division
Multiplexing (WDM, with typically 50~GHz frequency slots for WDM)
and can flexibly form optical transmission channels that span multiple
contiguous frequency slots.

Cvijetic~\cite{Cv13} has proposed a hierarchical flexi-grid infrastructure
for multiservice broadband optical access utilizing
centralized software-reconfigurable resource management and digital signal
processing. The proposed flexi-grid infrastructure incorporates mobile
backhaul, as well as SDN controlled transceivers~\ref{transc:sec}.
In follow-up work, Cvijetic et al.~\cite{CvTJSM14} have designed a
dynamic flexi-grid optical access and aggregation network.
They employ SDN to control tunable lasers in the OLT for flexible
downstream transmissions.
Flexi-grid wavelength selective switches are controlled through SDN
to dynamically tune the passband for the upstream transmissions
arriving at the OLT.
Cvijetic et al.~\cite{CvTJSM14} obtained good results for the upstream and
downstream bit error rate
and were able to provide 150~Mb/s per wireless network cell.

Oliveira et al.~\cite{OlSCH13} have demonstrated a testbed for a
Reconfigurable Flexible Optical Network (RFON), which was one of the
first physical layer SDN-based testbeds. The RFON testbed is
comprised of 4 ROADMs with flexi-grid Wavelength Selective Switching
(WSS) modules, optical amplifiers, optical channel monitors and
supervisor boards. The controller daemon implements a node
abstraction layer and provides configuration details for an overall
view of the network. Also, virtualization of the GMPLS control plane
with topology discovery and Traffic Engineering (TE)-link
instantiation have been incorporated.
Instead of using OpenFlow, the
RFON testbed uses the controller language YANG~\cite{Schonwalder2010}
to obtain the topology information and collect monitoring data  for the
lightpaths.

Zhao et al.~\cite{ZhZYY13} have presented an architecture with
OpenFlow-based optical interconnects for intra-data center networking
and OpenFlow-based flexi-grid optical networks for inter-data center
networking.
Zhao et al.~focus on the SDN benefits for inter-data center networking
with heterogeneous networks.
The proposed architecture includes a service controller, an IP controller,
and an optical controller based on the Father Network Operating System
(F-NOX)~\cite{gud2008nox,zha2013uni}.
The performance evaluations in~\cite{ZhZYY13} include results for
blocking probability, release latency, and bandwidth spectrum characteristics.

\subsection{Optical Performance Monitoring}  \label{opm:sec}
\label{sec:sdmonitoring}
\subsubsection{Cognitive Network Infrastructure}
A Cognitive Heterogeneous Reconfigurable Dynamic Optical Network (CHRON)
architecture has been outlined in~\cite{MiDJF13,cab2014cog,dur2016exp}.
CHRON senses the current network
conditions and adapts the network operation accordingly.
The three main components of CHRON are monitoring elements, software
adaptable elements, and cognitive processes. The monitoring elements
observe two main types of optical transmission impairments, namely
non-catastrophic impairments and catastrophic impairments.
Non-catastrophic impairments include the photonic impairments that
degrade the Optical Signal to Noise Ratio (OSNR), such as the
various forms of dispersion, cross-talk, and non-linear propagation
effects, but do not completely disrupt the communication. In
contrast, a catastrophic impairment, such as a fiber cut or
malfunctioning switch, can completely disrupt the communication.
Advances in optical performance monitoring allow for in-band OSNR
monitoring~\cite{dah2011opt,sui2010osn,sch2011osn,sai2012ban} at
midpoints in the communication path, e.g., at optical amplifiers and
ROADMs.

The cognitive processes involve the collection of the monitoring
information in the controller, executing control algorithms, and
instructing the software adaptable components to implement the
control decisions. SDN can provide the framework for implementing
these cognitive processes.
Two main types of software adaptable components have been considered
so far~\cite{Oliveira2015,Giglio2015}, namely control of transceivers and
control of wavelength selective switches/amplifiers.
For transceiver control,
the cognitive control adjusts the transmission parameters.
For instance, transmission bit
rates can be adjusted through varying the modulation format or the
number of signal carriers in multicarrier communication
(see Section~\ref{transc:sec}).

\subsubsection{Wavelength Selective Switch/Amplifier Control}
In general, ROADMs (see Section~\ref{ROADMs:par}) employ wavelength
selective switches based on filters to add or drop wavelength
channels for routing through an optical network. Detrimental
non-ideal filtering effects accumulate and impair the
OSNR~\cite{pao2015sup}. At the same time, Erbium Doped Fiber
Amplifiers (EDFAs)~\cite{zim2004amp} are widely deployed in optical
networks to boost optical signal power that has been depleted
through attenuation in fibers and ROADMs. However, depending on
their operating points, EDFAs can introduce significant noise. Moura
et al.~\cite{Moura2015,mou2016cog} have explored SDN based adaptation
strategies for EDFA operating points to increase the OSNR. In a
complementary study, Paolucci et al.~\cite{pao2015sup} have
exploited SDN control to reduce the detrimental filtering effects.
Paolucci group wavelength channels that jointly traverse a sequence
of filters at successive switching nodes. Instead of passing these
wavelength channels through individual (per-wavelength channel)
filters, the group of wavelength channels is jointly passed through
a superfilter that encompasses all grouped wavelength channels. This
joint filtering significantly improves the OSNR.

While the studies~\cite{Moura2015,mou2016cog,pao2015sup} have focused on
either the EDFA or the filters, Carvalho et al.~\cite{Carvalho2015} and
Wang et al.~\cite{Wang2015f} have jointly considered
the EDFA and filter control.
More specifically, the EDFA gain and the filter attenuation (and
signal equalization) profile were adapted to improve the OSNR.
Carvalho et al.~\cite{Carvalho2015} propose and evaluate a specific
joint EDFA and filter optimization approach that exploits the global
perspective of the SDN controller. The global optimization achieves
ONSR improvements close to 5~dB for a testbed consisting of four ROADMs
with 100~km fiber links.
Wang et al.~\cite{Wang2015f} explore different combinations of
EDFA gain control strategies and filter equalization strategies for a
simulated network with 14 nodes and 100~km fiber links.
They find mutual interactions between the EDFA gain control and the
filter equalization control as well as an additional wavelength assignment
module.
They conclude that global SDN control is highly useful for synchronizing
the EDFA gain and filter equalization in conjunction with wavelength
assignments so as to achieve improved OSNR.

\subsection{Infrastructure Layer: Summary and Discussion}
The research to date on the SDN controlled infrastructure layer has
resulted in a variety of SDN controlled transceivers as well as a few designs
of SDN controlled switching elements.
Moreover, the SDN control of switching paradigms and optical
performance monitoring have been examined.
The SDN infrastructure studies have paid close attention to the physical
(photonic) communication aspects. Principles of isolation of control plane and
data plane with the goals of simplifying network management and
making the networks more flexible have been explored.
The completed SDN infrastructure layer studies have indicated
that the SDN control of the infrastructure layer can reduce costs, facilitate
flexible reconfigurable resource management, increase utilizations,
and lower latency.
However, detailed comprehensive optimizations of the infrastructure components
and paradigms
that minimize capital and operational expenditures are an important
area for future research. Also, further refinements of the optical
components and switching paradigms are needed to ease the
deployment of SDONs and make the networks operating on
the SDON infrastructures more efficient.
Moreover, the cost reduction of implementations, easy adoption by network
providers, flexible upgrades to adopt new technologies, and reduced
complexity require thorough future research.

Most SDON infrastructure studies have
focused on a particular network component or networking aspect, e.g.,
a transceiver or the hybrid
packet-circuit switching paradigm, or a particular application context, e.g.,
data center networking. Future research should comprehensively
examine SDON infrastructure components and
paradigms to optimize their interactions
for a wide set of networking scenarios and application contexts.

The SDON infrastructure studies to date
have primarily focused on the optical transmission medium.
Future research should explore complementary infrastructure
components and paradigms to
support transmissions in hybrid fiber-wireless and other hybrid fiber-$X$
networks, such as
fiber-Digital Subscriber Line (DSL) or fiber-coax cable
networks~\cite{Fuentes2014,gur2014pon,luo2013act}.
Generally, the flexible SDN control can
be very advantageous for hybrid networks composed of heterogeneous
network segments. The OpenFlow protocol can facilitate the topology
abstraction of the heterogeneous physical transmission media, which
in turn facilitates control and optimization at the higher network
protocol layers.

\begin{figure*}[t!]
\footnotesize
\setlength{\unitlength}{0.10in} 
\centering
\begin{picture}(40,33)
\put(15,33){\textbf{SDN Control Layer, Sec.~\ref{sdnctl:sec}}}
\put(-10,30){\line(1,0){55}}
\put(21,30){\line(0,1){2}}
\put(15,30){\vector(0,-1){2}}
\put(-14,27){\makebox(0,0)[lt]{\shortstack[l]{			
\textbf{Ctl. of Infra. Comp.}, \\ \ Sec.~\ref{ctlinfra:sec}  \\ \\
		Transceiver Ctl.\\~\cite{YuZZG14,Chen2014,JiXia2014,liu2013field}    \\
		Circuit Sw. Ctl. \\ \cite{ChNS13,DaPMSG10,DaPM09}, \\
			\cite{DaPM09Jul,OFCircuitSwitch,ChNF13,Baik2014} \\
		Pkt. $+$ Burst Sw. Ctl.~\\ \cite{Cao2015,Harai2014,liu2013field}
		}}}
	\put(3,30){\vector(0,-1){2}}
	\put(-2.5,27){\makebox(0,0)[lt]{\shortstack[l]{			
              \textbf{Retro-fitting Devices},\\
                 Sec.~\ref{retfitctl:sec} \\
              \cite{ChNF13,ChNS13,Cao2015,liu2013field},\\
		\cite{Alvizu2014,LiTMG11,LiuTMGW11,LiCTM12,LiZTV12,ClSLT14}
	}}}
	\put(-10,30){\vector(0,-1){2}}
	\put(12,27){\makebox(0,0)[lt]{\shortstack[l]{			
              \textbf{Ctl. of Opt. Netw. Ops.}, \\
                  Sec.~\ref{ctlops:sec} \\ \\
		PON Ctl.~\cite{KhRS14,LeK15,PaP13,par2014fut,YaZWZ14,KaCTW13} \\
		Spectrum Defrag.~\\\cite{Ger2012,YuZZG14,Chen2014},\\
		\cite{Munoz2014a,Zhu2015b,Meloni2016} \\ \\
		\uline{Tandem Netw.}, Sec.~\ref{ctltan:sec} \\
		Metro$+$Access~\cite{wu2014glo,zha2015sof}	\\	
		Access$+$Wireless~\cite{boj2013adv,TaC14,Costa2015}	\\	
		Access$+$Metro$+$Core~\cite{SlR14} \\		
		DC~\cite{MaHSC14}	   \\	
		IoT~\cite{wan2015nov}
	}}}
	\put(30,30){\vector(0,-1){2}}		
	\put(25,27){\makebox(0,0)[lt]{\shortstack[l]{			
	      \textbf{Hybrid SDN-GMPLS}, \\
               Sec.~\ref{hybsdnctl:sec}  \\
		\cite{AzNEJ11,Alvizu2014,Munoz2014a}, \\
		\cite{LiCTMMM12,Casellas2013b}
	}}}
	\put(45,30){\vector(0,-1){2}}
	\put(39,27){\makebox(0,0)[lt]{\shortstack[l]{			
\textbf{Ctl. Perform.}, Sec.~\ref{cltperf:sec} \\ \\
		SDN vs. GMPLS~\cite{ChNF13}, \\
		\cite{LiTM12,ZhaoZYY13,CviAPJT13} \\
		Flow Setup Time~\cite{VeSBR14,liu2013field} \\
		Out of Band Ctl.~\cite{Sanchez2013} \\
		Clust. Ctl~\cite{Penna2014}
	}}}		
	\end{picture}
\vspace{-2.75cm}	
\caption{Classification of SDON control layer studies. }
                  \label{ctl_class:fig}
\end{figure*}
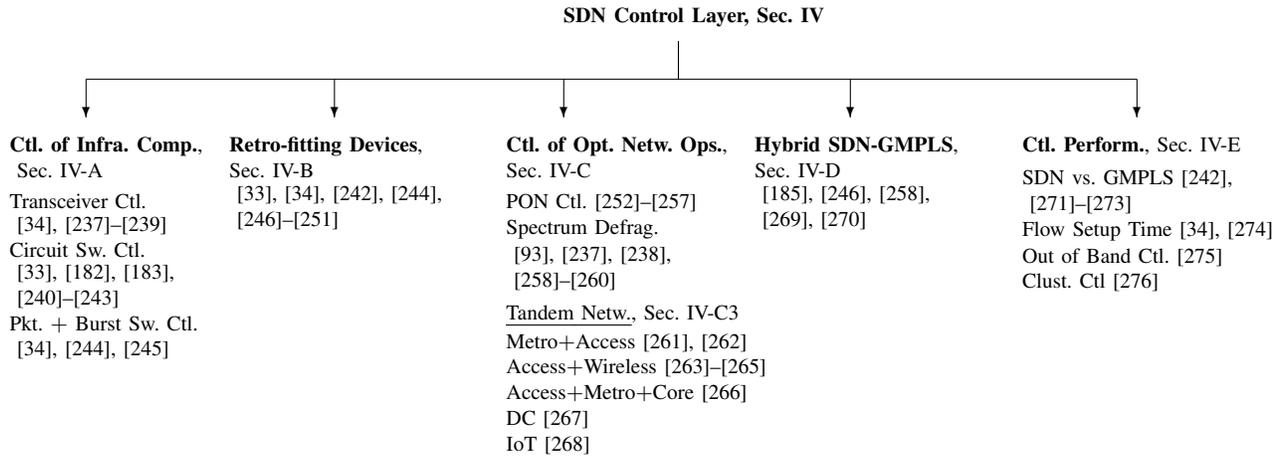
\section{SDN Control Layer}
\label{sdnctl:sec}
This section surveys the SDON studies that are focused on applying the
SDN principles at the SDN control layer to control the various optical network
elements and operational aspects.
The main challenges of SDON control include extensions of the OpenFlow
protocol for specifically controlling the optical transmission
and switching components surveyed in Section~\ref{sdninfra:sec}
and for controlling the optical
spectrum as well as for controlling optical networks spanning
multiple optical network tiers (see Section~\ref{optnetstruct:sec}).
As illustrated in Fig.~\ref{ctl_class:fig},
we first survey SDN control mechanisms and frameworks for controlling
infrastructure layer components, namely transceivers as well as
optical circuit, packet, and burst switches.
More specifically, we survey OpenFlow extensions for controlling the
optical infrastructure components.
We then survey mechanisms for retro-fitting non-SDN optical network elements
so that they can be controlled by OpenFlow.
The retro-fitting typically involves the insertion of an abstraction layer
into the network elements. The abstraction layer
makes the optical hardware controllable by OpenFlow.
The retro-fitting studies would also fit into Section~\ref{sdninfra:sec} as
the abstraction layer is inserted into the network elements; however,
the abstraction mechanisms closely relate to the OpenFlow extensions for
optical networking and we include the retro-fitting studies therefore in this
control layer section.
We then survey the various SDN control mechanisms for operational aspects of
optical networks, including the control
of tandem networks that include optical segments. Lastly, we survey
SDON controller performance analysis studies.

\subsection{SDN Control of Optical Infrastructure Components}
\label{ctlinfra:sec}

\subsubsection{Controlling Optical Transceivers with OpenFlow}
Recent generations of optical transceivers utilize digital signal
processing techniques that allow many parameters of the transceiver to
be software controlled (see Sections~\ref{SF-BVT:sec} and~\ref{MF-BVT:sec}).
These parameters include modulation scheme, symbol rate, and wavelength.
Yu et al.~\cite{YuZZG14} and Chen et al.~\cite{Chen2014} proposed adding a ``modulation format'' field to the OpenFlow cross-connect table entries to support this
programmable feature of some software defined optical transceivers.

Ji et al.~\cite{JiXia2014} created a testbed that places super-channel optical transponders and optical
amplifiers under SDN control. An OpenFlow extension is proposed to control these devices. The modulation
technique and FEC code for each optical subcarrier of the super-channel transponder
and the optical amplifier power level can be controlled via OpenFlow. Ji et al. do not discuss this explicitly
but the transponder subcarriers can be treated as OpenFlow switch ports that can be configured through the
OpenFlow protocol via port modification messages. It is unclear in \cite{JiXia2014} how the amplifiers would be
controlled via OpenFlow. However, doing so would allow the SDN controller to adaptively modify amplifiers to
compensate for channel impairments while minimizing energy consumption. Ji et al.~\cite{JiXia2014}
have established a testbed demonstrating the placement of transponders and EDFA optical amplifiers under SDN control.

Liu et al.~\cite{liu2013field} propose configuring optical transponder operation via flow table entries with
new transponder specific fields (without providing details). They also propose capturing failure alarms from optical
transponders and sending them to the SDN controller via OpenFlow Packet-In messages. These messages are normally
meant to establish new flow connections. Alternatively, a new OpenFlow message type could be created for the purpose
of capturing failure alarms~\cite{liu2013field}. With failure alarm information, the SDN controller can implement
protection switching services.

\subsubsection{Controlling Optical Circuit Switches with OpenFlow}
Circuit switching can be enabled by OpenFlow by adding new circuit switching flow table
entries~\cite{DaPM09Jul,DaPM09,DaPMSG10,Baik2014}. The OpenFlow circuit switching addendum~\cite{OFCircuitSwitch}
discusses the addition of cross-connect tables for this purpose.
These cross-connect tables are configured via OpenFlow messages inside the circuit switches. According to the addendum, a cross-connect table entry consists of the following
fields to identify the input:
\begin{itemize}
  \item Input Port
  \item Input Wavelength
  \item Input Time Slot
  \item Virtual Concatenation Group
\end{itemize}
and the following fields to identify the output:
\begin{itemize}
  \item Output Port
  \item Output Wavelength
  \item Output Time Slot
  \item Virtual Concatenation Group
\end{itemize}
These cross-connect tables cover circuit switching in space, fixed-grid wavelength, and time.

Channegowda et al.~\cite{ChNF13,ChNS13} extend the capabilities of the OpenFlow circuit switching addendum to
support flexible wavelength grid optical switching. Specifically, the wavelength identifier specified in the
circuit switching addendum to OpenFlow is replaced with two fields: \textit{center frequency}, and
\textit{slot width}. The \textit{center frequency} is an integer specifying the multiple of 6.25~GHz the center
frequency is away from 193.1~Thz and the \textit{slot width} is a positive integer specifying the spectral
width in multiples of 12.5~GHz.

An SDN controlled optical network testbed at the University of Bristol has been established to
demonstrate the OpenFlow extensions for flexible grid DWDM~\cite{ChNS13}. The testbed consists of
both fixed-grid and flexible-grid optical switching devices.
South Korea Telekom has also
built an SDN controlled optical network testbed~\cite{Shin2014}.

\subsubsection{Controlling Optical Packet and Burst Switches with OpenFlow}
OpenFlow flow tables can be utilized in
optical packet switches for expressing the forwarding table and its computation can be offloaded to an SDN
controller. This offloading can simplify the design of highly complex optical packet switches~\cite{Cao2015}.

Cao et al.~\cite{Cao2015} extend the OpenFlow protocol to work with Optical Packet Switching (OPS) devices by creating:
$(i)$ an abstraction layer that converts OpenFlow configuration messages to the native OPS configuration, $(ii)$ a process that
converts optical packets that do not match a flow table entry to the electrical domain for forwarding to the SDN
controller, and $(iii)$ a wavelength identifier extension to the flow table entries. To compensate for either the
lack of any optical buffering or limited optical buffering, an SDN controller, with its global view, can provide more
effective means to resolve contention that would lead to packet loss in optical packet switches.
Specifically, Cao et al. suggest to select the path with the most available resources among multiple available paths between two nodes~\cite{Cao2015}.
Paths can be re-computed periodically or on-demand to account for changes
in traffic conditions. Monitoring messages can be defined to keep the SDN controller updated of network traffic
conditions.

Engineers with Japan's National Institute of Information and Communications Technology~\cite{Harai2014}
have created an optical circuit and packet switched demonstration system in which the packet portion is SDN
controlled. The optical circuit switching is implemented with Wavelength Selective Switches (WSSs) and the
optical packet switching is implemented with an Semiconductor Optical Amplifier (SOA) switch.

OpenFlow flow tables can also be used to configure optical burst switching devices~\cite{liu2013field}. When there is
no flow table entry for a burst of packets, the optical burst switching device can send the Burst Header Packet (BHP) to
the SDN controller to process the addition of the new flow to the network~\cite{liu2013field} rather than the first
packet in the burst.

\subsection{Retro-fitting Devices to Support OpenFlow}  \label{retfitctl:sec}
\begin{figure}[t!]
    \centering
    \includegraphics[width=2in]{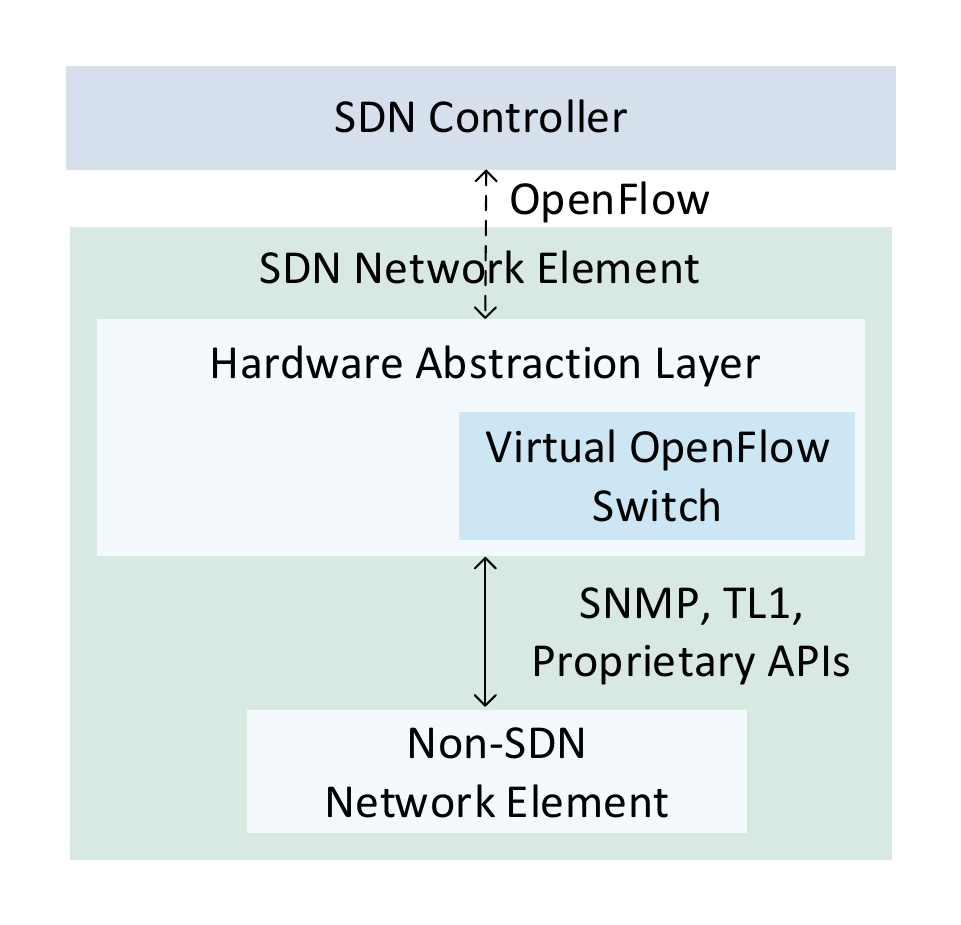}
    \caption{Traditional non-SDN network elements can be retro-fitted for control by an SDN controller using OpenFlow using
    a hardware abstraction layer~\cite{LiTMG11,LiuTMGW11,LiCTM12,LiZTV12,liu2013field}.}
    \label{fig:retrofit}
\end{figure}

An abstraction layer can be used to turn non-SDN optical switching
devices into OpenFlow controllable switching
devices~\cite{ChNF13,ChNS13,liu2013field,Alvizu2014,Cao2015}.
As illustrated in Fig.~\ref{fig:retrofit}, the
abstraction layer provides a conversion layer between OpenFlow
configuration messages and the optical switching devices' native
management interface, e.g., the Simple Network Management Protocol
(SNMP), the Transaction Language 1 (TL1) protocol, or a proprietary
(vendor-specific) API. Additionally, a virtual OpenFlow switch with
virtual interfaces that correspond to physical switching ports on the
non-SDN switching device completes the abstraction
layer~\cite{LiTMG11,LiuTMGW11,LiCTM12,LiZTV12,liu2013field}. When a
flow entry is added between two virtual ports in the virtual OpenFlow
switch, the abstraction layer uses the switching devices' native
management interface to add the flow entry between the two
corresponding physical ports.

\begin{figure}[t!]
    \centering
    \includegraphics[width=2.8in]{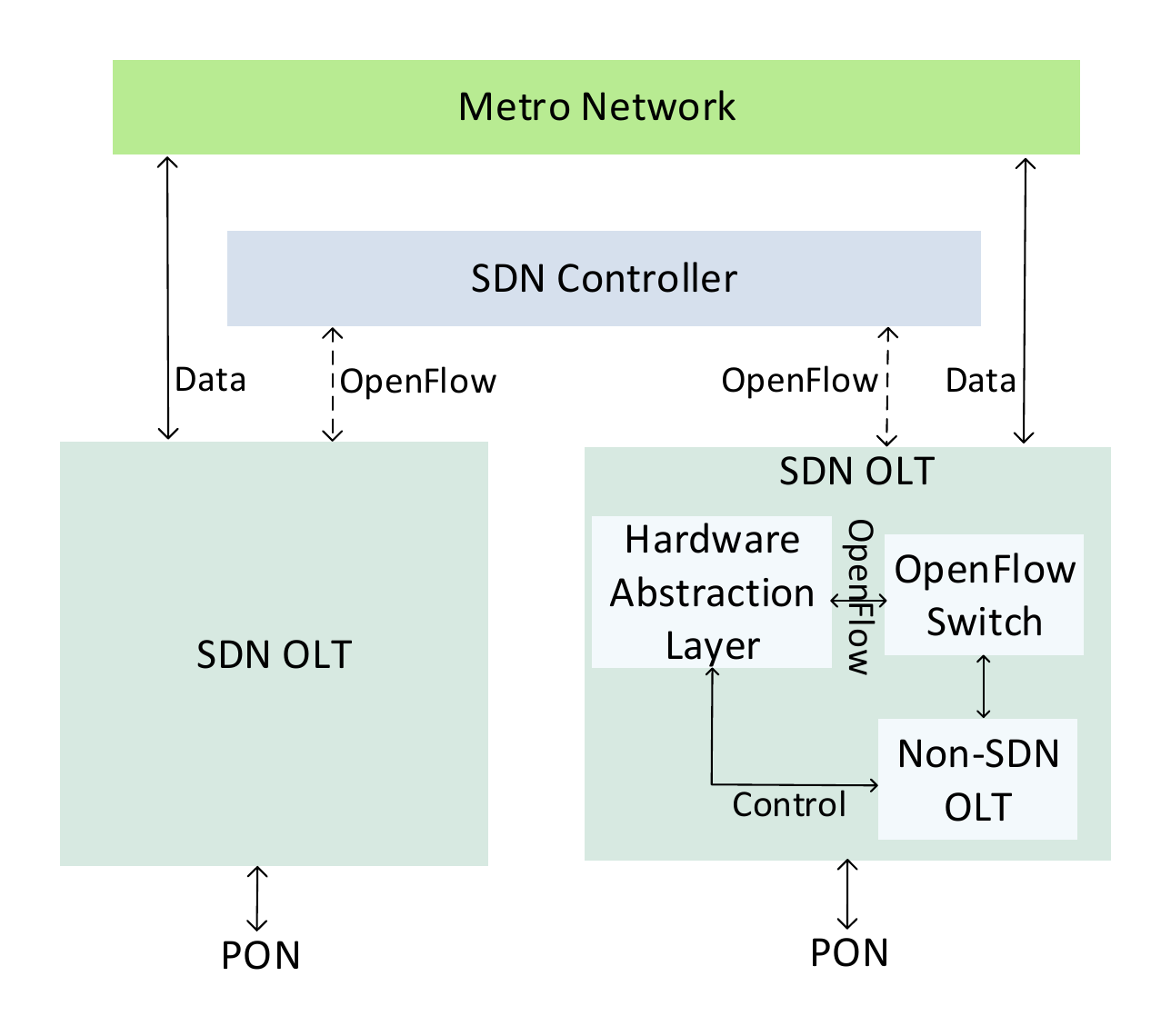}
    \caption{Non-SDN OLTs can be retro-fitted for control by an SDN controller using OpenFlow~\cite{ClSLT14}.}
    \label{fig:oltretrofit}
\end{figure}

A non-SDN PON OLT can be supplemented with a two-port OpenFlow switch and a hardware abstraction layer that
converts OpenFlow forwarding rules to control messages understood by the non-SDN OLT~\cite{ClSLT14}. Fig.~\ref{fig:oltretrofit} illustrates this OLT retro-fit for SDN control via OpenFlow. In this way the
PON has its switching functions controlled by OpenFlow.

\subsection{SDN Control of Optical Network Operation}  \label{ctlops:sec}

\subsubsection{Controlling Passive Optical Networks with OpenFlow}
\label{ponctl:sec}
An SDN controlled PON can be created by upgrading OLTs to SDN-OLTs that can be
controlled using a Southbound Interface, such as OpenFlow~\cite{KhRS14,LeK15}. A centralized PON controller,
potentially executing in a data center, controls one or more SDN-OLTs. The advantage of using SDN is the
broadened perspective of the PON controller as well as the potentially reduced cost of the SDN-OLT
compared to a non-SDN OLT.

Parol and Pawlowski~\cite{PaP13,par2014fut} define OpenFlowPLUS to extend the OpenFlow SBI for GPON. OpenFlowPLUS extends SDN programmability to both OLT and ONU devices whereby each act
as an OpenFlow switch through a programmable flow table. Non-switching functions (e.g., ONU registration,
dynamic bandwidth allocation) are outside the scope of OpenFlowPLUS. OpenFlowPLUS extends OpenFlow
by channeling OpenFlow messages through the GPON ONU Management and Control Interface (OMCI) control channel and adding PON specific action
instructions to flow table entries. The PON specific action instructions defined in OpenFlowPLUS are:
\begin{itemize}
  \item (new \textit{gpon} action type): map matching packets to PON specific traffic identifiers, e.g.,
  GPON Encapsulation Method (GEM) ports and GPON Traffic CONTainers (T-CONTs)
  \item (\textit{output} action type): activate PON specific framing of matching packets
\end{itemize}

Many of the OLT functions operate at timescales that are problematic for the controller due
to the latency between the controller and OLTs. However, Khalili et al.~\cite{KhRS14} identify ONU
registration policy and coarse timescale DBA policy as functions that operate at timescales that allow
effective offloading to an SDN controller. Yan et al.~\cite{YaZWZ14} further identify OLT and ONU power
control for energy savings as a function that can be effectively offloaded to an SDN controller.

There is also a movement to use PONs in edge networks to provide
connectivity inside a multitenant building or on a campus with multiple
buildings~\cite{PaP13,par2014fut}. The use of PONs in this edge scenario
requires rapid re-provisioning from the OLT. A software controlled PON
can provide this needed rapid reprovisioning~\cite{PaP13,par2014fut}.

Kanonakis et al. \cite{KaCTW13} propose leveraging the
broad perspective that SDN can provide to perform dynamic bandwidth
allocation across several Virtual PONs (VPONs). The VPONs are
separated on a physical PON by the wavelength bands that they
utilize. Bandwidth allocation is performed at the granularity of OFDMA
subcarriers that compose the optical spectrum.

\subsubsection{SDN Control of Optical Spectrum Defragmentation}
In a departure from the fixed wavelength grid
(ITU-T G.694.1), elastic optical networking allows flexible use of the optical spectrum. This flexibility can permit higher spectral efficiency by avoiding consuming an entire fixed-grid
wavelength channel when unnecessary and avoiding unnecessary guard bands in certain circumstances~\cite{Ger2012}.
However, this flexibility causes fragmentation of the optical spectrum as flexible grid lightpaths are established
and terminated over time.

Spectrum fragmentation leads to the circumstance in which there is enough spectral capacity to satisfy a demand but
that capacity is spread over several fragments rather than being consolidated in adjacent spectrum as required. If
the fragmentation is not counter-acted by a periodic defragmentation process than overall spectral utilization will
suffer. This resource fragmentation problem appears in computer systems in main memory and long term storage. In those
contexts the problem is typically solved by allowing the memory to be allocated using non-adjacent segments. Memory and
storage is partitioned into pages and blocks, respectively. The allocations of pages to a process or blocks to a file do
not need to be contiguous. With communication spectrum this would mean combining multiple small bandwidth channels
through inverse multiplexing to create a larger channel~\cite{Munoz2014a}.

An SDN controller can provide a broad network perspective to empower the periodic optical spectrum defragmentation
process to be more effective~\cite{Munoz2014a}. In general, optical spectrum defragmentation operations can reduce
lightpath blocking probabilities from 3\%~\cite{YuZZG14} up to as much as 75\%~\cite{Chen2014,Zhu2015b}. Multicore
fibers provide additional spectral resources through additional transmission cores to permit quasi-hitless
defragmentation~\cite{Meloni2016}.

\subsubsection{SDN Control of Tandem Networks}  \label{ctltan:sec}

\paragraph{Metro and Access}
Wu et al.~\cite{wu2014glo,zha2015sof} propose leveraging the broad perspective that SDN can provide to
improve bandwidth allocation. Two cooperating stages of SDN controllers: $(i)$ access stage that controls each SDN
OLT individually, and $(ii)$ metro stage that controls global bandwidth allocation strategy, can coordinate bandwidth
allocation across several physical PONs~\cite{wu2014glo,zha2015sof}. The bandwidth allocation is managed cooperatively
among the two stages of SDN controllers to optimize the utilization of the access and metro network bandwidth. Simulation
experiments indicate a 40\% increase in network bandwidth utilization as a result of the global coordination compared
to operating the bandwidth allocation only within the individual PONs~\cite{wu2014glo,zha2015sof}.

\paragraph{Access and Wireless}
Bojic et al. \cite{boj2013adv} expand on the concept of SDN controlled OFDMA enabled VPONs~\cite{KaCTW13} to
provide mobile backhaul service. The backhaul service can be provided for wireless small-cell sites (e.g., micro and femto cells)
that utilize millimeter wave frequencies. Each small-cell site contains an OFDMA-PON ONU that provides the backhaul service
through the access network over a VPON. An SDN controller is utilized to assign bandwidth to each small-cell site through
OFDMA subcarrier assignment in a VPON to the constituent ONU. The SDN controller leverages its broad view of the network
to provide solutions to the joint bandwidth allocation and routing across several network segments. With this broad
perspective of the network, the SDN controller can make globally rather than just locally optimal bandwidth allocation and
routing decisions. Efficient optimization algorithms, such as genetic algorithms, can be used to provide computationally
efficient competitive solutions, mitigating computational complexity issues associated with optimization for large networks.
Additionally, network partitioning with an SDN controller for each partition can be used to mitigate unreasonable
computational complexity that arises when scaling to large networks. Tanaka and Cvijetic~\cite{TaC14} presented one
such optimization formulation for maximizing throughput.

Costa-Requena et al.~\cite{Costa2015} described a proof-of-concept LTE testbed they have constructed
whereby the network consists of software defined base stations and various network functions
executing on cloud resources. The testbed is described in broad qualitative terms, no technical
details are provided. There was no mathematical or experimental analysis provided.

\paragraph{Access, Metro, and Core}
Slyne and Ruffini~\cite{SlR14} provide a use case for SDN switching control across network segments: use Layer 2 switching
across the access, metro, and core networks. Layer 2 (e.g., Ethernet) switching does not scale well due to a lack of hierarchy
in its addresses. That lack of hierarchy does not allow for switching rules on aggregates of addresses thereby limiting the
scaling of these networks. Slyne and Ruffini~\cite{SlR14} propose using SDN to create hierarchical pseudo-MAC addresses that
permit a small number of flow table entries to configure the switching of traffic using Layer 2 addresses across network
segments. The pseudo-MAC addresses encode information about the device location to permit simple switching rules. At the entry
of the network, flow table entries are set up to translate from real (non-hierarchical) MAC addresses to hierarchical pseudo-MAC
addresses. The reverse takes place at the exit point of the network.

\paragraph{DC Virtual Machine Migration}
Mandal et al.~\cite{MaHSC14} provided a cloud computing use case for SDN bandwidth allocation across network segments:
Virtual Machine (VM) migration between data centers. VM migrations require significant network bandwidth. Bandwidth
allocation that utilizes the broad perspective that SDN can provide is critical for reasonable VM migration latencies
without sacrificing network bandwidth utilization.

\paragraph{Internet of Things}
Wang et al.~\cite{wan2015nov} examine another use case for SDN bandwidth allocation across network segments: the Internet
of Things (IoT). Specifically, Wang et al. have developed a Dynamic Bandwidth Allocation (DBA) protocol that exploits
SDN control for multicasting and suspending flows. This DBA protocol is studied in the context of a virtualized
WDM optical access network that provides IoT services through the distributed ONUs to individual devices.
The SDN controller employs multicasting and flow suspension to efficiently prioritize the IoT service requests.
Multicasting allows multiple requests to share resources in the central nodes that are responsible for processing a
prescribed wavelength in the central office (OLT). Flow suspension allows high-priority requests (e.g., an emergency
call) to suspend ongoing low-priority traffic flows (e.g., routine meter readings). Performance results for a real-time
SDN controller implementation indicate that the proposed bandwidth (resource) allocation with multicast and flow
suspension can improve several key performance metrics, such as request serving ratio, revenue, and delays
by 30--50~\%~\cite{wan2015nov}.

\subsection{Hybrid SDN-GMPLS Control}  \label{hybsdnctl:sec}
\subsubsection{Generalized MultiProtocol Label Switching (GMPLS)}
Prior to SDN, MultiProtocol Label Switching (MPLS) offered a mechanism to separate the control and data planes
through label switching. With MPLS, packets are forwarded in a connection-oriented manner through Label Switched
Paths (LSPs) traversing Label Switching Routers (LSRs). An entity in the network establishes an LSP through a network
of LSRs for a particular class of packets and then signals the label-based forwarding table entries to the LSRs. At
each hop along an LSP, a packet is assigned a label that determines its forwarding rule at the next hop. At the next
hop, that label determines that packet's output port and label for the next hop; the process repeats until the packet
reaches the end of the LSP. Several signalling protocols for programming the label-based forwarding table entries
inside LSRs have been defined, e.g., through the Resource Reservation Protocol (RSVP). Generalized MPLS (GMPLS) extends MPLS to offer circuit switching
capability. Although never commercially deployed~\cite{liu2013field}, GMPLS and a centralized Path Computation Element
(PCE)~\cite{mun2014pce,oki2005dyn,pao2013sur,Casellas2013} have been considered for control of optical networks.

\subsubsection{Path Computation Element (PCE)}  \label{PCE:sec}
A PCE is a concept developed by the IETF (see RFC 4655) to refer to an entity that computes network paths given a topology
and some criteria. The PCE concept breaks the path computation action from the forwarding action in switching devices.
A PCE could be distributed in every switching element in a network domain or there could be a single centralized PCE
for an entire network domain. The network domain could be an area of an Autonomous System (AS), an AS, a conglomeration
of several ASes, or just a group of switching devices relying on one PCE. Some of an SDN controller's functionality
falls under the classification of a centralized PCE. However, the PCE concept does not include the external
configuration of forwarding tables. Thus, a centralized PCE device does not necessarily have a means to configure the
switching elements to provision a computed path.

When the entity requesting path computation is not co-located with the PCE, a PCE Communication Protocol (PCEP) is used
over TCP port 4189 to facilitate path computation requests and responses. The PCEP consists of the following message
types:
\begin{itemize}
  \item Session establishment messages (open, keepalive, close)
  \item PCReq -- Path computation request
  \item PCRep -- Path computation reply
  \item PCNtf -- event notification
  \item PCErr -- signal a protocol error
\end{itemize}

The path computation request message must include the end points of the path and can optionally include the requested
bandwidth, the metric to be optimized in the path computation, and a list of links to be included in the path. The Path
computation reply includes the computed path expressed in the Explicit Route Object format (see RFC 3209) or an
indication that there is no path. See RFC 5440 for more details on PCEP.

A PCE has been proposed as a central entity to manage a GMPLS-enabled optical circuit switched network. Specifically,
the PCE maintains the network topology in a structure called the Traffic Engineering Database (TED). The traffic
engineering modifier (see RFC 2702) signifies that the path computations are made to relieve congestion that is caused
by the sub-optimal allocation of network resources. This modifier is used extensively in discussions of MPLS/GMPLS
because their use case is for traffic engineering; in acronym form the modifier is TE (e.g., TE LSP, RSVP-TE).

If the PCE is stateful with complete control over its network domain, it will also maintain an LSP database recording the provisioned GMPLS lightpaths. A lightpath request can be sent to the PCE, it will use the topology and LSP
database to find the optimal path and then configure the GMPLS-controlled optical circuit switching nodes using NETCONF
(see RFC 6241) or proprietary command line interfaces (CLIs)~\cite{Munoz2014a}. This stateful PCE with instantiation
capabilities (capabilities to provision lightpaths) operates similarly to an SDN controller. For that reason, GMPLS with a
centralized stateful PCE with instantiation capabilities can provide a baseline for performance analysis of an SDN
controller as well as provide a mechanism to be blended with an SDN controller for hybrid control~\cite{ChNF13,ChNS13,Alvizu2014}.

\subsubsection{Approaches to Hybrid SDN-GMPLS Control}
\begin{figure}[t!]
	\centering
	\includegraphics[width=3.5in]{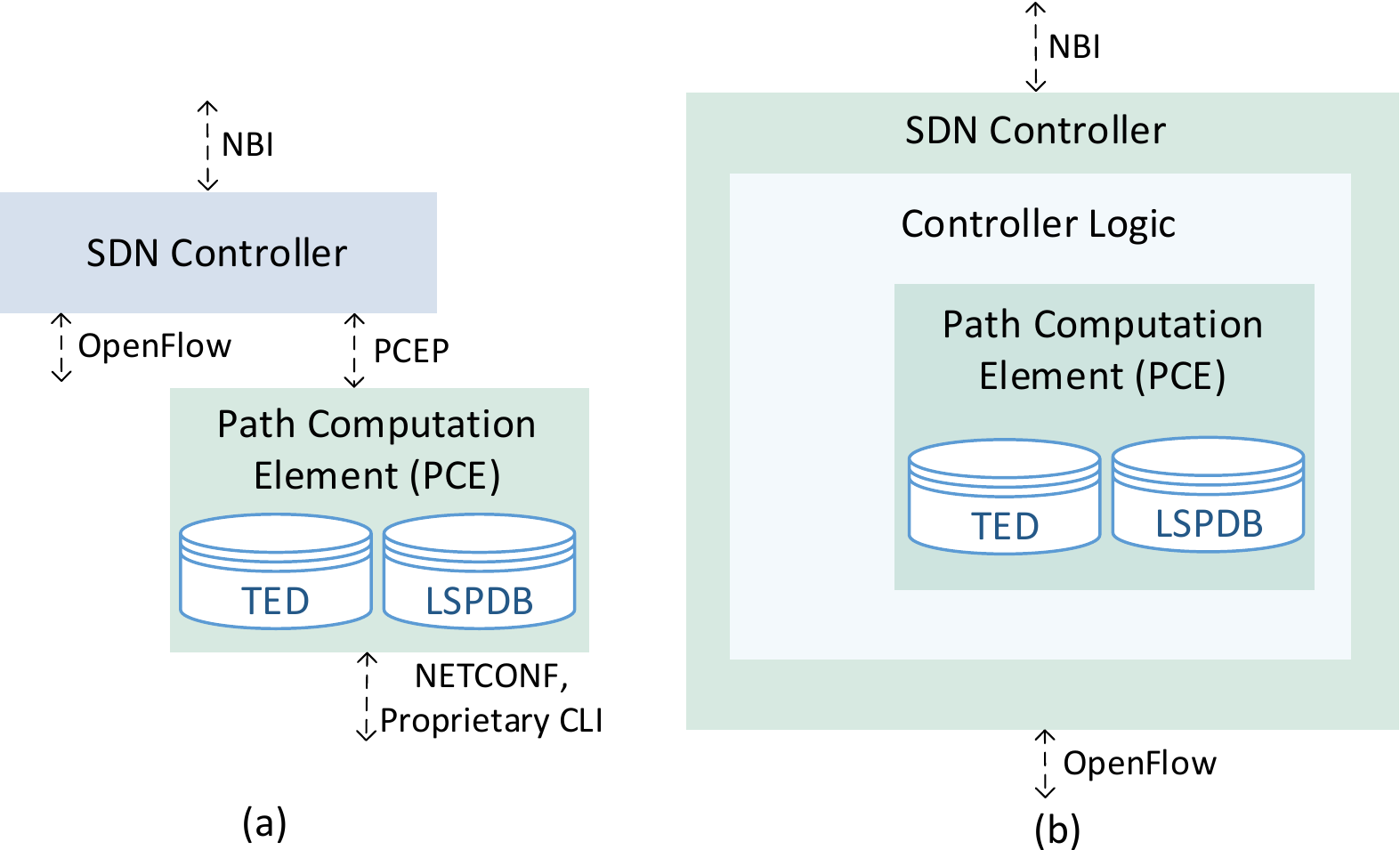}
\caption{Hybrid GMPLS/PCE and SDN network control:  (a) SDN controller
  utilizes a PCE to control a portion of the
  network~\cite{AzNEJ11,Munoz2014a} through the NETCONF protocol or a
  proprietary command line interface (CLI).  (b) SDN controller
  utilizes the path computation ability of the
  PCE~\cite{LiCTMMM12,Casellas2013b,Alvizu2014} and controls network through
  OpenFlow protocol.}
	\label{fig:hybridctl}
\end{figure}

Hybrid GMPLS/PCE and SDN control can be formed by allowing an SDN controller to leverage a centralized PCE to control a
portion of the infrastructure using PCEP as the SBI~\cite{AzNEJ11,Munoz2014a}; see illustration a) in Fig.~\ref{fig:hybridctl}.
The SDN controller builds higher functionality above what the PCE provides and can possibly control a large network that
utilizes several PCEs as well as OpenFlow controlled network elements.

Alternatively, the SDN controller can leverage a PCE for its path computation abilities with the SDN controller handling
the configuration of the network elements to establish a path using an SBI protocol, such as
OpenFlow~\cite{LiCTMMM12,Casellas2013b,Alvizu2014};
see illustration b) in Fig.~\ref{fig:hybridctl}.

\subsection{SDN Performance Analysis}  \label{cltperf:sec}
\subsubsection{SDN vs. GMPLS}
Liu et al.~\cite{LiTM12} provided a qualitative comparison of GMPLS,
GMPLS/PCE, and SDN OpenFlow for control of wavelength switched optical
networks. Liu et al.~noted that there is an evolution of centralized
control from GMPLS to GMPLS/PCE to OpenFlow. Whereas GMPLS offers
distributed control, GMPLS/PCE is commonly regarded as having centralized path
computation but still distributed provisioning/configuration; while
OpenFlow centralizes all of the network control. In
our discussion in Section \ref{hybsdnctl:sec} we noted that a
stateful PCE with instantiation capabilities centralizes all
network control and is therefore very similar to SDN. Liu et al.~have
also pointed out that GMPLS/PCE is more technically mature compared to
OpenFlow with IETF RFCs for GMPLS (see RFC 3471) and PCE (see RFC
4655) that date back to 2003 and 2006, respectively. SDN has just
recently, in 2014, received standardization attention from the IETF
(see RFC 7149).

A comparison of GMPLS and OpenFlow has been conducted by Zhao et al.~\cite{ZhaoZYY13} for large-scale optical networks.
Two testbeds were built, based on GMPLS and on Openflow, respectively. Performance metrics, such as
blocking probability, wavelength utilization, and lightpath setup time were evaluated for a 1000 node topology. The
results indicated that GMPLS gives slightly lower blocking probability. However, OpenFlow gives higher wavelength
utilization and shorter average lightpath setup time. Thus, the results suggest that OpenFlow is overall advantageous
compared to GMPLS in large-scale optical networks.

Cvijetic et al.~\cite{CviAPJT13} conducted a numerical analysis to compare the computed shortest path
lengths for non-SDN, partial-SDN, and full-SDN optical networks. A full-SDN network enables path lengths
that are approximately a third of those computed on a non-SDN network. These path lengths can also translate
into an energy consumption measure, with shortest paths resulting in reduced energy consumption. An SDN
controlled network can result in smaller computed shortest paths that translates to smaller network latency
and energy consumption~\cite{CviAPJT13}.

Experiments conducted on the testbed described in \cite{ChNF13} show a 4~\% reduction in lightpath blocking probability
using SDN OpenFlow compared to GMPLS for lightpath provisioning. The same experiments show that lightpath setup times
can be reduced to nearly half using SDN OpenFlow compared to GMPLS. Finally, the experiments show that an Open vSwitch
based controller can process about three times the number of flows per second as a NOX~\cite{gud2008nox} based controller.

\subsubsection{SDN Controller Flow Setup}
Veisllari et al.~\cite{VeSBR14} evaluated the use of SDN to support both circuit and packet
switching in a metropolitan area ring network that interconnects access network segments
with a backbone network. This network is assumed to be controlled by a single SDN controller.
The objective of the study~\cite{VeSBR14} was to determine the effect of packet service flow
size on the required SDN controller flow service time to meet stability conditions at the
controller. Toward this end, Veisllari et al.~produced a mean arrival rate function of new packet and
circuit flows at that controller. This arrival rate function was visualized by varying the length
of short-lived (``mice'') flows, the fraction of long-lived (``elephant'') flows, and the volume
of traffic consumed by ``elephant'' flows. Veisllari et al.~discovered, through these visualizations,
that the length of ``mice'' flows is the dominating parameter in this model.

Veisllari et al.~translated the arrival rate function analysis to an analysis of the ring MAN network
dimensions that can be supported by a single SDN controller. The current state-of-the-art Beacon
controller can handle a flow request every 571~ns. Assuming mice flows sizes of 20~kB and average circuit
lifetimes of 1 second, as the fraction of packet traffic increases from 0.1 to 0.9, the network
dimension supported by a single Beacon SDN controller decreases from 14 nodes with 92 wavelengths per
node to 5 nodes with 10 wavelengths per node.

Liu et al.~\cite{liu2013field} use a multinational (Japan, China, Spain) NOX:OpenFlow controlled
four-wavelength optical circuit and burst switched network to study path setup/release times as well as path
restoration times. The optical transponders that can generate failure alarms were also under NOX:OpenFlow
control and these alarms were used to trigger protection switching. The single SDN controller was located
in the Japanese portion of the network. The experiments found the path setup time to vary from 250--600 ms and the path release times to vary from 130--450 ms. Path restoration times varied
from 250--500~ms. Liu et al.~noted that the major contributing factor to these times was the OpenFlow message delivery time~\cite{liu2013field}.

\subsubsection{Out of Band Control}
Sanchez et al.~\cite{Sanchez2013} have qualitatively compared four SDN controlled ring metropolitan network architectures. The architectures vary in whether the SDN control traffic is carried
in-band with the data traffic or out-of-band separately from the data traffic. In a single wavelength ring
network, out-of-band control would require a separate physical network that would come at a high cost,
but provide reliability of the network control under failure of the ring network. In a multiwavelength
ring network, a separate wavelength can be allocated to carry the control traffic. Sanchez et al.~\cite{Sanchez2013}
focused on a Tunable Transceiver Fixed Receiver (TTFR) WDM ring node architecture. In this architecture each
node receives data on a home wavelength channel and has the capability to transmit on any of the available
wavelengths to reach any other node. The addition of the out-of-band control channel on a separate wavelength
requires each node to have an additional fixed receiver, thereby increasing cost. Sanchez et al. identified a clear tradeoff between cost and reliability when comparing the four architectures.

\subsubsection{Clustered SDN Control}
Penna et al.~\cite{Penna2014} described partitioning a wavelength-switched optical network into administrative
domains or clusters for control by a single SDN controller. The clustering should meet certain
performance criteria for the SDN controller. To permit lightpath establishment across clusters, an inter-cluster
lightpath establishment protocol is established. Each SDN controller provides a lightpath establishment function
between any two points in its associated cluster. Each SDN controller also keeps a global view of the network
topology. When an SDN controller receives a lightpath establishment request whose computed path traverses
other clusters, the SDN controller requests lightpath establishment within those clusters via a WBI.

The formation of clusters can be performed such that for a specified number of clusters the average distance to
each SDN controller is minimized~\cite{Penna2014}. The lightpath establishment time decreases exponentially
as the number of clusters increases.

\subsection{Control Layer: Summary and Discussion}
A very large body of literature has explored how to expand the OpenFlow protocol to support various optical network
technologies (e.g., optical circuit switching, optical packet switching, passive optical networks). A significant
body of literature has investigated methodologies for retro-fitting non-SDN network elements for OpenFlow control
as well as integrating SDN/OpenFlow with the GMPLS/PCE control framework. A variety of SDN controller use cases have
been identified that motivate the benefits of the centralized network control made possible with SDN (e.g., bandwidth
allocation over large numbers of subscribers, controlling tandem networks).

However, analyzing the performance of SDN controllers for optical network applications is still in a state of infancy.
It will be important to understand the connection between the implementation of the SDN controller (e.g., processor
core architecture, number of threads, operating system) and the network it can effectively control (e.g., network
traffic volume, network size) to meet certain performance objectives (e.g., maximum flow setup time). At present there
are not enough relevant studies to gain an understanding of this connection. With this understanding network service
providers will be able to partition their networks into control domains in a manner that meets their performance
objectives.

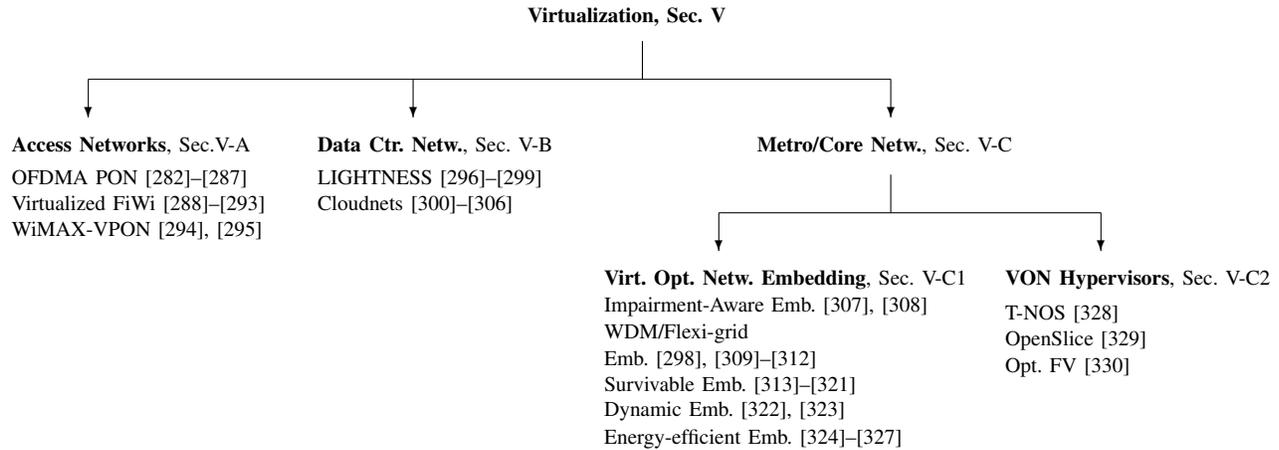
\begin{figure*}[t!]
\footnotesize
\setlength{\unitlength}{0.10in} 
\centering
\begin{picture}(40,33)
\put(15,33){\textbf{Virtualization, Sec.~\ref{virt:sec}}}
\put(-8,30){\line(1,0){42}}
\put(25,23){\line(1,0){20}}
\put(27,26.2){\textbf{Metro/Core Netw.}, Sec.~\ref{virt_core:sec}}
\put(34,25){\line(0,-1){2}}
\put(21,30){\line(0,1){2}}
\put(-8,30){\vector(0,-1){2}}
\put(-12,27){\makebox(0,0)[lt]{\shortstack[l]{			
\textbf{Access Networks}, Sec.\ref{virt_access:sec}	\\	\\
OFDMA PON~\cite{wei2009pon,wei2009pro,wei2010ada,li2016pro,jin2009vir,zhou2015dem}\\		
Virtualized FiWi~\cite{QiZSH14,QiGYZ13,ShGYZ13,he2013int,men2014eff,QiZHG14}\\		
WiMAX-VPON~\cite{dha2010wim,DhPX10}
}}}

\put(9,30){\vector(0,-1){2}}
\put(4,27){\makebox(0,0)[lt]{\shortstack[l]{			
\textbf{Data Ctr. Netw.}, Sec.~\ref{virt_dc:sec} \\ \\
LIGHTNESS~\cite{miao2015sdn,pen2015mul,pag2015opt,Saridis2016} \\
Cloudnets~\cite{kan2015res,ahm2014enh,dov2015usi,xie2014dyn,vel2014tow,zha2015ren,tza2014con}
}}}				
\put(34,30){\vector(0,-1){2}}
\put(25,23){\vector(0,-1){2}}
\put(19,20){\makebox(0,0)[lt]{\shortstack[l]{
\textbf{Virt. Opt. Netw. Embedding}, Sec.~\ref{virt_emb:sec} \\
Impairment-Aware Emb.~\cite{pen2011imp,PeNS13} \\
WDM/Flexi-grid \\ Emb.~\cite{zha2013net,zha2013vir,gon2014vir,wan2015vir,pag2015opt} \\
Survivable Emb.~\cite{hu2013sur,ye2015sur,xie2014sur,che2016cos,jia2015ava,son2011ban,pao2014mul,ass2016net,Hong2015} \\
Dynamic Emb.~\cite{ye2014upg,zha2015dyn} \\
Energy-efficient Emb.~\cite{non2015ene,che2016pow,she2014fol,wan2014hie}
}}}							
\put(45,23){\vector(0,-1){2}}		
\put(40,20){\makebox(0,0)[lt]{\shortstack[l]{			
\textbf{VON Hypervisors}, Sec.~\ref{virt_hv:sec} \\ \\
T-NOS~\cite{siq2015pro} \\
OpenSlice~\cite{LiuMCTMM13} \\
Opt. FV~\cite{Azod2012}
}}}
\end{picture}
\vspace{-2.8cm}	
\caption{Classification of SDON virtualization studies.}
\label{virt_class:fig}
\end{figure*}
\section{Virtualization}  \label{virt:sec}
This section surveys control layer mechanisms for virtualizing
SDONs.
As optical infrastructures have typically high costs, creating
multiple VONs over the optical network infrastructure is
especially important for access networks, where the costs need to
be amortized over relatively few users. Throughout, accounting
for the specific optical transmission and signal propagation characteristics
is a key challenge for SDON virtualization.
Following the classification structure illustrated in
Fig.~\ref{virt_class:fig},
we initially survey virtualization mechanisms for access networks
and data center networks, followed by virtualization mechanisms for
optical core networks.

\begin{figure*}[t!]
    \centering
  \vspace{-.1cm}
\begin{tabular}{cc}
\includegraphics[width=3in]{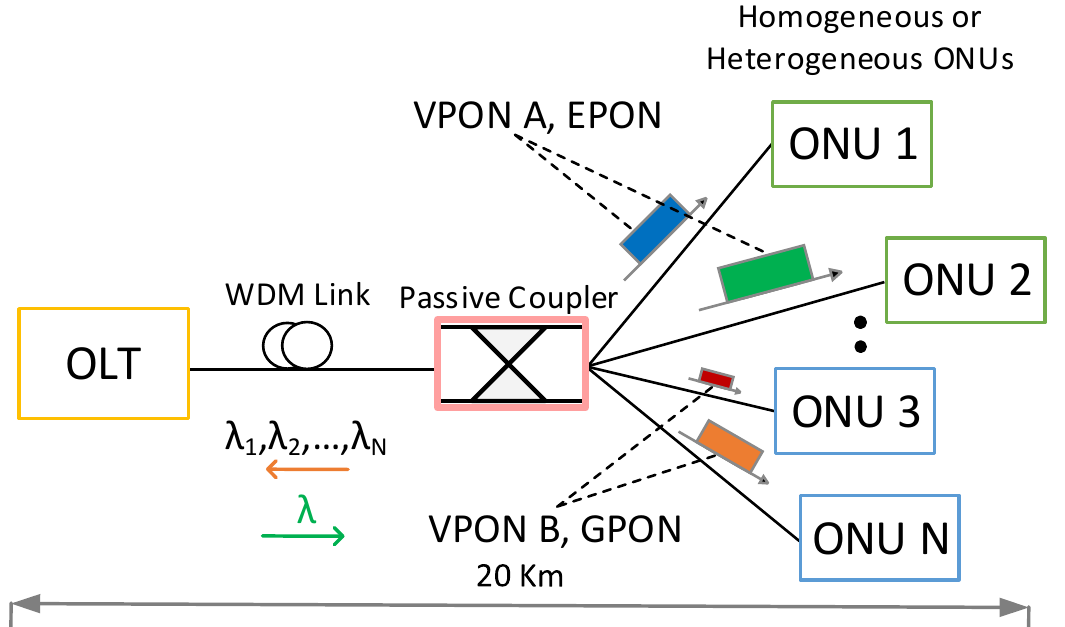} &
           \includegraphics[width=4in]{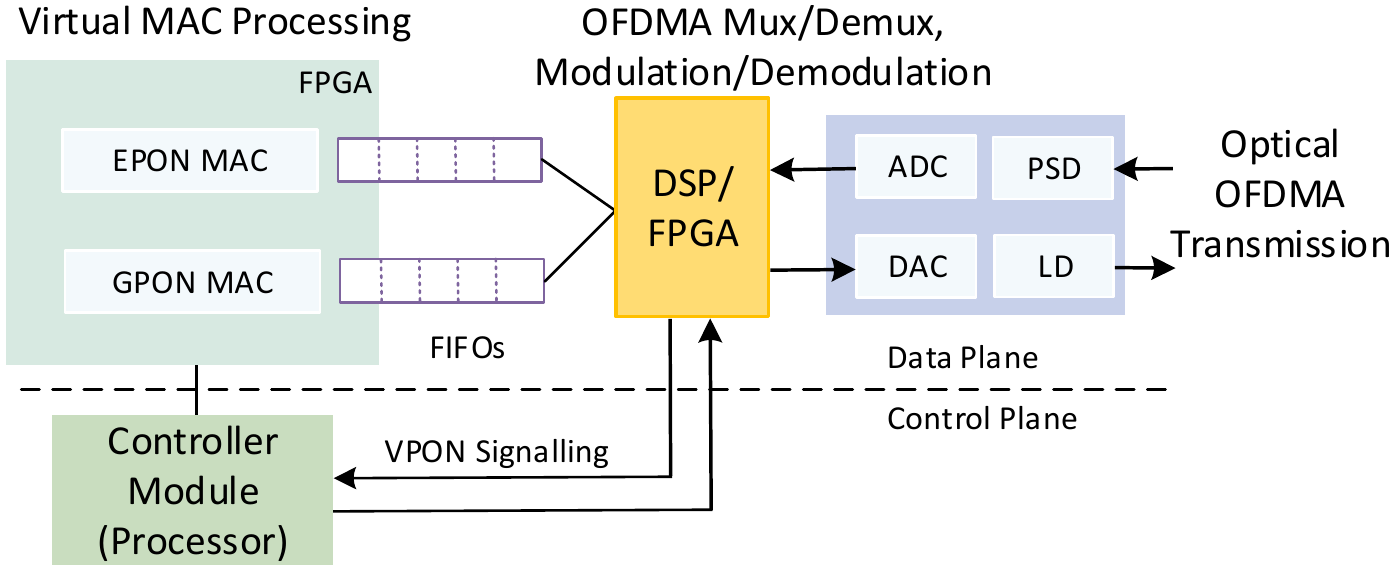} \\
\footnotesize{(a) Overall virtualization structure} &
\footnotesize{(b) Virtualization of OLT}
\end{tabular}
   \vspace{-.1cm}
    \caption{Illustration of OFDMA based virtual access
network~\cite{wei2009pon}: The different VPONs operate on isolated OFDMA
sub carriers allowing different MAC standards, such as EPON and GPON,
to operate on the same physical PON infrastructure, as illustrated in
part (a). A central SDN control module controls the OFDMA
transmissions and receptions as well as the virtual MAC processing,
see part (b).}
    \vspace{-.1cm}
    \label{fig_poniard}
\end{figure*}
\subsection{Access Networks}  \label{virt_access:sec}
\subsubsection{OFDMA Based PON Access Network Virtualization}
Wei et al.~\cite{wei2009pon,wei2009pro,wei2010ada} have developed
a link virtualization mechanism that can span from
optical access to backbone networks
based on Orthogonal Frequency Division Multiple Access (OFDMA).
Specifically, for access networks, a Virtual PON (VPON) approach
based on multicarrier OFDMA over WDM has been proposed. Distinct
network slices (VPONs) utilize distinct OFDMA subcarriers, which
provide a level of isolation between the VPONs. Thus, different
VPONs may operate with different MAC standards, e.g., as illustrated
in Fig.~\ref{fig_poniard}(a), VPON A may operate as an Ethernet PON
(EPON) while VPON~B operates as a Gigabit PON (GPON). In addition,
virtual MAC queues and processors are isolated to store
  and process the data from multiple VPONs, thus creating virtual
MAC protocols, as illustrated in Fig.~\ref{fig_poniard}(b).
The OFDMA transmissions and receptions are processed in a
DSP module that is controlled by a central SDN control module.
The central SDN control module also controls the different
virtual MAC processes in Fig.~\ref{fig_poniard}(b), which feed/receive
data to/from the DSP module.
Additional bandwidth partitioning between VPONs can be achieved through
Time Division Multiple Access (TDMA).
Simulation studies compared a static allocation of subcarriers to
VPONs with a dynamic allocation based on traffic demands.
The dynamic allocation achieved significantly higher numbers of
supported VPONs on a given network infrastructure as well as lower
packet delays than the static allocation.
A similar strategy for flexibly employing different dynamic
bandwidth allocation modules for different groups of ONU queues has been
examined in~\cite{li2016pro}.

Similar OFDMA based slicing strategies for supporting
cloud computing have been examined by Jinno et al.~\cite{jin2009vir}.
Zhou et al.~\cite{zhou2015dem} have explored a FlexPON with similar
virtualization capabilities. The FlexPON employs OFDM for adaptive
transmissions. The isolation of different VPONs is
mainly achieved through separate MAC processing.
The resulting VPONs allow for flexible port
assignments in ONUs and OLT, which have been demonstrated
in a testbed~\cite{zhou2015dem}.

\subsubsection{FiWi Access Network Virtualization}  \label{virt_fiwi:sec}
\paragraph{Virtualized FiWi Network}
Dai et al.~\cite{QiZSH14,QiGYZ13,ShGYZ13} have examined the
virtualization of FiWi networks~\cite{Bock2014,Effenberger2015}
to eliminate the differences between
the heterogeneous segments (fiber and wireless). The virtualization
provides a unified homogenous (virtual) view of the FiWi network.
The unified network view simplifies flow control and other
operational algorithms for traffic transmissions over the
heterogeneous network segments. In particular, a virtual resource
manager operates the heterogeneous segments. The resource manager
permits multiple routes from a given source node to a given
destination node. Load balancing across the multiple paths has been
examined in~\cite{he2013int,men2014eff}. Simulation results indicate
that the virtualized FiWi network with load balancing significantly
reduces packet delays compared to a conventional FiWi network. An
experimental OpenFlow switch testbed of the virtualized FiWi
network has been presented in~\cite{QiZHG14}. Testbed measurements
demonstrate the seamless networking across the heterogeneous fiber
and wireless networks segments. Measurements for nodal throughput,
link bandwidth utilization, and packet delay indicate performance
improvements due to the virtualized FiWi networking approach.
Moreover, the FiWi testbed performance is measured for a video
service scenario indicating that the virtualized FiWi networking approach
improves the Quality of Experience
(QoE)~\cite{che2014qos,seu2014sur} of the video streaming. A
mathematical performance model of the virtualized FiWi network has
been developed in~\cite{QiZHG14}.

\paragraph{WiMAX-VPON}
WiMAX-VPON~\cite{dha2010wim,DhPX10} is a Layer-2 Virtual Private Network
(VPN) design for FiWi access networks.
WiMAX-VPON executes a common MAC protocol across the
wireless and fiber network segments.
A VPN based admission control mechanism in conjunction with a VPN
bandwidth allocation ensures per-flow Quality of Service (QoS).
Results from discrete event simulations demonstrate that the
proposed WiMAX-VPON achieves favorable performance.
Also, Dhaini et al.~\cite{dha2010wim,DhPX10}
demonstrate how the WiMAX-VPON design can be extended
to different access network types with polling-based wireless and optical
medium access control.

\subsection{Data Centers}  \label{virt_dc:sec}

\subsubsection{LIGHTNESS}
LIGHTNESS~\cite{miao2015sdn,pen2015mul,pag2015opt,Saridis2016} is a
European research project examining an optical Data
Center Network (DCN) capable of providing dynamic,
programmable, and highly available DCN connectivity services.
Whereas conventional DCNs have rigid control and management
platforms, LIGHTNESS strives to introduce flexible control and
management through SDN control.
The LIGHTNESSS architecture comprises server racks that are
interconnected through optical packet
switches, optical circuit switches, and hybrid Top-of-the-Rack
(ToR) switches. The server racks and switches  are all
controlled and managed by an SDN controller.
LIGHTNESS control consists of an SDN controller above the optical
physical layer and OpenFlow agents that interact with the optical
network and server  elements. The SDN controller in cooperation with
the OpenFlow-agents provides a programmable data plane to the
virtualization modules.
The virtualization creates  multiple
Virtual Data Centers (VDCs), each with its own virtual computing
and memory resources, as well as virtual networking resources, based on
a given physical data center.
The virtualization is achieved through a VDC planner module and an
NFV application that directly interact with
the SDN controller.
The VDC planner composes the VDC slices through mapping of the VDC requests
to the physical SDN-controlled switches and server racks.
The VDC slices are monitored by the NFV application, which interfaces
with the VDC planner. Based on monitoring data, the NFV application
and VDC planner may revise the VDC composition, e.g., transition from
optical packet switches to optical circuit switches.

\subsubsection{Cloudnets}
Cloudnets~\cite{azo2013sdn,ban2013mer,bari2013data,FerLR13,WoRSV11,WoRVS10}
exploit network virtualization for
pooling resources among distributed data centers. Cloudnets support
the migration of virtual machines across networks to achieve
resource pooling. Cloudnet designs can be supported through
optical networks~\cite{ShNS13}.
Kantarci and Mouftah~\cite{kan2015res} have examined
designs for a virtual cloud backbone network that interconnects
distributed backbone nodes, whereby each backbone node is
associated with one data center. A network resource manager
periodically executes a virtualization algorithm to accommodate
traffic demands through appropriate resource provisioning. Kantarci
and Mouftah~\cite{kan2015res} have developed and evaluated algorithms for
three provisioning objectives: minimize the outage probability of
the cloud, minimize the resource provisioning, and minimize a
tradeoff between resource saving and cloud outage probability. The
range of performance characteristics for outage probability,
resource consumption, and delays of the provisioning approaches have
been evaluated through simulations. The outage probability of
optical cloud networks has been reduced in~\cite{ahm2014enh} through
optimized service re-locations.

Several complementary aspects of optical cloudnet networks have
recently been investigated.
A multilayer network architecture with an SDN based
network management structure for cloud services has been
developed in~\cite{dov2015usi}.
A dynamic variation of the sharing of optical network resources
for intra- and inter-data center networking has been examined
in~\cite{xie2014dyn}.
The dynamic sharing does not statically assign optical network resources
to virtual optical networks; instead, the network resources are
dynamically assigned according to the time-varying traffic demands.
An SDN based optical transport mode for data center traffic has been
explored in~\cite{vel2014tow}.
Virtual machine migration mechanisms that take the characteristics
of renewable energy into account have been examined in~\cite{zha2015ren}
while general energy efficiency mechanisms for optically networked
could computing resources have been examined in~\cite{tza2014con}.

\subsection{Metro/Core Networks}  \label{virt_core:sec}

\subsubsection{Virtual Optical Network Embedding}  \label{virt_emb:sec}
Virtual optical network embedding seeks to map requests for virtual
optical networks to a given physical optical network infrastructure (substrate).
A virtual optical network consists of both a set of virtual
nodes and a set of interconnecting links that need to
be mapped to the network substrate.
This mapping of virtual networks consisting of both
network nodes and links is fundamentally different from the
extensively studied virtual topology design for optical wavelength
routed networks~\cite{dut2000sur}, which only considered network links
(and did not map nodes).
Virtual network embedding of both nodes and link has already been
extensively studied in general network graphs~\cite{fis2013vir,rah2013svn}.
However, virtual optical network embedding requires additional
constraints to account for the special optical transmission characteristics,
such as the wavelength continuity constraint and the transmission reach
constraint.
Consequently, several studies have begun to examine virtual network
embedding algorithms specifically for optical networks.

\paragraph{Impairment-Aware Embedding}
Peng et al.~\cite{pen2011imp,PeNS13} have modeled the optical
transmission impairments to facilitate the embedding of isolated
VONs in a given
underlying physical network infrastructure.
Specifically, they model the physical (photonic)
layer impairments of both single-line
rate and mixed-line rates~\cite{nag2010opt}.
Peng et al.~\cite{PeNS13} consider intra-VON impairments
from Amplified Spontaneous Emission (ASE) and inter-VON impairments
from non-linear impairments and four wave mixing.
These impairments are captured in a $Q$-factor~\cite{azo2009sur,sar2009phy},
which is considered in the mapping of
virtual links to the underlying physical link resources,
such as wavelengths and wavebands.

\paragraph{Embedding on WDM and Flexi-grid Networks} \label{wdm_flexi_emb:sec}
Zhang et al.~\cite{zha2013net} have considered the embedding of
overall virtual networks encompassing both virtual nodes and virtual links.
Zhang et al. have considered both conventional WDM networks as well as
flexi-grid networks.
For each network type, they formulate the virtual node and virtual link
mapping as a mixed integer linear program.
Concluding that the mixed integer
linear program is NP-hard, heuristic solution approaches
are developed. Specifically, the overall embedding (mapping) problem
is divided into a node mapping problem and a link mapping problem.
The node mapping problem is heuristically solved through a greedy
MinMapping strategy that maps the largest computing resource demand to the
node with the minimum remaining computing capacity (a complementary
MaxMapping strategy that maps the largest demand to the
node with the maximum remaining capacity is also considered).
After the node mapping, the link mapping problem is solved
with an extended grooming graph~\cite{zhu2003nov}.
Comparisons for a small network indicate that
the MinMapping strategy approaches the optimal mixed integer linear program
solution quite closely;
whereas the MaxMapping strategy gives poor results.
The evaluations also indicate that the flexi-grid network requires only about
half the spectrum compared to an equivalent WDM network for several
evaluation scenarios.

The embedding of virtual optical networks in the context of
elastic flexi-grid optical networking has been further examined in several
studies.
For a flexi-grid network based on OFDM~\cite{zha2013surflexi},
Zhao et al.~\cite{zha2013vir} have compared a greedy heuristic that maps
requests in decreasing order of the required resources with an arbitrary
first-fit benchmark.
Gong et al.~\cite{gon2014vir} have considered flexi-grid networks with
a similar overall strategy of node mapping followed by link mapping
as Zhang et al.~\cite{zha2013net}.
Based on the local resource constraints at each node,
Gong et al.~have formed a layered auxiliary graph for the node mapping.
The link mapping is then solved with a shortest path routing approach.
Wang et al.~\cite{wan2015vir} have examined an embedding approach
based on candidate mapping patterns that could provide the requested
resources. The VON is then embedded according to a shortest path
routing.
Pages et al.~\cite{pag2015opt} have considered embeddings that
minimize the required optical transponders.

che2016cos

\paragraph{Survivable Embedding}  \label{surv_emb:sec}
Survivability of a virtual optical network, i.e., its continued
operation in the face of physical node or link failures, is
important for many applications that require dependable service. Hu
et al.~\cite{hu2013sur} developed an embedding that can survive the
failure of a single physical node. Ye et al.~\cite{ye2015sur} have
examined the embedding of virtual optical networks so as to survive
the failure of a single physical node or a physical link.
Specifically, Ye et al. ensure that each virtual
node request is mapped to a primary physical node as well as a
distinct backup physical node. Similarly, each virtual link is
mapped to a primary physical route as well as a node-disjoint backup
physical route. Ye et al. mathematically formulate an optimization
problem for the survivable embedding and then propose a Parallel
Virtual Infrastructure (VI) Mapping (PAR) algorithm. The PAR
algorithm finds distinct candidate physical nodes (with the highest
remaining resources) for each virtual node request. The candidate
physical nodes are then jointly examined with pairs of
 shortest node-disjoint paths.
The evaluations in~\cite{ye2015sur} indicate that the parallel
PAR algorithm reduces the blocking probabilities of virtual network requests
by 5--20~\% compared to a sequential algorithm benchmark.
A limitation of the survivable embedding~\cite{ye2015sur} is that it
protects only from a single link or node failure.
As the optical infrastructure is expected to penetrate
deeper in the access network deployments (e.g., mobile backhaul),
it will become necessary to consider multiple failure points.
Similar survivable network embedding algorithms that employ node-disjoint
shortest paths in conjunction with specific cost metrics for
node mappings have been investigated by Xie et al.~\cite{xie2014sur}
and Chen et al.~\cite{che2016cos}.
Jiang et al.~\cite{jia2015ava} have examined a solution variant based
on maximum-weight maximum clique formation.

The studies~\cite{son2011ban,pao2014mul,ass2016net} have examined
so-called bandwidth squeezed restoration for virtual topologies.
With bandwidth squeezing, the back-up path
bandwidths of the surviving virtual topologies are generally lower than
the bandwidths on the working paths.

Survivable virtual topology design in the context of multidomain optical
networks has been studied by Hong et al.~\cite{Hong2015}.
Hong et al.~focused on minimizing the total network link cost
for a given virtual traffic demand. A heuristic algorithm
for partition and contraction
mechanisms based on cut set theory has been proposed for the
mapping of virtual links onto multidomain optical networks.
A hierarchical SDN control plane is split between local controllers
that to manage individual domains and a
global controller for the overall management.
The partition and contraction
mechanisms abstract inter- and intra-domain information as a
method of contraction. Survivability conditions are ensured
individually for inter- and intra-domains such that survivability is
met for the entire network.
The evaluations in~\cite{Hong2015} demonstrate successful virtual
network mapping at the scale required by commercial Internet
service providers and infrastructure providers.

\paragraph{Dynamic Embedding}
The embedding approaches surveyed so far have mainly focused on the
offline embedding of a static set of virtual network requests.
However, in the ongoing network operation the dynamic embedding of
modifications (upgrades) of existing virtual networks, or the
addition of new virtual networks are important. Ye et
al.~\cite{ye2014upg} have examined a variety of strategies for
upgrading existing virtual topologies. Ye et al. have considered
both scenarios without advance planning (knowledge) of virtual
network upgrades and scenarios that plan ahead for possible
(anticipated) upgrades. For both scenarios, a divide-and-conquer
strategy and an integrate-and-cooperate strategy are
examined. The divide-and conquer strategy sequentially maps all the
virtual nodes
and then the virtual links. In contrast, the integrate-and-cooperate
strategy jointly
considers the virtual node and virtual link mappings. Without
advance planning, these strategies are applied sequentially, as the
virtual network requests arrive over time, whereas, with planning,
the initial and upgrade requests are jointly considered. Evaluation
results indicate that the integrate-and-cooperate strategy
slightly increases a revenue
measure and request acceptance ratio compared to the divide-and-conquer
strategy.
The results also indicate that planning has the potential to
substantially increase the revenue and acceptance ratio. In a
related study, Zhang et al.~\cite{zha2015dyn} have examined
embedding algorithms for virtual network requests that arrive
dynamically to a multilayer network consisting of electrical and
optical network substrates.

\paragraph{Energy-efficient Embedding}
Motivated by the growing importance of green networking and information
technology~\cite{BiPCR12}, a few studies have begun to consider
the energy efficiency of the embedded virtual optical networks.
Nonde et al.~\cite{non2015ene} have developed and evaluated
mechanisms for embedding virtual cloud networks so as to minimize
the overall power consumption, i.e., the aggregate of the
power consumption for communication and computing (in the data centers).
Nonde et al. have incorporated the power consumption of the
communication components, such as transponders and optical switches,
as well as the power consumption characteristics of data center servers
into a mathematical power minimization model.
Nonde et al. then develop a real-time heuristic for energy-optimized
virtual network embedding.
The heuristic strives to consolidate computing requests in the
physical nodes with the least residual computing capacity.
This consolidation strategy is motivated by the typical
power consumption characteristic of a compute server that
has a significant idle power consumption and then grows linearly with
increasing computing load; thus a fully loaded server is more
energy-efficient than a lightly loaded server.
The bandwidth demands are then routed between the nodes according to
 a minimum hop algorithm.
The energy optimized embedding is compared with a cost optimized
 embedding that only seeks to minimize the number of utilized wavelength
 channels.
 The evaluation results in~\cite{non2015ene} indicate
 that the energy optimized embedding significantly reduces the overall energy
 consumption for low to moderate loads on the physical infrastructure;
 for high loads, when all physical resources need to be utilized,
 there are no significant savings. Across the entire load range, the
 energy optimized embedding saves on average 20~\% energy compared to the
 benchmark minimizing the wavelength channels.

Chen~\cite{che2016pow} has examined a similar energy-efficient
virtual optical network embedding that considers primary and
link-disjoint backup paths, similar to the survivable embeddings in
Section~\ref{surv_emb:sec}. More specifically, virtual link requests
are mapped in decreasing order of their bandwidth requirements to
the shortest physical transmission distance paths, i.e., the highest
virtual bandwidth demands are allocated to the shortest physical
paths. Evaluations indicate that this link mapping approach roughly
halves the power consumption compared to a random node mapping
benchmark. Further studies focused on energy savings have examined
virtual link embeddings that maximize the usage of nodes with
renewable energy~\cite{she2014fol} and the traffic
grooming~\cite{wan2014hie} onto sliceable BVTs~\cite{zha2015ene}.

\subsubsection{Hypervisors for VONs} \label{virt_hv:sec}
The operation of VONs over a given
underlying physical (substrate) optical network requires an
intermediate hypervisor. The hypervisor presents the physical
network as multiple isolated VONs to the corresponding VON
controllers (with typically one VON controller per VON). In turn,
the hypervisor intercepts the control messages issued by a VON
controller and controls the physical network to effect the control
actions desired by the VON controller for the corresponding VON.

Towards the development of an optical network hypervisor, Siquera et
al.~\cite{siq2015pro} have developed a SDN-based controller for an
optical transport architecture. The controller implements a
virtualized GMPLS control plane with offloading to facilitate the
implementation of hypervisor functionalities, namely the creation
optical virtual private networks, optical network slicing, and
optical interface management. A major contribution of Siquera et
al.~\cite{siq2015pro} is a Transport Network Operating System
(T-NOS), which abstracts the physical layer for the controller and
could be utilized for hypervisor functionalities.

OpenSlice~\cite{LiuMCTMM13} is a comprehensive
OpenFlow-based hypervisor that creates VONs over underlying elastic
optical networks~\cite{cha2015rou,tal2014spe}. OpenSlice
dynamically provisions end-to-end paths and offloads IP traffic by
slicing the optical communications spectrum. The paths are set up
through a handshake protocol that fills in cross-connection table
entries. The control messages for slicing the optical communications
spectrum, such as slot width and modulation format, are carried in
extended OpenFlow protocol messages. OpenSlice relies on special
distributed network elements, namely bandwidth variable wavelength
cross-connects~\cite{jin2009spe} and multiflow optical
transponders~\cite{jin2012mul} that have been extended for control
through the extended OpenFlow messages. The OpenSlice evaluation
includes an experimental demonstration. The evaluation results
include path provisioning latency comparisons with a GMPLS-based
control plane and indicate that OpenFlow outperforms GMPLS for paths
with more than three hops.
OpenSlice extension and refinements to
multilayer and multidomain networks are surveyed in
Section~\ref{orch:sec}.
An alternate centralized Optical FlowVisor that does not require
extensions to the distributed network elements has been investigated
in~\cite{Azod2012}.

\subsection{Virtualization: Summary and Discussion}
The virtualization studies on access
networks~\cite{wei2009pon,wei2009pro,wei2010ada,jin2009vir,zhou2015dem,QiZSH14,QiGYZ13,ShGYZ13,he2013int,men2014eff,QiZHG14,dha2010wim,DhPX10} have primarily
focused on exploiting and manipulating the specific properties of the
optical physical layer (e.g., different OFDMA subcarriers)
and MAC layer (e.g., polling based MAC protocol) of the optical
access networks for virtualization.
In addition, to virtualization studies on
purely optical PON access networks, two sets of studies, namely
sets~\cite{QiZSH14,QiGYZ13,ShGYZ13,he2013int,men2014eff,QiZHG14}
and WiMAX-VPON~\cite{dha2010wim,DhPX10} have examined
virtualization for two forms of FiWi access networks.
Future research needs to consider virtualization of a wider set of
FiWi network technologies, i.e., FiWi networks that consider
optical access networks with a wider variety of wireless access
technologies, such as different forms of cellular access or combinations
of cellular with other forms of wireless access.
Also, virtualization of integrated access and metropolitan area
networks~\cite{ahm2012rpr,seg2007all,val2015exp,woe2013sdn} is
an important future research direction.

A set of studies has begun to explore optical networking support for
SDN-enabled cloudnets that exploit virtualization to
dynamically pool resources across distributed data centers.
One important direction for
future work on cloudnets is to examine moving data center resources
closer to the users and the subsequent resource pooling across edge
networks~\cite{man2013clo}. Also, the exploration of the benefits of
FiWi networks for decentralized
cloudlets~\cite{din2013sur,sat2009cas,ScSF12,ver2012clo} that
support mobile wireless network services is an important future
research direction~\cite{mai2015inv}.

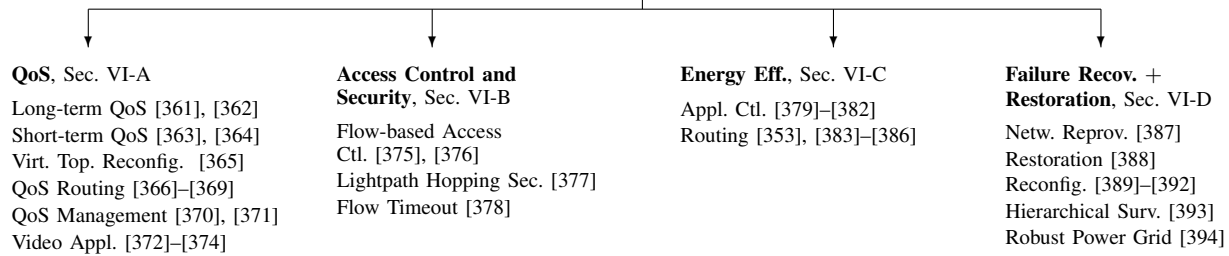
\begin{figure*}[t!]
	\footnotesize
	\setlength{\unitlength}{0.10in} 
	\centering
	\begin{picture}(40,33)
	\put(15,33){\textbf{Applications, Sec.~\ref{sdnapp:sec}}}
	\put(-8,30){\line(1,0){53}}
	\put(21,30){\line(0,1){2}}
	
	\put(-8,30){\vector(0,-1){2}}
	\put(-12,27){\makebox(0,0)[lt]{\shortstack[l]{			
\textbf{QoS}, Sec.~\ref{app_qos:sec} \\ \\
				Long-term QoS~\cite{ZhZZ13,kho2016qua} \\
				Short-term QoS~\cite{Li2014,PatelJiWang2013} \\
				Virt. Top. Reconfig.~~\cite{WetteKarl2013} \\
				QoS Routing~\cite{Tariq2015,SgPCVC13,Ilchmann2015,ChangLi2015} \\
				QoS Management~\cite{RuBRK14,TeMAD14} \\
				Video Appl.~\cite{chi2015app,Chitimalla2015,Li2014video}
			}}}
			
		\put(10,30){\vector(0,-1){2}}
		\put(5,27){\makebox(0,0)[lt]{\shortstack[l]{		
\textbf{Access Control and} \\
\textbf{Security}, Sec.~\ref{app_secur:sec}\\ \\  	
				Flow-based Access \\ Ctl.~\cite{MaGMT14,nay2009res}	\\			
				Lightpath Hopping Sec.~\cite{Li2016c}	\\			
				Flow Timeout~\cite{ZhuFan2015}
			}}}
					
		\put(31,30){\vector(0,-1){2}}
		\put(23,27){\makebox(0,0)[lt]{\shortstack[l]{			
\textbf{Energy Eff.}, Sec.~\ref{ene_eff:sec}	\\ \\  			
				Appl. Ctl.~\cite{Ji2014,yan2013mul,yan2015per,yan2016per}			\\	
				Routing~\cite{Tego2014,Wang2015,Yevsieieva2015,Yoon2015,val2015exp}
			}}}
	
		\put(45,30){\vector(0,-1){2}}		
		\put(40,27){\makebox(0,0)[lt]{\shortstack[l]{			
\textbf{Failure Recov. $+$} \\
\textbf{Restoration}, Sec.~\ref{app_fail_rec:sec} \\ \\
				Netw. Reprov.~\cite{sav2015bac} \\
				Restoration~\cite{Giorgetti2015} \\
		Reconfig.~\cite{Aguado2016,aib2016sof,SlKMPR14,Kim2015} \\
				Hierarchical Surv.~\cite{ZhangSong2014} \\
				Robust Power Grid~\cite{Rastegarfar2016}
			}}}
									
\end{picture}
\vspace{-4.6cm}	
\caption{Classification of application layer SDON studies.}
\label{app_class:fig}
\end{figure*}
A fairly extensive set of studies has examined virtual network embedding
for metro/core networks.
The virtual network embedding studies have considered the
specific limitations and constraints of optical networks and have
begun to explore specialized embedding strategies that strive to
meet a specific optimization objective,
such as survivability, dynamic adaptability,
or energy efficiency.
Future research should seek to develop a comprehensive framework of
embedding algorithms that can be tuned with weights to achieve
prescribed degrees of the different optimization objectives.

A relatively smaller set of studies has developed and refined hypervisors
for creating VONs over metro/core optical networks.
Much of the SDON hypervisor research has centered on the OpenSlice
hypervisor concept~\cite{LiuMCTMM13}. While OpenSlice accounts
for the specific characteristics of the optical transmission medium,
it is relatively complex as it requires a distributed implementation
with specialized optical networking components.
Future research should seek to achieve the hypervisor functionalities
with a wider set of common optical components so as to reduce cost
and complexity.
Overall, SDON hypervisor research should examine the performance-complexity/cost
tradeoffs of distributed versus centralized approaches.
Within this context of examining the spectrum of distributed to
centralized hypervisors,
future hypervisor research should further refine and optimize the
virtualization mechanisms so as to achieve strict isolation between
virtual network slices, as well as low-complexity hypervisor
deployment, operation, and maintenance.

\section{SDN Application Layer}  \label{sdnapp:sec}
In the SDN paradigm, applications interact with the controllers to
implement network services. We organize the survey of the studies on
application layer aspects of SDONs according to the main application
categories of quality of service (QoS), access control and security,
energy efficiency, and failure recovery, as illustrated in
Fig.~\ref{app_class:fig}.

\subsection{QoS}  \label{app_qos:sec}
\subsubsection{Long-term QoS: Time-Aware SDN}
Data Center (DC) networks move data back and forth between
DCs to balance the computing load and the data storage usage
(for upload)~\cite{DeCusatis2014}.
These data movements between DCs can span large geographical areas and help
ensure DC service QoS for the end users.
Load balancing algorithms can exploit the characteristics
of the user requests.
One such request characteristic is the high degree of time-correlation over
various time scales
ranging from several hours of a day (e.g., due to a sporting event)
to several days in a year (e.g., due to a political event).
Zhao et al.~\cite{ZhZZ13} have proposed a time-aware SDN application using
OpenFlow extensions to dynamically balance the load across the
DC resources so as to improve the QoS.
Specifically, a time correlated PCE algorithm based on flexi-grid optical
transport (see Section~\ref{PCE:sec})
has been proposed. An SDN application monitors
the DC resources and applies network rules to preserve the QoS.
Evaluations of the algorithm indicate improvements
in terms of network blocking probability, global blocking probability, and
spectrum consumption ratio.
This study did not consider short time scale traffic bursts,
which can significantly affect the load conditions.

We believe that in order to avoid pitfalls in the operation of load balancing
through PCE algorithms implemented with SDN, a wide range of
traffic conditions needs to be considered.
The considered traffic range should include short and long term traffic
variations, which should be traded off with various QoS aspects,
such as type of application
and delay constraints, as well as the resulting costs and control overheads.
Khodakarami et al.~\cite{kho2016qua} have taken steps in this direction by
forming a traffic forecasting model for both long-term and short-term forecasts
in a wide-area mesh network.
Optical lightpaths are then configured based on the overall traffic forecast,
while electronic switching capacities are allocated based on short-term
forecasts.

\begin{figure}[t!]
    \centering
    \vspace{-.1cm}
    \includegraphics[width=3.2in]{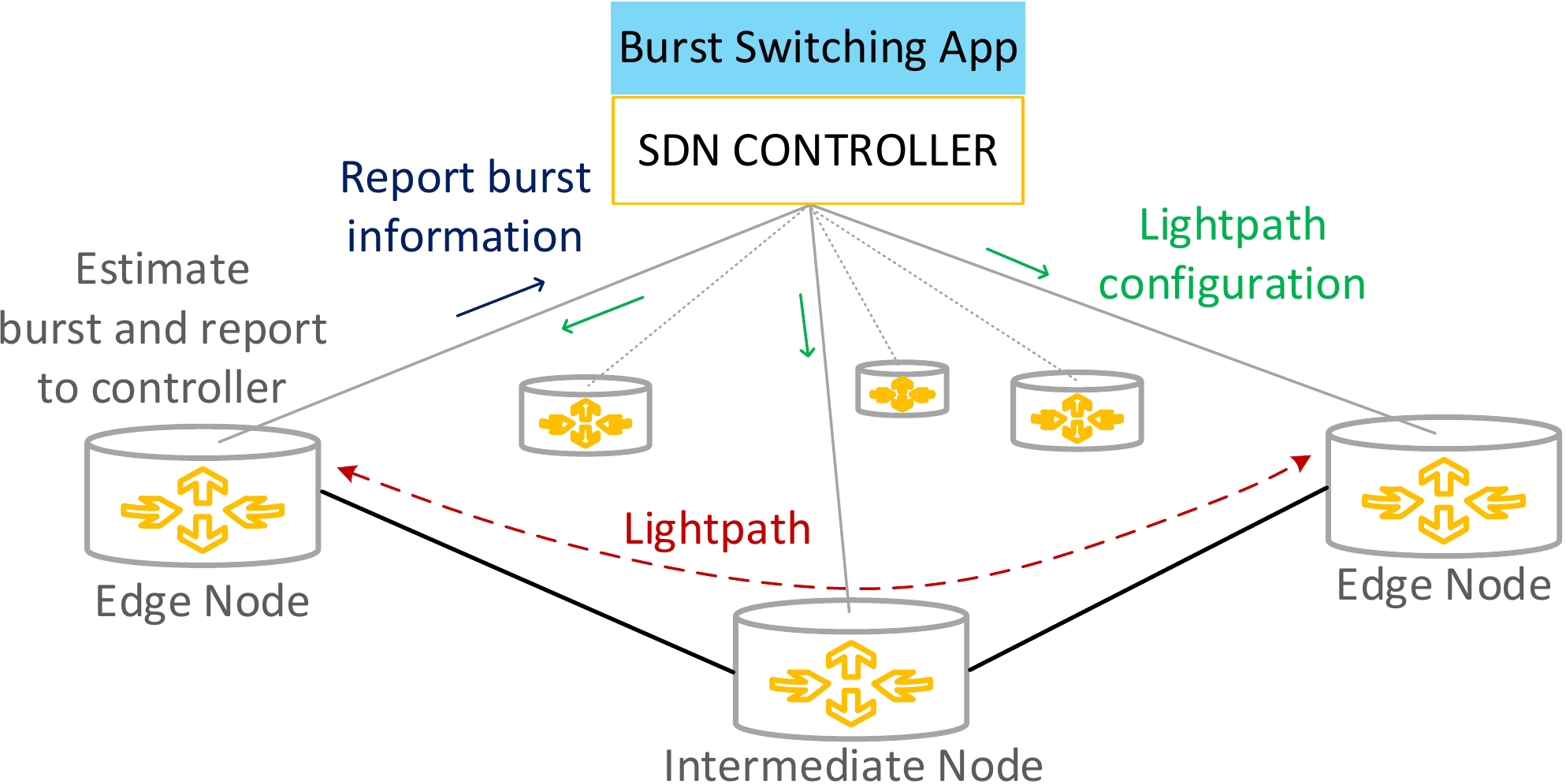}
    \vspace{-.1cm}
    \caption{Optical SDN-based QoS-aware burst switching
      application~\cite{PatelJiWang2013}:
      Based on short-term traffic burst estimates of an edge node,
the SDN controller configures end-to-end light paths ensuring QoS.}
   \vspace{-.05cm}
    \label{fig_app_qos}
\end{figure}
\subsubsection{Short Term QoS}
Users of a high-speed FTTH access network may request
very large bandwidths
due to simultaneously running applications that require high data rates.
In such a scenario,
applications requiring very high data rates may affect each other.
For instance, a video conference running simultaneously with
the streaming of a sports video may result in
call drops in the video conference application and
in stalls of the sports video.
Li et al.~\cite{Li2014} proposed an SDN based bandwidth provisioning
application in the broadband remote access server~\cite{Dietz2015}
network. They defined and assigned the minimum bandwidth, which they named
``sweet point'', required
for each application to experience good QoE.
Li et al. showed that maintaining the ``sweet point'' bandwidth for each
application can significantly improve the QoE while other
applications are being served
according to their bandwidth requirements.

In a similar study, Patel et al.~\cite{PatelJiWang2013} proposed
a burst switching mechanism based on a software
defined optical network. Bursts typically originate
at the edge nodes and the aggregation
points due to statistical multiplexing of high speed
optical transmissions. To ensure QoS for multiple traffic classes,
bursts at the edge nodes have to be managed by deciding their end-to-end
path to meet their QoS requirements, such as minimum delay and data rate.
In non-SDN based mechanisms, complicated
distributed protocols, such as GMPLS~\cite{mun2014pce,pao2013sur},
are used to route the burst traffic.
In the proposed application,
the centralized unified control plane decides the routing
path for the burst based on latency and QoS requirements.
A simplified procedure involves $(i)$ burst evaluation at the edge node,
$(ii)$ reporting burst information to the SDN controller, and
$(iii)$ sending of configurations to the optical nodes by the controller
to set up a lightpath as illustrated in Fig.~\ref{fig_app_qos}.
Simulations indicate an increase of performance in terms of
throughput, network blocking probability, and latency
along with improved QoS when compared to non-SDN GMPLS methods.

\subsubsection{Virtual Topology Reconfigurations}
The QoS experienced by traffic flows greatly depends on their
route through a network.
Wette et al.~\cite{WetteKarl2013} have examined an application algorithm that
reconfigures WDM network virtual topologies
(see Section~\ref{wdm_flexi_emb:sec})
according to the traffic levels.
The algorithm considers the localized traffic information and
optical resource availability at the nodes.
The algorithm does not require synchronization,
thus reducing the overhead while simplifying the network design.
In the proposed architecture, optical switches are connected to ROADMs.
The reconfiguration application manages
and controls the optical switches through the SDN controller.
A new WDM controller is introduced to configure the lightpaths taking
 wavelength conversion and lightpath switching at the ROADMs into consideration.
The SDN controller operates on the optical network which appears as a
static network, while the WDM controller configures (and re-configures)
the ROADMs to create
multiple virtual optical networks according to the traffic levels.
Evaluation results indicate improved utilization and throughput.
The results indicate that virtual topologies reconfigurations can
significantly increase the flexibility of the network while achieving
the desired QoS.  However, the control overhead and the delay aspects
due to virtualization and separation of control and lightwave paths
needs to be carefully considered.

\subsubsection{End-to-End QoS Routing}
Interconnections between DCs involve typically
multiple data paths. All the interfaces existing
between DCs can be utilized by MultiPath TCP (MPTCP).
Ensuring QoS in such an MPTCP setting
while preserving throughput efficiency in a
reconfigurable underlying burst switching optical network is a
challenging task. Tariq et al.~\cite{Tariq2015} have
proposed QoS-aware bandwidth reservation for
MPTCP in an SDON.
The bandwidth reservation proceeds in two stages
$(i)$ path selection for MPTCP, and
$(ii)$ OBS wavelength reservation to assign the priorities for
latency-sensitive flows. Larger portions of a wavelength reservation are
assigned to high priority flows, resulting in reduced
burst blocking probability
while achieving the higher MPTCP throughput.
The simulation results in~\cite{Tariq2015} validate the two-stage algorithm
for QoS-aware  MPTCP over an SDON, indicating decreased dropping probabilities,
and increased throughputs.

\begin{figure}[t!]
    \centering
    \vspace{-.1cm}
    \includegraphics[width=3in]{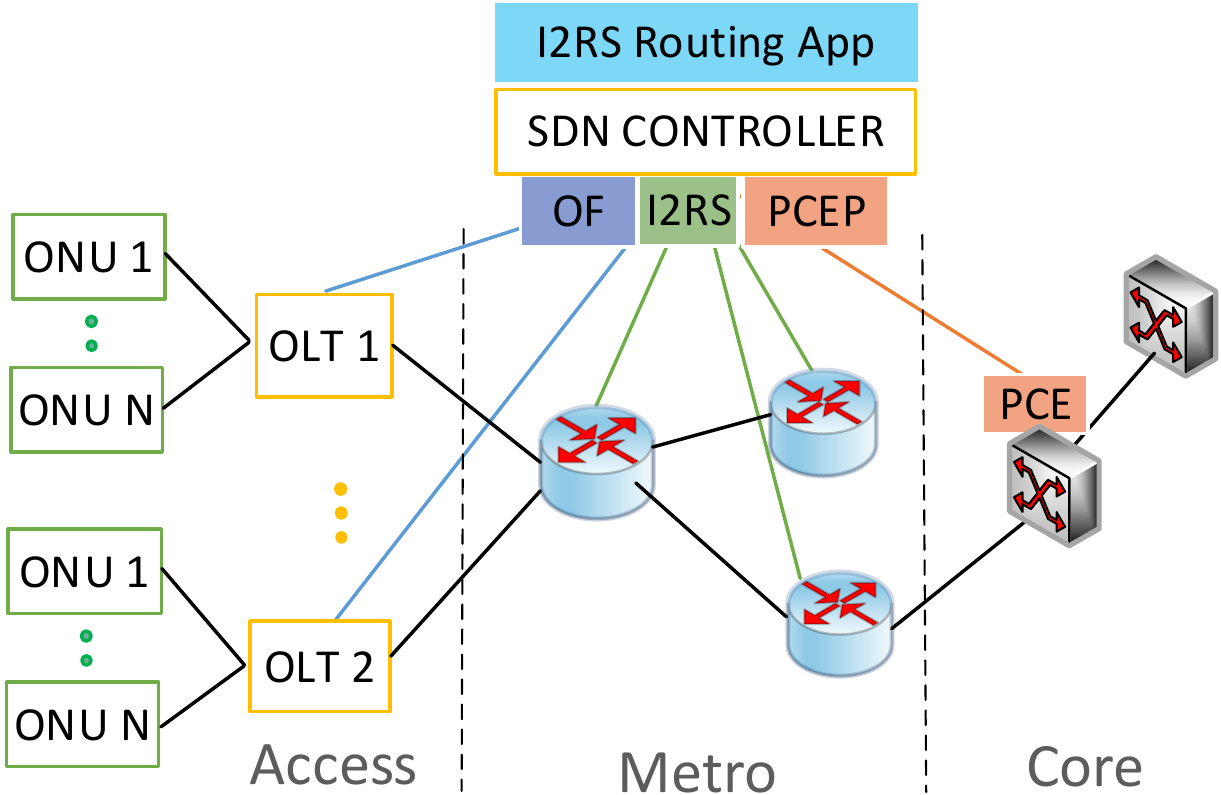}
    \vspace{-.1cm}
    \caption{Illustration of routing application
 with integrated control of access, metro, and core networks using SDN and the
Information To the Routing System (I2RS)~\cite{SgPCVC13}: The SDN controller
interacts with the access network, e.g., through the OpenFlow protocol, the
metro network, e.g., through the I2RS, and the core network, e.g., through
the Path Computation Elements (PCEs). }
   \vspace{-.05cm}
    \label{fig_app_i2rs}
\end{figure}
Information To the Routing System (I2RS)~\cite{I2RSietf} is a high-level
architecture for communicating and interacting with routing systems,
such as BGP routers.
A routing system may consists of several complex
functional entities, such as a Routing Information Base (RIB),
an RIB manager, topology and policy databases,
along with routing and signalling units.
The I2RS provides a programmability platform that enables
access and modifications of the configurations of the routing system elements.
The I2RS can be extended with SDN principles to achieve global network
management and reconfiguration~\cite{har2013sof}.
Sgambelluri et al.~\cite{SgPCVC13} presented an SDN based routing application
within the I2RS framework to integrate the control of the
access, metro, and core networks as illustrated in Fig.~\ref{fig_app_i2rs}.
The SDN controller communicates with the Path Computation
Elements (PCEs) of the core network
to create Label Switched Paths (LSPs)
based on the information received by the OLTs.
Experimental demonstrations validated the routing optimization
based on the current traffic status and previous load as well as
the unified control interface for access, metro, and core networks.

Ilchmann et al.~\cite{Ilchmann2015} developed an SDN application that communicates to an SDN controller via
an HTTP-based REST API. Over time, lightpaths in an optical network can become inefficient for a number of
reasons (e.g., optical spectrum fragmentation). For this reason, Ilchmann et al. developed an SDN application
that evaluates existing lightpaths in an optical network and offers an application user the option to
re-optimize the lightpath routing to improve various performance metrics (e.g., path length). The application
is user-interactive in that the user can see the number of proposed lightpath routing changes before they are made
and can potentially select a subset of the proposed changes to minimize network down-time.

At the ingress and egress routers of optical networks (e.g., the
edge routers between access and metro networks), buffers are highly
non-economical to implement, as they require large buffers sizes to
accommodate the channel rates of 40~Mb/s or more. To reduce the buffer
requirements at the edge routers, Chang et al.~\cite{ChangLi2015}
have proposed a backpressure application referred to as Refill and
SDN-based Random Early Detection (RS-RED).  RS-RED implements a
refill queue at the ingress device and a droptail
queue at the egress device, whereby both queues are centrally
managed by the RS-RED algorithm running on the SDN
controller. Simulation results showed that at the expense of small delay
increases, edge router buffer sizes can be significantly reduced.

\subsubsection{QoS Management}
Rukert et al.~\cite{RuBRK14} proposed SDN based controlled home-gateway
supporting heterogeneous wired technologies, such as DSL, and
wireless technologies, such as LTE and WiFi.
SDN controllers managed by the ISPs optimize the traffic flows
to each user while accommodating large numbers of users  and ensuring their
minimum QoS.
Additionally, Tego et al.~\cite{TeMAD14} demonstrated an experimental
SDN based QoS management setup to optimize the energy utilization.
GbE links are switched on and off based on the traffic levels.
The QoS management reroutes the traffic to avoid congestion and achieve
efficient throughput. SDN applications conduct active QoS
probing to monitor the network QoS characteristics.
Evaluations have indicated that the SDN based techniques
achieve significantly higher throughput than non-SDN techniques~\cite{TeMAD14}.

\subsubsection{Video Applications}
The application-aware SDN-enabled resource allocation
application has been introduced by Chitimalla et al.~\cite{chi2015app}
to improve the video QoE in a PON access network.
The resource allocation application
uses application level feedback to schedule the optical
resources. The video resolution is incrementally increased or
decreased based on the buffer utilization statistics that the client
sends to the controller. The scheduler at the OLT schedules the
packets based on weights calculated by the SDN controller, whereby
the video applications at the clients communicate with the
controller to determine the weights. If the network is congested,
then the SDN controller communicates to the clients to reduce the
video resolution so as to reduce the stalls and to improve the QoE.

\begin{figure}[t!]
    \centering
    \vspace{-.1cm}
    \includegraphics[width=3.2in]{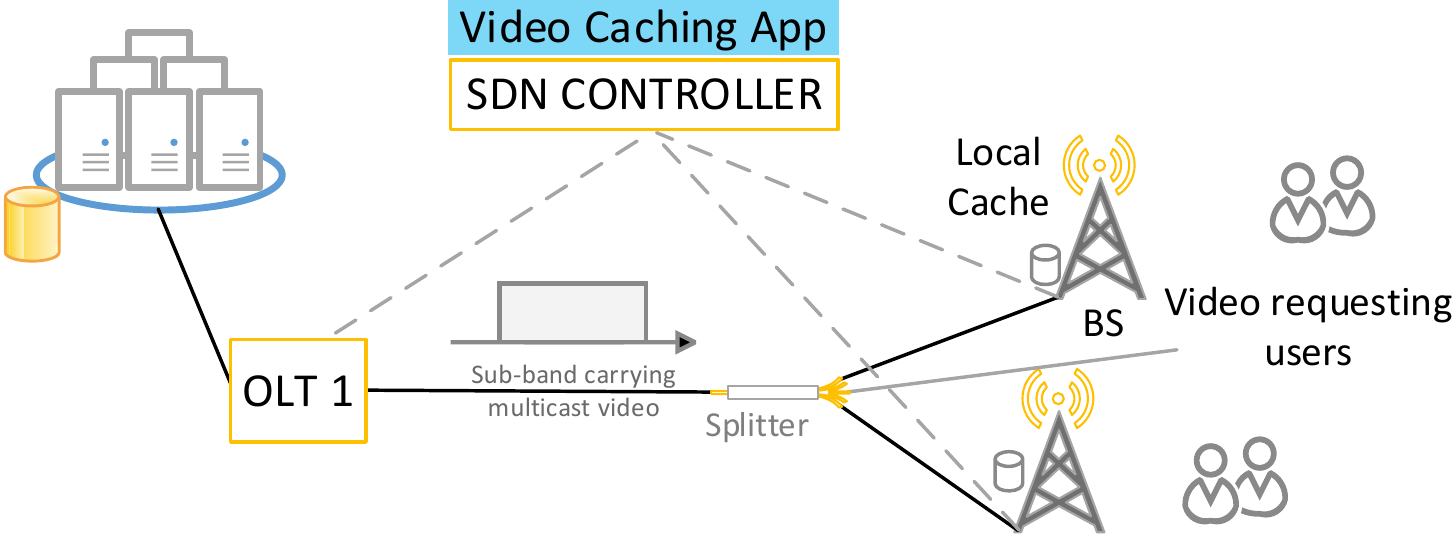}
    \vspace{-.1cm}
    \caption{SDN based video caching application in PON
        for mobile users~\cite{Li2014video}:
    The SDN controller pushes frequently requested videos to base station
    (BS) caches, whereby multicast can reach several BS caches.}
    \label{fig_app_video}      \vspace{-.05cm}
\end{figure}
Caching of video data close the users is generally beneficial for
improving the QoE of video services~\cite{Ahlgren2012,Choi2012}.
Li et al.~\cite{Li2014video} have introduced
caching mechanisms for software-defined PONs.
In particular, Li et al.~have proposed
joint provisioning of the bandwidth to service the video and
the cache management, as illustrated in Fig.~\ref{fig_app_video}.
Based on the request frequency for specific
video content, the Base Station (BS) caches the content with the
assistance of the SDN controller.
The proposed \textit{push}-based mechanism delivers (pushes) the video
to the BS caches when the PON is not congested.
A specific PON transmission sub-band can be used to multicast video
content that needs to be cached at multiple BSs.
The simulation evaluation in~\cite{Li2014video} indicate
that up to 30\% additional videos can be serviced while the service
response delay is reduced to 50\%.

\subsection{Access Control and Security} \label{app_secur:sec}
\subsubsection{Flow-based Access Control}
Network Access Control (NAC) is a networking application that
regulates the access to network services~\cite{cas2007eth,par2014fut}.
A NAC based on traffic flows has been developed by Matias~\cite{MaGMT14}.
FlowNAC exploits the forwarding rules of OpenFlow switches, which are set by a
central SDN controller, to control the access of traffic flows to
network services.
FlowNAC can implement the access control based on various flow identifiers,
such as MAC addresses or IP source and destination addresses.
Performance evaluations measured
the connection times for flows on a testbed and found average connection
times on the order of 100~ms for completing the flow access control.

In a related study, Nayak et al.~\cite{nay2009res} developed the
Resonance flow based access control system for an enterprise
network. In the Resonance system, the network elements, such as the
routers themselves, dynamically enforce access control policies. The
access control policies are implemented through real-time alerts and
flow based information that is exchanged with SDN principles. Nayak
et al. have demonstrated the Resonance system on a production
network at Georgia Tech. The Resonance design can be readily
implemented in SDON networks and can be readily extended to wide
area networks. Consider for example multiple heterogeneous DCs of
multiple organizations that are connected by an optical backbone
network. The Resonance system can be extended to provide access
control mechanisms, such as authentication and authorization,
through such a wide area SDON.

\subsubsection{Lightpath Hopping Security}
\begin{figure}[t!]
    \centering     \vspace{-.1cm}
    \includegraphics[width=3in]{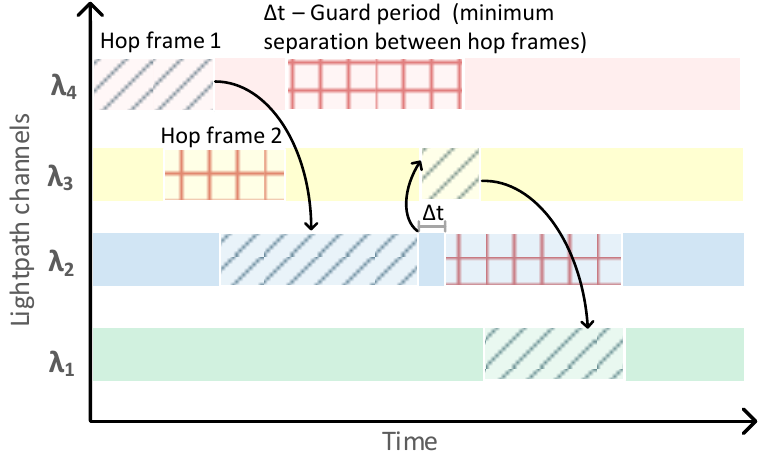}   \vspace{-.1cm}
\caption{Overview of optical light path hopping mechanism to secure link from
    eavesdropping and jamming~\cite{Li2016c}: The flow marked by
the diagonal shading hops from lightpath channel $\lambda_4$ to
$\lambda_2$, then to $\lambda_3$ and on to $\lambda_1$.
Transmissions by distinct flows on a given lightpath channel must
be separated by at least a guard period.}   \label{fig_app_hop}
\vspace{-.05cm}  \end{figure}
The broad network perspective of SDN controllers facilitates
the implementation of security functions that require this broad
perspective~\cite{ahm2015sec, sco2015sur, ShJJJ14}.
However, SDN may also be vulnerable to a wide range of attacks and
vulnerabilities, including unauthorized access, data leakage,
data modification, and misconfiguration.
Eavesdropping and jamming are security threats on the physical layer
and are especially relevant for the optical layer of SDONs.
In order to prevent
eavesdropping and jamming in an optical lightpath, Li et al.~\cite{Li2016c}
have proposed an SDN based fast lightpath hopping mechanism.
As illustrated in Fig.~\ref{fig_app_hop}, the hopping mechanism operates
over multiple lightpath channels. Conventional optical
lightpath setup times range from several hundreds of milliseconds
to several seconds and would result in a very low hopping frequency.
To avoid the
optical setup times during each hopping period, an SDN based
high precision time synchronization has been proposed.
As a result, a fast hopping
mechanism can be implemented and executed in a coordinated manner. A hop frame
is defined and guard periods are added in between hop frames.
The experimental evaluations indicate that a maximum hopping frequency
of~1~MHz can be achieved with a BER of $1 \times 10^{-3}$.
However, shortcomings of such mechanisms are the secure exchange of hopping
sequences between the transmitter and
the receiver. Although, centralized SDN control provides authenticated
provisioning of the hopping sequence, additional mechanisms to secure
the hopping sequence from
being obtained through man-in-the-middle attacks should be investigated.

\subsubsection{Flow Timeout}
SDN flow actions on the forwarding and switching elements have generally a
validity period.
Upon expiration of the validity period, i.e., the flow action timeout,
the forwarding or switching
element drops the flow action from the forwarding information base or the flow
table. The switching
element  CPU must be able to access the flow action information with very
low latency so as to perform switching actions at the line rate.
Therefore, the flow actions are commonly stored in Ternary Content
Addressable Memories (TCAMs)~\cite{Pagiamtzis2006}, which are limited
to storing on the order of thousands of distinct entries.
In SDONs, the optical network elements perform
the actions set by the SDN controller. These actions have to be stored in
a finite memory space.  Therefore, it is important to utilize the finite
memory space as efficiently as
possible~\cite{Bull2015, Liang2015, ngu2016rul, Xie2014d, Zhang2015timeout}.
In the dynamic timeout approach~\cite{ZhuFan2015},
the SDN controller tracks the TCAM occupancy levels in the switches and adjusts
timeout durations accordingly.
However, a shortcoming of such techniques
is that the bookkeeping processes at the SDN controllers can become
cumbersome for a large network.  Therefore, autonomous
timeout management techniques that are implemented at the hypervisors
can reduce the controller processing load and are
an important future research direction.

\subsection{Energy Efficiency}  \label{ene_eff:sec}
The separation of the control plane from the data plane
and the global network perspective are unique advantages of SDN for
 improving the energy efficiency of networks,
which is an important goal~\cite{tuc2011gre,zha2010ene}.

\subsubsection{Power-saving Application Controller}
Ji et al.~\cite{Ji2014} have proposed an all optical
energy-efficient network centered around an application
controller~\cite{yan2013mul,yan2015per} that monitors power consumption
characteristics and enforces power savings policies.
Ji et al. first
introduce energy-efficient variations of Digital-to-Analog
Converters (DACs) and wavelength selective ROADMs as components for
their energy-efficient network. Second, Jie et al. introduce
an energy-efficient switch architecture that consists of multiple
parallel switching planes, whereby each plane consists of three
stages with optical burst switching employed in the second (central)
switching stage.
Third, Jie et al. detail a multilevel SDN based control architecture
for the network built from the introduced components and switch.
The control structure accommodates multiple networks domains,
whereby each network domain can involve multiple switching
technologies, such as time-based and frequency-based optical switching.
All controllers for the various domains and technologies are placed
under the control of an application controller.
Dedicated power monitors that are distributed throughout the network
update the SDN based application controller about the energy consumption
characteristics of each network node.
Based on the received energy consumption updates,
the application controller executes power-saving strategies.
The resulting control actions are signalled by the application
controller to the various controllers for the different network domains
and technologies.
An extension of this multi-level architecture to cloud-based
radio access networks has been examined in~\cite{yan2016per}.

\subsubsection{Energy-Saving Routing}
Tego et al.~\cite{Tego2014} have proposed an energy-saving
application that switches off under-utilized GbE network links.
Specifically, Tego et al. proposed two methods: Fixed Upper Fixed
Lower (FUFL) and Dynamic Upper and Fixed Lower (DLFU). In FUFL, the
IP routing and the connectivity of the logical topology are
\textit{fixed}. The utilization of physical GbE links (whereby
multiple parallel physical links form a logical link) is compared
with a threshold to determine whether to switch off or on individual
physical links (that support a given logical link). The traffic on a
physical link that is about to be switched off is rerouted on a
parallel physical GbE link (within the same logical link). In
contrast, in the DLFU approach, the energy saving application
monitors the load levels on the virtual links. If the load level on
a given virtual link falls below a threshold value, then the virtual
link topology is reconfigured to eliminate the virtual link with the
low load. A general pitfall of such link switch-off techniques is
that energy savings may be achieved at the expense of deteriorating
QoS. The QoS should therefore be closely monitored when switching
off links and re-routing flows.

A similar SDN based routing strategy that strives to save energy while
preserving the QoS has been examined in the context of a GMPLS
optical networks in~\cite{Wang2015}.
Multipath routing optimizing applications that strive to save
energy in an SDN based transport optical network have been presented
in~\cite{Yevsieieva2015}.
A similar SDN based optimization approach
for reducing the energy consumption in
data centers has been examined by Yoon et al.~\cite{Yoon2015}.
Yoon et al. formulated a mixed integer linear program that models the switches
and hosts as queues. Essentially, the optimization decides on the
switches and hosts that could be turned off.
As the problem is NP-hard,
annealing algorithms are examined.
Simulations indicate that energy savings of more than 80\% are possible for
low data center utilization rates, while the energy savings decrease to less
than 40\% for high data center utilization rates.
Traffic balancing in the metro optical access networks through the
SDN based reconfiguration of optical subscriber units
in a TWDM-PON systems for energy
savings has been additionally demonstrated in~\cite{val2015exp}.

\subsection{Failure Recovery and Restoration}  \label{app_fail_rec:sec}
\subsubsection{Network Reprovisioning}
\begin{figure}[t!]
    \centering
    \vspace{-.1cm}
    \includegraphics[width=2.5in]{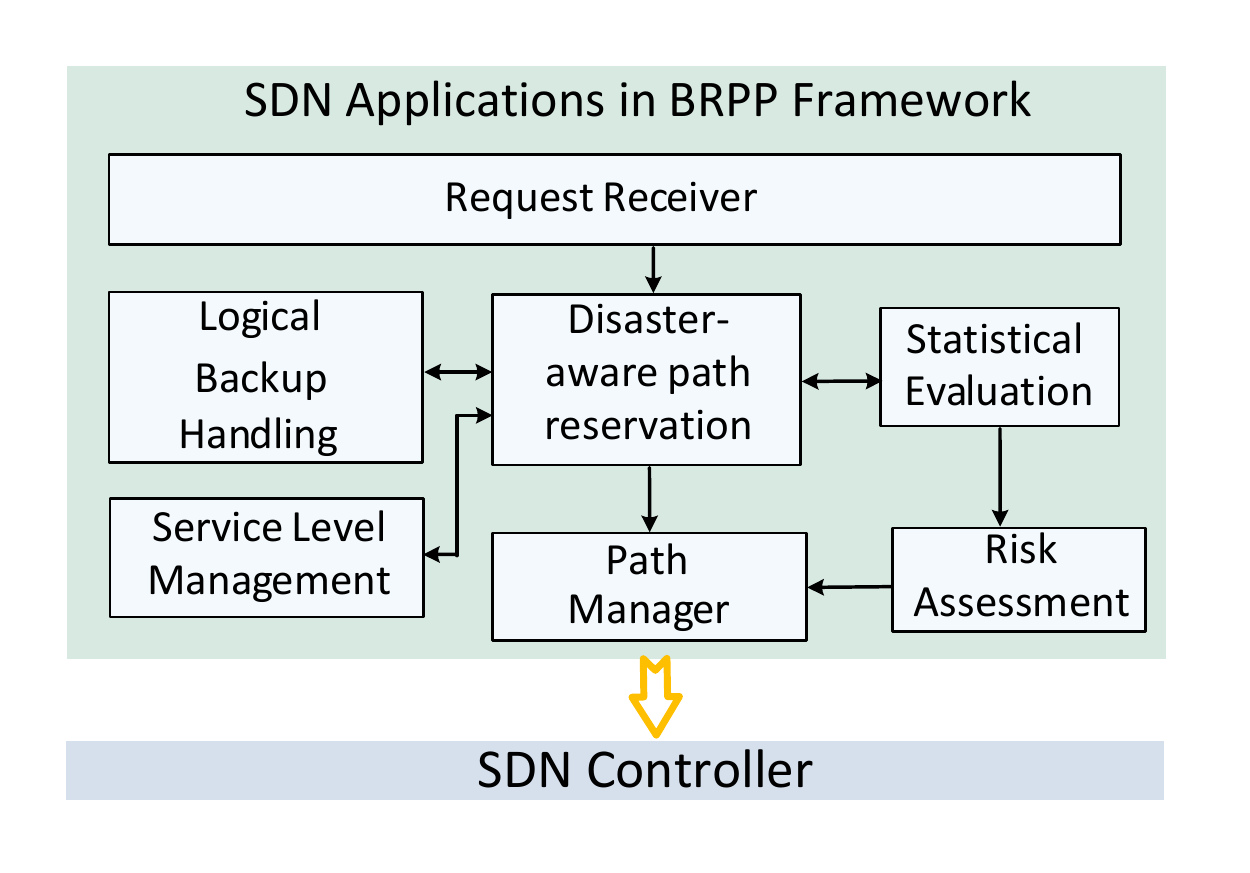}
    \vspace{-.1cm}
    \caption{Illustration of application layer
        modules of SDN based network reprovisioning framework
 for disaster aware networking~\cite{sav2015bac}.}
    \vspace{-.05cm}
    \label{fig_app_reprov}
\end{figure}
Network disruptions can occur due to various natural and/or man-made
factors.
Network resource reprovisioning is a process to
change the network configurations, e.g., the network topology and routes,
to recover from failures.
A Backup Reprovisioning with Path Protection (BRPP), based on SDN for
optical networks has been presented by Savas et al.~\cite{sav2015bac}.
An SDN application framework as illustrated in
Fig.~\ref{fig_app_reprov} was designed to support the
reprovisioning with services, such
as provisioning the new connections, risk assessment, as well as service level
and backup management.
When new requests are received by the BRPP application
framework, the statistics module evaluates the network state to find the
primary path and a link-disjoint backup path.  The computed backup
paths are stored as logical links without being
provisioned on the physical network. The logical backup module manages
and recalculates the logical links when a new backup path cannot be
accommodated or to optimize the existing backup paths (e.g., minimize the
backup path distance). Savas et al.~introduce a degraded backup path mechanism
that reserves not the full, but a lower (degraded) transmission capacity
on the backup paths, so as to accommodate more requests.
Emulations of the proposed mechanisms indicate improved
network utilization while effectively provisioning the backup paths
for restoring the network after network failures.

As a part of DARPA's core
optical networks CORONET project, a non-SDN based Robust Optical Layer
End-to-end X-connection (ROLEX)
protocol has been demonstrated and presented along with the lessons
learned~\cite{VonLehmen2015}.
ROLEX is a distributed protocol for failure recovery
which requires a considerable amount of signaling between nodes
for the distributed management.
Therefore to avoid the pitfall of excessive signalling,
it may be worthwhile to examine a ROLEX version with
centralized SDN control in future research
to reduce the recovery time and
signaling overhead, as well as the costs of restored
paths while ensuring the user QoS.

\subsubsection{Restoration Processing}
During a restoration, the network control plane simultaneously triggers backup
provisioning of all disrupted paths.
In GMPLS restoration, along with signal flooding,
there can be contention of signal messages at the network nodes.
Contentions may arise due to spectrum conflicts of the lightpath,
or node-configuration overrides, i.e., a new configuration request arrives
while a preceding reconfiguration is under way.
Giorgetti et al.~\cite{Giorgetti2015} have proposed
dynamic restoration in the elastic optical
network to avoid signaling contention in SDN (i.e., of OpenFlow messages).
Two SDN restoration mechanisms were presented:
$(i)$ the independent restoration scheme (SDN-ind),
and $(ii)$ the bundle restoration scheme (SDN-bund).
In SDN-ind, the controller triggers simultaneous independent
flow modification (Flow-Mod) messages for each backup path
to the switches involved in the reconfigurations.
During contention, switches enqueue the multiple received Flow-Mod messages
and process them sequentially. Although SDN-ind achieves reduced
recovery time as compared to non-SDN GMPLS, the waiting of messages in
the queue incurs a delay. In SDN-bund,
the backup path reconfigurations are bundled into a single message,
i.e., a Bundle Flow-Mod message, and sent to each involved switch.
Each switch then configures the flow modifications in
one reconfiguration, eliminating the delay incurred by the queuing of
Flow-Mod messages. A similar OpenFlow enabled restoration
in Elastic Optical Networks (EONs) has been studied in~\cite{Liu2015d}.

\subsubsection{Reconfiguration}
\begin{figure}[t!]
    \centering
    \vspace{-.1cm}
    \includegraphics[width=2in]{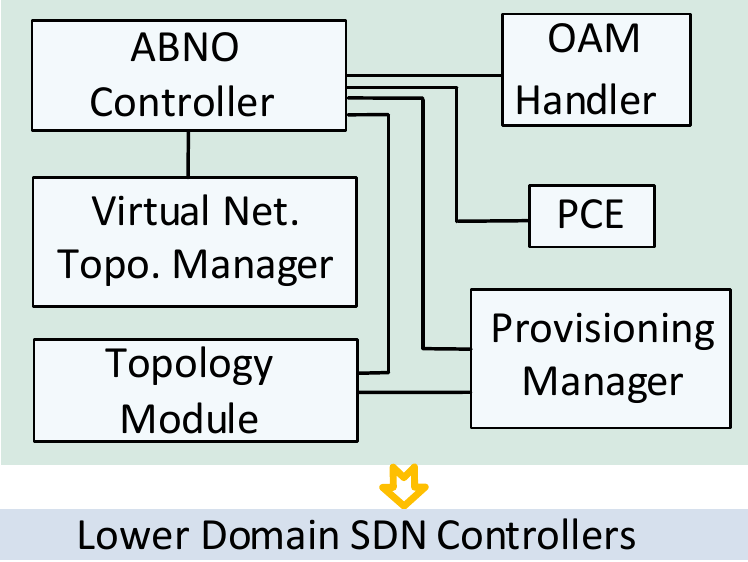}
    \vspace{-.1cm}
    \caption{Illustration of Application-Based Network Operation (ABNO)
         architecture: The ABNO controller
        communicates with the Operation, Administration, and
Maintenance (OAM) module, the Path Computation Element (PCE) module as well as
the topology modules and the provisioning manager to control the
lower domain SDN controllers so as to
        recover from network failures~\cite{Aguado2016}.}
      \vspace{-.05cm}
    \label{fig_app_abno}
\end{figure}
Aguado et al.~\cite{Aguado2016} have demonstrated
a failure recovery mechanism as part of the EU FP7 STRAUSS project with
dynamic virtual reconfigurations using SDN.
They considered multidomain hypervisors and
domain-specific controllers to virtualize the multidomain networks.
The Application-Based Network Operations (ABNO) framework
illustrated in Fig.~\ref{fig_app_abno} enables network automation
and programmability.
ABNO can compute end-to-end optical paths and delegate the
configurations to lower layer domain SDN controllers.
Requirements for fast recovery from network failures
would be in the order of tens of milliseconds, which is challenging to
achieve in large scale networks. ABNO reduces the recovery
times by pre-computing the backup connections after the first failure,
while the Operation, Administration and Maintenance (OAM)
module~\cite{Paolucci2015}
communicates with the ABNO controller to configure the new end-to-end
connections in response to a failure alarm. Failure alarms are triggered
by the domain SDN controllers monitoring the traffic via the optical
power meters when power is below $-20$~dBm.
In order to ensure survivability, an adaptive survivability scheme that
takes routing as well as spectrum assignment and modulation into
consideration has been explored in~\cite{aib2016sof}.

A similar design for end-to-end protection and failure
recovery has been demonstrated by Slyne et al.~\cite{SlKMPR14} for a long-reach
(LR) PON. LR-PON failures are highly
likely due to physical breaks in the long feeder fibers. Along with the
high impact of connectivity break down or degraded service,
physical restoration time can be very long. Therefore, 1:1 protection for
LR-PONs based on SDN has been proposed, where primary and secondary (backup)
OLTs are used without traffic duplication.
More specifically, Slyne et al. have
devised and demonstrated an OpenFlow-Relay located at the switching unit.
The OpenFlow-Relay
detects and reports a failure along with fast updating of forwarding rules.
Experimental demonstration show the backup OLT carrying protected traffic
within $7.2$ ms after a failure event.

An experimental demonstration utilizing multiple paths in optical
transport networks for failure recovery has been discussed by
Kim et al.~\cite{Kim2015}.
Kim et al. have used commercial grade IP WDM network equipment and
implemented multipath TCP
in an SDN framework to emulate inter-DC communication.
They developed an SDN application, consisting of an
cross-layer service manager module and a
cross-layer multipath transport module to
reconfigure the optical paths
for the recovery from connection impairments.
Their evaluations show increased bandwidth
utilization and reduced cost while being resilient to network impairments
as the cross-layer multipath transport module does not reserve the backup
path on the transport network.

\subsubsection{Hierarchical Survivability}
Networks can be made survivable by introducing resource redundancy.
However, the cost of the network increases with increased redundancy.
Zhang et al.~\cite{ZhangSong2014} have demonstrated a highly survivable
IP-Optical multilayered transport network.
Hierarchal controllers are placed for multilayer
resource provisioning. Optical nodes are controlled by
Transport Controllers (TCs), while higher
layers (IP) are controlled by unified controllers (UCs).
The UCs communicate with the TCs
to optimize the routes based on cross-layer information. If a fiber causes a
service disruption, TCs may directly set up alternate routes or
ask the UCs for optimized routes. A pitfall of such
hierarchical control techniques can be long restoration times. However,
the cross layer restorations
can recover from high degrees of failures, such as
multipoint and concurrent failures.

\subsubsection{Robust Power Grid}
The lack of a reliable communication infrastructure for power grid
management was one the many reasons for the widespread blackout in
the Northeastern U.S.A. in the year 2003, which affected the lives
of 50 million people~\cite{Parandehgheibi2014}. Since then building
a reliable communication infrastructure for the power grid has
become an important priority. Rastegarfar et
al.~\cite{Rastegarfar2016} have proposed a communication
infrastructure that is focused on monitoring and can react to and recover from
failures so as to reliably support power grid applications.
More specifically, their
architecture was built on SDN based optical networking for
implementing robust power grid control applications. Control and
infrastructure in the SDN based power grid management exhibits an
interdependency i.e., the physical fiber relies on the control plane
for its operations and the logical control plane relies on the
same physical fiber for its signalling communications. Therefore,
they only focus on optical protection switching instead of IP layer
protection, for the resilience of the SDN control. Cascaded failure
mechanisms were modeled and simulated for two geographical
topologies (U.S. and E.U.). In addition, the impacts of cascaded
failures were studied for two scenarios $(i)$ static optical layer
(static OL), and $(ii)$ dynamic optical layer (dynamic OL). Results
for a static OL illustrated that the failure cascades are persistent
and are closely dependent on the network topology. However, for a
dynamic OL (i.e., with reconfiguration of the physical layer),
failure cascades were suppressed by an average of 73\%.

\subsection{Application Layer: Summary and Discussion}
The SDON QoS application studies have mainly examined traffic and
network management mechanisms that are supported through the
OpenFlow protocol and the central SDN controller.
The studied SDON QoS applications are structurally very similar in
that the traffic conditions or
network states (e.g., congestion levels) are probed
or monitored by the central SDN controller.
The centralized knowledge of the traffic and network is then utilized
to allocate or configure resources, such as DC resources in~\cite{ZhZZ13},
application bandwidths in~\cite{Li2014}, and
topology configurations or routes
in~\cite{WetteKarl2013,Tariq2015,SgPCVC13,ChangLi2015}.
Future research
 on SDON QoS needs to further optimize the interactions of the
controller with the network applications and data plane
to quickly and correctly react to changing user demands and
network conditions, so as to assure consistent QoS.
The specific characteristics and requirements
of video streaming applications have
been considered in the few studies on video
QoS~\cite{chi2015app,Chitimalla2015,Li2014video}.
Future SDON QoS research should consider a wider range of
specific prominent application traffic types with
specific characteristics and requirements, e.g., Voice over IP (VoIP)
traffic has relatively low bit rate requirements, but requires low
end-to-end latency.

\begin{figure*}[t!]
	\centering
	\vspace{-.1cm}
	\includegraphics[width=6in]{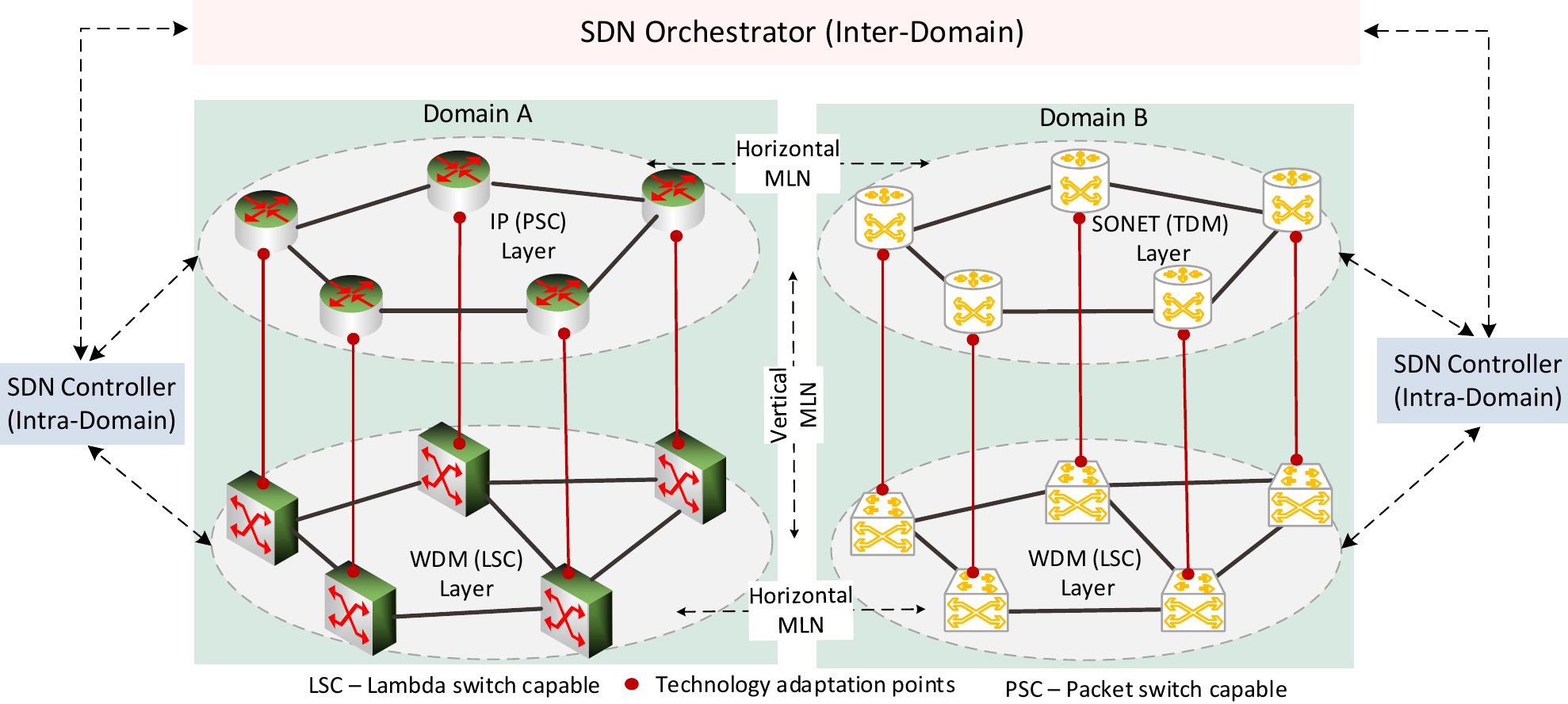}
	\vspace{-.1cm}
	\caption{Illustration of SDN orchestration of multilayer networking:
Vertical MultiLayer Networking (MLN) spans layers at different horizontal
	positions within a given domain.
	Horizontal MLN spans multiple layers at the same horizontal position
	(or in different horizontal positions) across
	multiple domains.
	The inter-domain SDN orchestrator coordinates the individual
	intra-domain SDN controllers.}
	\vspace{-0.05cm}
	\label{fig_control_multilay}
\end{figure*}
Very few studies have considered security and access control for SDONs.
The thorough study of the broad topic area of security and privacy is
an important future research direction in SDONs, as outlined in
Section~\ref{futworksec:sec}
Energy efficiency is similarly a highly important topic
within the SDON research area that has received relatively little
attention so far and presents overarching research challenges,
see Section~\ref{futworkene:sec}.

One common theme of the SDON application layer studies focused
on failure recovery and restoration has been to exploit
the global perspective of the SDN control.
The global perspective has been exploited for
for improved planning of the recovery and
restoration~\cite{sav2015bac,Aguado2016,ZhangSong2014} as well as
for improved coordination of the execution of the restoration
processes~\cite{Giorgetti2015,Liu2015d}.
Generally, the existing failure recovery and restoration studies
have focused on network (routing) domain that is owned by a particular
organizational entity. Future research should seek to examine the
tradeoffs when exploiting the
global perspective of orchestration of multiple routing domains, i.e.,
the failure recovery and restoration techniques surveyed in this section
could be combined with the multidomain orchestration techniques
surveyed in Section~\ref{orch:sec}.
One concrete example of multidomain orchestration could be to coordinate
the specific LR-PON access network protection and
failure recovery~\cite{SlKMPR14}
with protection and recovery techniques for metropolitan and core
network domains, e.g.,~\cite{sav2015bac,Aguado2016,Kim2015,ZhangSong2014},
for improved end-to-end protection and recovery.

\section{Orchestration}  \label{orch:sec}
As introduced in Section~\ref{intro:orch:sec}, orchestration
accomplishes higher layer abstract coordination of network services
and operations. In the context of SDONs, orchestration has mainly
been studied in support of multilayer networking.
Multilayer networking in the context of SDN and network virtualization
generally refers to networking across multiple network layers
and their respective technologies, such as
IP, MPLS, and WDM, in combination with networking across multiple
routing domains~\cite{LeZGF11,leo2003vir,rui2011sur,tou2001net,vig2005mul}.
The concept of multilayer networking is generally an abstraction of
providing network services with multiple networking layers (technologies) and
multiple routing domains.
The different network layers and their technologies are sometimes classified into
Layer~0 (e.g., fiber-switch capable), Layer~1 (e.g., lambda switching
capable), Layer~1.5 (e.g., TDM SONET/SDH), Layer~2 (e.g., Ethernet),
Layer~2.5 (e.g., packet switching capable using MPLS), and Layer~3
(e.g., packet switching capable using IP routing)~\cite{YoMKN14}.
Routing domains are also commonly referred to as
network domains, routing areas, or levels~\cite{LeZGF11}.

The recent multilayer
networking review article~\cite{LeZGF11} has introduced a range of capability
planes to represent the grouping of related functionalities for a
given networking technology. The capability planes include the data
plane for transmitting and switching data. The control plane and the
management plane directly interact with the data plane for
controlling and provisioning data plane services as well as for
trouble shooting and monitoring the data plane. Furthermore, an
authentication and authorization plane, a service plane, and an
application plane have been introduced for providing network
services to users.

Multilayer networking can involve vertical layering or
horizontal layering~\cite{LeZGF11}, as illustrated in
Fig.~\ref{fig_control_multilay}. In vertical layering, a given
layer, e.g., the routing layer, which may employ a
particular technology, e.g., the Internet Protocol (IP),
uses another (underlying) layer, e.g., the Wavelength
Division Multiplexing (WDM) circuit switching layer, to provide
services to higher layers. In horizontal layering, services are
provided by ``stitching'' together a service path across multiple
routing domains.

SDN provides a convenient control framework for these
flexible multilayer networks~\cite{LeZGF11}. Several research
networks, such as ESnet, Internet2, GEANT, Science DMZ
(Demilitarized Zone) have experimented with these multilayer
networking concepts~\cite{KiSTP13,RoMSS14}.
In particular, SDN based multilayer network
architectures, e.g.,~\cite{woe2013sdn, iiz2016mul, mun2016nee},
are formed by conjoining
the layered technology regions
$(i)$ in vertical fashion i.e., multiple technology layers internetwork
within a single domain, or
$(ii)$ in horizontal layering fashion across multiple domains,
i.e., technology layers internetwork across distinct domains.
Horizontal multilayer networking can be viewed as a generalization of
vertical multilayer networking in that the horizontal networking
may involve the same or different (or even multiple) layers in the
distinct domains. As illustrated in
Fig.~\ref{fig_control_multilay}, the formed SDN based multilayer network
architecture is controlled by an SDN orchestrator.
\begin{figure*}[t!]
\footnotesize
\setlength{\unitlength}{0.10in} 
\centering
\begin{picture}(40,33)
\put(13,33){\textbf{Orchestration, Sec.~\ref{orch:sec}}}
\put(18,30){\line(0,1){2}}
\put(0,30){\line(1,0){34}}
\put(0,30){\vector(0,-1){2}}
\put(-4,26.2){\textbf{Multilayer Orch.}, Sec.~\ref{mullayorch:sec} }
\put(0,25){\line(0,-1){2}}

\put(25,23){\line(1,0){20}}
\put(27,26.2){\textbf{Multidomain Orch.}, Sec.~\ref{muldomorch:sec}}
\put(34,25){\line(0,-1){2}}

\put(-10,23){\line(1,0){20}}
\put(-10,23){\vector(0,-1){2}}
\put(10,23){\vector(0,-1){2}}

\put(-14,20){\makebox(0,0)[lt]{\shortstack[l]{			
\textbf{Frameworks}, Sec.~\ref{mullayorchfr:sec} \\ \\
Hier. Multilay. Ctl.~\cite{Felix2014,ger2013dem,sal2013inf,sun2014des,ShZBLT12}\\
Appl. Centric Orch.~\cite{Gerstel2015}
}}}

\put(2,20){\makebox(0,0)[lt]{\shortstack[l]{			
\textbf{Appl.-Specific Orch.}, Sec.~\ref{applorch:sec} \\ \\
Failure Rec.~\cite{Khaddam2015} \\
Res. Util.~\cite{Liu2015e} \\
Virt. Opt. Netws.~\cite{vil2015net}
}}}				
\put(34,30){\vector(0,-1){2}}
\put(25,23){\vector(0,-1){2}}
\put(20,20){\makebox(0,0)[lt]{\shortstack[l]{
\textbf{General Netw.}, Sec.~\ref{muldomorchgen:sec} \\ \\
Opt. Multidom. Multitechn.~\cite{YoMKN14} \\
Hier. Multidom. Ctl.~\cite{Jing2015} \\
IDP~\cite{Zhu2015} \\
Multidom. Net. Hyperv.~\cite{vil2015mul} \\
ABNO~\cite{Munoz2015}
}}}							
\put(45,23){\vector(0,-1){2}}		
\put(40,20){\makebox(0,0)[lt]{\shortstack[l]{			
\textbf{Data Center Orch.}, Sec.~\ref{muldomorchdc:sec} \\ \\
Control Arch.~\cite{Liu2015b, may2016sdn} \\
H-PCE~\cite{cas2015sdn} \\
Virt.-SDN Ctl.~\cite{mun2015int,Vilalta2016}
}}}
\end{picture}
\vspace{-3.3cm}	
\caption{Classification of SDON orchestration studies:
Multilayer orchestration studies focus on vertical multilayer networking
within a single domain.
Multidomain orchestration studies focus on horizontal
multilayer networking across multiple domains and may involve
multiple vertical layers in the various domains.}
\label{orch_class:fig}
\end{figure*}
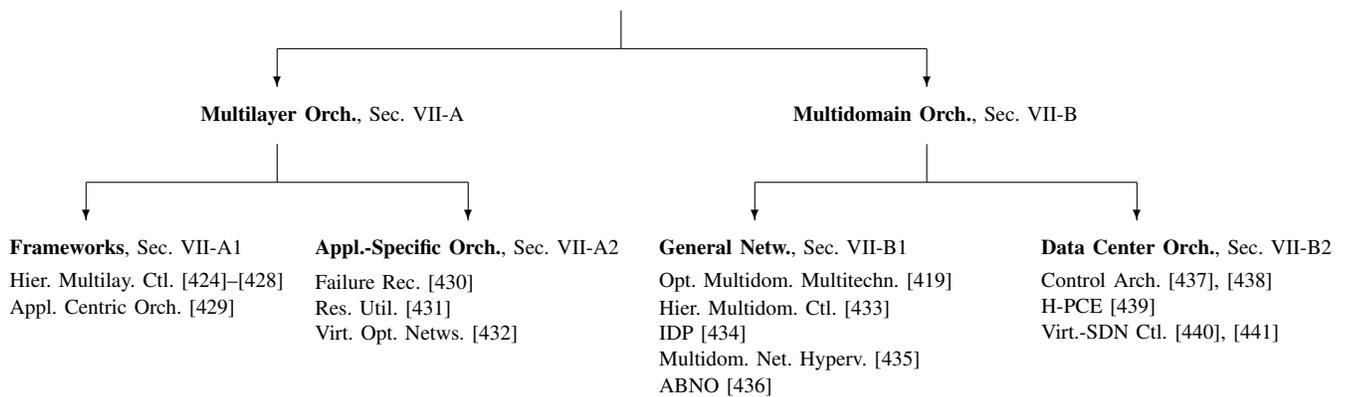
As illustrated in Fig.~\ref{orch_class:fig} we organize
the SDON orchestration studies according to their focus into studies
that primarily address the orchestration of vertical multilayer
(multitechnology) networking,
i.e., the vertical networking across multiple layers (that typically
implement different technologies) within a given domain,
and into studies that primarily
address the orchestration of horizontal multilayer (multidomain)
networking, i.e., the horizontal networking across multiple
routing domains (which may possibly involve different or multiple
vertical layers in the different domains).
We subclassify the vertical multilayer studies into general (vertical)
multilayer networking frameworks and studies focused on supporting specific
applications through vertical multilayer networking.
We subclassify the multidomain (horizontal multilayer) networking studies into
studies on general network domains and studies focused on internetworking
with Data Center (DC) network domains.

\subsection{Multilayer Orchestration}  \label{mullayorch:sec}

\subsubsection{Multilayer Orchestration Frameworks} \label{mullayorchfr:sec}

\paragraph{Hierarchical Multilayer Control}
Felix et al.~\cite{Felix2014} presented an hierarchical
SDN control mechanism for packet optical networks.
Multilayer optimization techniques
are employed at the SDN orchestrator
to integrate the optical transport technology
with packet services by provisioning end-to-end Ethernet services.
Two aspects are investigated, namely
$(i)$ bandwidth optimization for the optical transport services,
and $(ii)$ congestion control for packet network services
in an integrated packet optical network.
More specifically, the SDN controller initially allocates the
minimum available bandwidth required
for the services and then dynamically scales
allocations based on the availability.
Optical-Virtual Private Networks (O-VPNs) are created
over the physical transport network. Services are then mapped
to O-VPNs based on class of service requirements.
When congestion is detected for a service,
the SDN controller switches the service to another O-VPN, thus
balancing the traffic to maintain the required class of service.

Similar steps towards the orchestration of multilayer networks
have been taken within the OFELIA
project~\cite{ger2013dem,sal2013inf,sun2014des}. Specifically,
Shirazipour et al.~\cite{ShZBLT12} have explored
extensions to OpenFlow version 1.1 actions to enable
multitechnology transport layers, including Ethernet transport and
optical transport. The explorations of the extensions include
justifications of the use of SDN in circuit-based transport
networks.

\paragraph{Application Centric Orchestration}
Gerstel et al.~\cite{Gerstel2015} proposed an application centric network
service provisioning approach based on multilayer orchestration.
This approach enables the network applications
to directly interact with the physical layer
resource allocations to achieve the desired
service requirements.
Application requirements for
a network service may include
maximum end-to-end latency, connection setup and hold times,
failure protection, as well as security and encryption.
In traditional IP networking, packets from multiple applications
requiring heterogeneous services
are simply aggregated and sent over a common transport link (IP services).
As a result, network applications are typically
assigned to a single (common) transport service within an optical link.
Consider a failure recovery process with multiple available paths.
IP networking typically selects the single path
with the least end-to-end delay.
However, some applications may tolerate higher
latencies and therefore, the traffic can be split
over multiple restoration paths achieving better traffic management.
The orchestrator needs to interact with multiple
network controllers operating across multiple (vertical) layers
supported by north/south bound interfaces
to achieve the application centric control.
Dynamic additions of new IP links are demonstrated
to accommodate the requirements of multiple application services
with multiple IP links when
the load on the existing IP link was increased.

\subsubsection{Application-specific Orchestration}  \label{applorch:sec}
\paragraph{Failure Recovery}
Generally, network CapEx and OpEx increase as more protection against
network failures is added.
Khaddam et al.~\cite{Khaddam2015} propose
an SDN based integration of multiple layers, such as WDM and IP, in a failure
recovery mechanism to improve the utilization
(i.e., to eventually reduce CapEx and OpEx while maintaining
high protection levels).
An observation study was conducted
over a five year period to understand the impact of
network failures on the real deployment
of backbone networks.
Results showed $75$ distinct failures following a Pareto distribution,
in which, $48\%$ of the total deployed capacity was affected
by the top (i.e., the highest impact) $20\%$ of the failures.
And, $10\%$ of the total deployed
capacity was impacted by the top two failure instances.
These results emphasize the significance of backup
capacities in the optical links for restoration processes.
However, attaining the optimal protection capacities while
achieving a high utilization of the optical links is challenging.
A failure recovery mechanism is proposed based
on a ``hybrid'' (i.e., combination of optical transport and IP)
multilayer optimization.
The hybrid mechanism improved the optical link utilization up to 50~\%.
Specifically, 30~\% increase of the transport
capacity utilization is achieved by dynamically reusing the remainder
capacities in the optical links, i.e.,
the capacity reserved for failure recoveries.
The multilayer optimization technique was validated on an experimental
testbed utilizing central path-computation (PCE)~\cite{rfc5440}
within the SDN framework.
Experimental verification of failure
recovery mechanism resulted in recovery times
on the order of sub-seconds for MPLS restorations and
several seconds for optical WSON restorations.

\paragraph{Resource Utilization}
Liu et al.~\cite{Liu2015e} proposed a method to improve
resource utilization and to reduce transmission latencies
through the processes of virtualization and service abstraction.
A centralized SDN control implements the service abstraction
layer (to enable SDN orchestrations) in order to integrate the
network topology management (across both IP and WDM),
and the spectrum resource allocation in a single control platform.
The SDN orchestrator also achieves dynamic and simultaneous
connection establishment across both IP and OTN layers
reducing the transmission latencies. The control plane design is split
between local (child) and root (parent) controllers.
The local controller realizes the label switched paths on the optical nodes
while the root controller realizes the forwarding rules
for realizing the IP layer.
Experimental evaluation of average transfer time measurements
showed IP layer
latencies on the order of several milliseconds, and
several hundreds of milliseconds for the OTN latencies, validating
the feasibility of control plane unification for IP over
optical transport networks.

\paragraph{Virtual Optical Networks (VONs)}
Vilalta et al.~\cite{vil2015net} presented
controller orchestration to integrate multiple transport network technologies,
such as IP and GMPLS. The proposed architectural framework devises
VONs to enable the virtualization
of the physical resources within each domain.
VONs are managed by lower level physical controllers (PCs), which
are hierarchically managed by an SDN network orchestrator (NO). Network
Virtualization Controllers (NVC) are introduced (on top of the NO) to
abstract the virtualized multilayers across multiple domains.
End-to-end provisioning of VONs is facilitated through
hierarchical control interaction over three levels, the
customer controller, the NO\&NVCs, and the PCs.
An experimental evaluation demonstrated average VON
provisioning delays on the order of
several seconds (5~s and 10~s), validating the flexibility of
dynamic VON deployments over the optical transport networks.
Longer provisioning delays may impact the network
application requirements, such as failure recovery processes,
congestion control, and traffic engineering.
General pitfalls of such hierarchical structures are
increased control plane complexity, risk of controller failures, and
maintenance of reliable communication links between control plane entities.

\subsection{Multidomain Orchestration}  \label{muldomorch:sec}
Large scale network deployments typically involve multiple domains,
which have often heterogeneous layer technologies.
Achieve high utilization of the networking resources
while provisioning end-to-end network paths and services across multiple
domains and their respective layers and respective technologies is
highly challenging~\cite{Mayoral2015,Munoz2015b,Yu2015a}.
Multidomain SDN orchestration studies have sought to exploit
the unified SDN control plane to aid the resource-efficient
provisioning across the multiple domains.

\subsubsection{General Multidomain Networks}  \label{muldomorchgen:sec}

\paragraph{Optical Multitechnologies Across Multiple Domains}
Optical nodes are becoming
increasingly reconfigurable (e.g., through variable BVTs and
OFDM transceivers, see Section~\ref{sdninfra:sec}),
adding flexibility to the switching elements. When a single
end-to-end service
establishment is considered, it is more likely that a service is supported by
different optical technologies that operate across multiple domains.
Yoshida et al.~\cite{YoMKN14} have demonstrated SDN based orchestration with
emphasis on the physical interconnects between multiple domains and multiple
technology specific controllers so as to realize end-to-end services.
OpenFlow capabilities have been extended for fixed-length variable capacity
optical packet switching~\cite{Losada2015}.
That is, when an optical switch matches the label on an
incoming optical packet, if a rule
exists in the switch (flow entry in the table) for a specific label,
a defined action is performed on the optical packet by the switch.
Otherwise, the optical packet is dropped
and the controller is notified. Interconnects between optical packet
switching networks and elastic optical networks are enabled
through a novel OPS-EON interface card.
The OPS-EON interface is designed as an extension
to a reconfigurable, programmable
and flexi-grid EON supporting the OpenFlow protocol.
The testbed implementation of OPS-EON interface cards demonstrated
the orchestration of multiple domain controllers and the reconfigurability
of FL-VC OPS across multidomain, multilayer, multitechnology scenarios.

\paragraph{Hierarchical Multidomain Control}
Jing et al.~\cite{Jing2015} have also examined
the integration of multiple optical transport
technologies from multiple
vendors across multiple domains, focusing on
the control mechanisms across multiple domains.
Jing et al.~proposed hierarchical SDN orchestration with parent and
domain controllers.
Domain controllers abstract the physical layer by
virtualizing the network resources.
A Parent Controller (PC) encompasses a Connection Controller (CC) and
a Routing Controller (RC) to process the abstracted virtual network.
When a new connection setup request is
received by the PC, the RC (within the PC) evaluates the end-to-end
routing mechanisms and forwards the
information to the CC.
The CC breaks the end-to-end routing information into shorter
link segments belonging to a domain.
Segmented routes are then sent to the respective domain
controllers for link provisioning over the physical infrastructures.
The proposed mechanism was experimentally verified on a testbed
built with the commercial OTN equipment.

\paragraph{Inter-Domain Protocol}
\begin{figure}[t!]
	\centering
	\vspace{-.1cm}
	\includegraphics[width=3.6in]{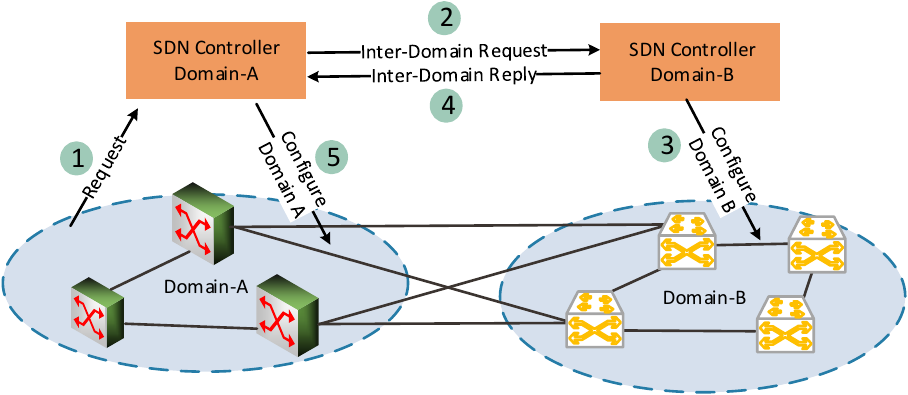}
	\vspace{-.1cm}
	\caption{Inter-domain lightpath provisioning mechanism
		facilitated by an Inter-Domain Protocol (IDP, which provides
		inter-domain request and inter-domain reply messages)
		by employing the Routing and Spectrum Allocation (RSA)
                 algorithm proposed in~\cite{Zhu2015}.
		Steps 1 through
	5 provision an end-to-end path across multiple domains.}
	\label{fig_multilay_idp}
\end{figure}
Zhu et al.~\cite{Zhu2015} followed a different approach for the
SDN multidomain control mechanisms by considering the
flat arrangement of controllers as shown in Fig.~\ref{fig_multilay_idp}.
Each domain is autonomously managed by
an SDN controller specific to the domain.
An Inter-Domain Protocol (IDP) was devised to establish the communication
between domain specific controllers
to coordinate the lightpath setup across multiple domains.
Zhu et al.~also proposed a Routing and
Spectrum Allocation (RSA) algorithm for the end-to-end provisioning
of services in the SD-EONs.
The distributed RSA algorithm operates on the domain specific controllers
using the IDP protocol. The RSA considers both transparent lightpath
connections, i.e., all-optical lightpath,
and translucent lightpath connections, i.e.,
optical-electrical-optical connections.
The benefit of such techniques is privacy, since
the domain specific policies and topology information are not
shared among other network entities.
Neighbor discovery is  independently conducted by the domain specific controller
or can initially be configured.
A domain appears as an abstracted virtual node to all other domain specific
controllers. Each controller then assigns the shortest path
routing within a domain between its border nodes.
An experimental setup validating the
proposed mechanism was demonstrated across
geographically-distributed domains in the USA and China.

\paragraph{Multidomain Network Hypervisors}
\begin{figure}[t!]
	\centering
	\vspace{-.1cm}
	\includegraphics[width=2.5in]{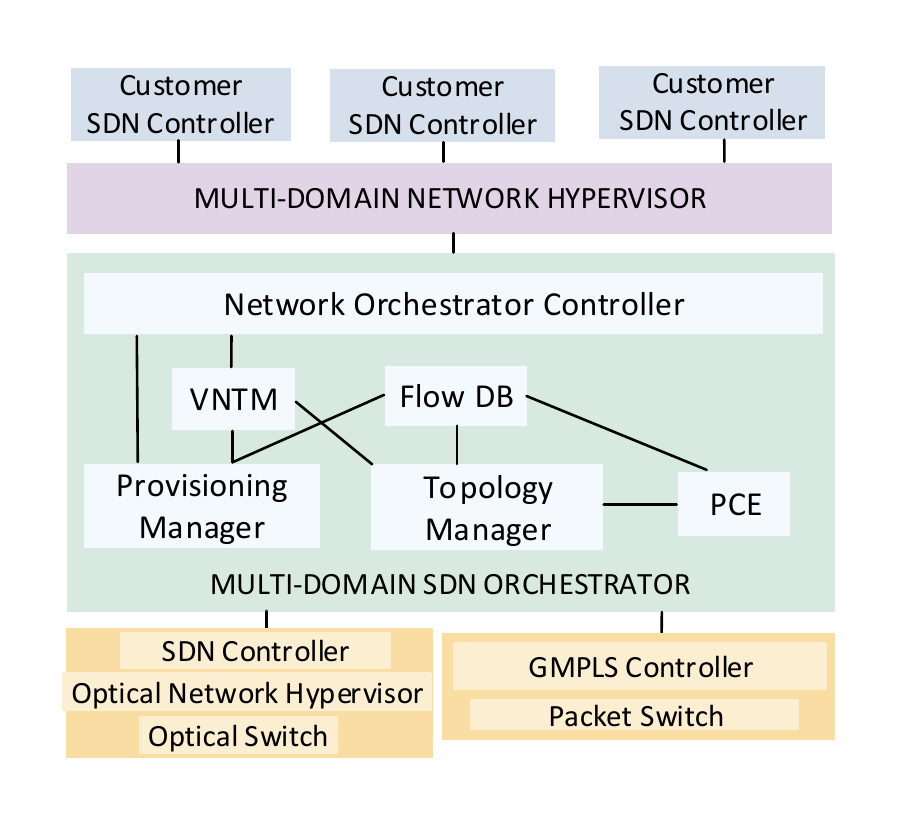}
	\vspace{-.1cm}
	\caption{Illustration of multilevel virtualization enabled by the
	  Multidomain Network Hypervisor (MNH)~\cite{vil2015mul}
          operating over a network orchestrator controller and domain
          specific SDN controllers to provide the
                multidomain end-to-end virtualization.}
	\vspace{-0.05cm}
	\label{fig_multilay_mnh}
\end{figure}
Vilalta et al.~\cite{vil2015mul} presented a mechanism for virtualizing
multitechnology optical, multitenant networks.
The Multidomain Network Hypervisor (MNH) creates customer specific
virtual network slices managed by the customer specific SDN controllers
(residing at the customers' locations) as illustrated
in Fig.~\ref{fig_multilay_mnh}.
Physical resources are managed by their domain specific
physical SDN controllers.
The MNH operates over the network orchestrator and
physical SDN controllers for provisioning VONs
on the physical infrastructures.
The MNH abstracts both $(i)$ multiple optical transport technologies,
such as optical packet switching and Elastic Optical Networks (EONs),
and $(ii)$ multiple control domains, such as GMPLS and OpenFlow.
Experimental assessments on a testbed achieved VON provisioning
within a few seconds (5~s), and
control overhead delay on the order of several tens of milliseconds.
Related virtualization mechanisms for multidomain optical SDN networks
with end-to-end provisioning have been
investigated in~\cite{SzAEK14,vil2016hie}.

\paragraph{Application-Based Network Operations}
\begin{figure}[t!]
	\centering
	\vspace{-.1cm}
	\includegraphics[width=3.4in]{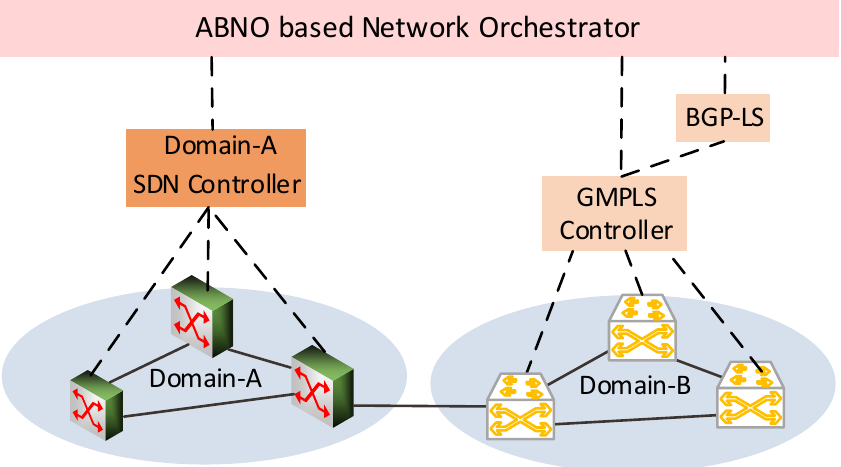}
	\vspace{-.1cm}
	\caption{The application-based network operations (ABNO)
          based SDN multilayer orchestrator~\cite{Munoz2015}
          receives the physical topology information from
          the OpenFlow/GMPLS controllers.
                The orchestrator centrally computes paths
                and sends the path information to the lower level
                controllers for path provisioning.}
	\vspace{-0.05cm}
	\label{fig_multilay_abno}
\end{figure}
Mu{\~{n}}oz et al.~\cite{Munoz2015}, have presented an SDN
orchestration mechanism based on the
application-based network operations (ABNO) framework, which
is being defined by the IETF~\cite{rfc7491}.
The ABNO based SDN orchestrator integrates
OpenFlow and GMPLS in transport networks.
Two SDN orchestration designs have been presented: $(i)$ with centralized
physical network topology aware path computation
(illustrated in Fig.~\ref{fig_multilay_abno}), and
$(ii)$ with topology abstraction and distributed path computation.
In the centralized design, OpenFlow and GMPLS controllers (lower level control)
expose the physical
topology information to the ABNO-orchestrator (higher level control).
The PCE in the ABNO-orchestrator has the global view of the network
and can compute end-to-end paths with complete knowledge of the network.
Computed paths are then provisioned through
the lower level controllers. The pitfalls of such centralized designs
are $(i)$ computationally intensive path computations,
$(ii)$ continuous updates of topology and traffic information,
and $(iii)$ sharing of confidential network information and policies
with other network elements.
To reduce the computational load at the orchestrator,
the second design implements distributed path computation
at the lower level controllers (instead of path computation at the
centralized orchestrator). However, such distributed mechanisms may
lead to suboptimal solutions due to the limited network knowledge.

\subsubsection{Multidomain Data Center Orchestration} \label{muldomorchdc:sec}
\paragraph{Control Architectures}
Geographically distributed DCs are typically
interconnected by links traversing multiple domains.
The traversed domains may be homogeneous i.e., have the same type of network
technology, e.g., OpenFlow based ROADMs,
or may be heterogeneous, i.e., have different types of network
technologies, e.g.,
OpenFlow based ROADMs and GMPLS based WSON. The SDN control structures
for a multidomain network can be broadly classified into the categories of
$(i)$ single SDN orchestrator/controller, $(ii)$ multiple mesh SDN
controllers, and
$(iii)$ multiple hierarchical SDN controllers~\cite{Liu2015b, may2016sdn}.
The single SDN orchestrator/controller has to support heterogeneous SBIs
in order to operate with multiple heterogeneous domains, e.g.,
the Path Computation Element Protocol (PCEP) for GMPLS network domains
and the OpenFlow protocol for OpenFlow supported ROADMs.
Also, domain specific details, such as topology, as well as
network statistics and configurations, have to be exposed to an external
entity, namely the single SDN orchestrator/controller,
raising privacy concerns. Furthermore, a single
controller may result in scalability issues.
Mesh SDN control connects the domain-specific controllers side-by-side
by extending the east/west bound interfaces.
Although mesh SDN control addresses the scalability and privacy issues,
the distributed nature of the control mechanisms may lead to sub-optimal
solutions.
With hierarchical SDN control, a logically centralized controller
(parent SDN controller) is placed above the domain-specific controllers
(child SDN controllers), extending the north/south bound interfaces.
Domain-specific controllers virtualize
the underlying networks inside their domains, exposing only the abstracted
view of the domains to the parent controller, which addresses the privacy
concerns. Centralized
path computation at the parent controller can achieve optimal solutions.
Multiple hierarchical levels can address the scalability issues.
These advantages of hierarchal SDN control are achieved at the expense of
an increased number of network entities,
resulting in the operational complexities.

\paragraph{Hierarchical PCE}
\begin{figure}[t!]
	\centering
	\vspace{-.1cm}
	\includegraphics[width=3.5in]{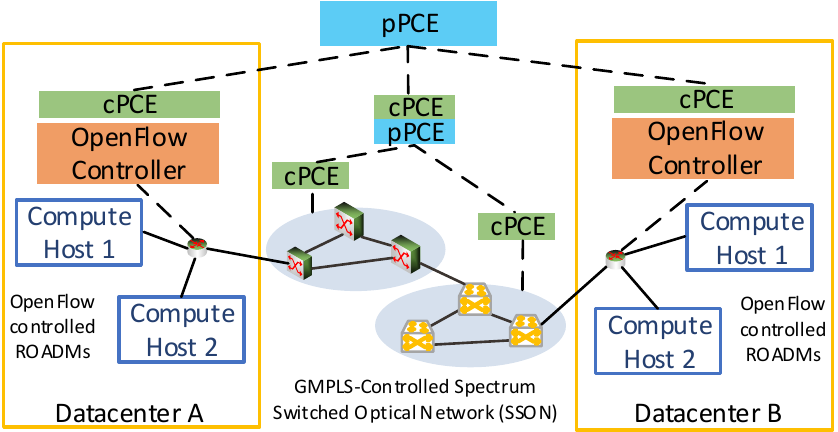}
	\vspace{-.1cm}
\caption{Illustration of SDN orchestration based on Hierarchical
  Path Computation Element (H-PCE)~\cite{cas2015sdn}:
  The H-PCE internetworks GMPLS inter-DC communication and
  OpenFlow intra-DC communication.
The parent-PCE~(pPCE) aggregates the active PCE states from the
child-PCEs~(cPCEs) of both GMPLS and OpenFlow.}
	\vspace{-0.05cm}
	\label{fig_multilay_HPCE}
\end{figure}
Casellas et al.~\cite{cas2015sdn} considered
DC connectivities involving both intra-DC and inter-DC communications.
Intra-DC communications enabled through OpenFlow networks
are supported by an OpenFlow controller.
The inter-DC communications
are enabled by optical transport
networks involving more complex control,
such as GMPLS, as illustrated in Fig.~\ref{fig_multilay_HPCE}. To
achieve the desired SDN benefits of flexibility and scalability,
a common centralized control platform spanning across heterogeneous
control domains is proposed.
More specifically, an Hierarchical PCE (H-PCE) aggregates PCE states
from multiple domains.
The end-to-end path setup between DCs is orchestrated by a parent-PCE (pPCE)
element, while the paths are provisioned
by the child-PCEs (cPCEs) on the physical resources, i.e.,
the OpenFlow and GMPLS domains.
The proposed mechanism utilizes existing protocol interfaces,
such as BGP-LS and PCEP, which are extended with OpenFlow to support the H-PCE.

\paragraph{Virtual-SDN Control}
Mu{\~n}oz et al.~\cite{mun2015int,Vilalta2016} proposed a
mechanism to virtualize the SDN control functions
in a DC/cloud by integrating SDN with Network Function Virtualization (NFV).
In the considered context, NFV refers to realizing network functions
by software modules running on
generic computing hardware inside a DC; these network functions were
conventionally implemented on specialized hardware modules.
The orchestration of Virtual Network Functions (VNFs)
is enabled by an integrated SDN and NFV management which
dynamically instantiates virtual SDN controllers.
The virtual SDN controllers
control the Virtual Tenant Networks (VTNs), i.e., virtual
multidomain and multitechnology networks.
Multiple VNFs running on a Virtual Machine (VM)
in a DC are managed by a VNF manger.
A virtual SDN controller is responsible for creating, managing, and tearing
down the VNF achieving the flexibility in the control plane management
of the multilayer and the multidomain networks.
Additionally, as an extension to the proposed mechanism,
the virtualization of the control functions of the
LTE Evolved Packet Core (EPC) has been discussed in~\cite{mar2016int}.

\subsection{Orchestration: Summary and Discussion}
Relatively few SDN orchestration studies to date have focused on
vertical multilayer networking within a given domain. The few studies
have developed two general orchestration frameworks and have examined
a few orchestration strategies for some specific applications.
More specifically, one orchestration framework has
focused on optimal bandwidth allocation based mainly on
congestion~\cite{Felix2014}, while the other
framework has focused on exploiting application traffic tolerances for
delays for efficiently routing traffic~\cite{Gerstel2015}.
SDN orchestration of vertical multilayer optical networking is thus
still a relatively little explored area.
Future research can develop orchestration frameworks that accommodate
the specific optical communication technologies in the various layers and
rigorously examine their performance-complexity tradeoffs.
Similarly, relatively few applications have been
examined to date in the application-specific orchestration
studies for vertical multilayer
networking~\cite{Khaddam2015,Liu2015e,vil2015net}.
The examination of the wide range of existing applications and
any newly emerging network application in the context of
SDN orchestrated vertical multilayer networking presents rich research
opportunities.
The cross-layer perspective of the SDN orchestrator over a given
domain could, for instance, be exploited for strengthening security and
privacy mechanisms or for accommodating demanding real-time multimedia.

Relatively more SDN orchestration studies to date have
examined multidomain networking than multilayer networking (within a single
domain). As the completed multidomain orchestration studies
have demonstrated, the SDN orchestration can help greatly in coordinating
complex network management decisions across multiple distributed
routing domains.
The completed studies have illustrated the fundamental
tradeoff between centralized decision making in a hierarchical
orchestration structure and distributed decision making in a flat
orchestration structure.
In particular, most studies have focused on hierarchical
structures~\cite{Jing2015,cas2015sdn,mun2015int}, while only one study
has mainly focused on a flat orchestration structure~\cite{Zhu2015}.
In the context of DC internetworking, the
studies~\cite{Liu2015b, may2016sdn} have sought to bring out
the tradeoffs between these two structures by examining
a range of structures from centralized to distributed.
While centralized orchestration can make decisions with
a wide knowledge horizon across the states in multiple domains,
distributed decision making preserves the privacy of network status
information, reduces control traffic, and can make fast localized
decisions.
Future research needs to shed further light on these complex
tradeoffs for a wide range of combinations of optical technologies employed
in the various domains.
Throughout, it will be critical to abstract and convey the key
characteristics of optical physical layer components and switching nodes to
the overall orchestration protocols. Optimizing each abstraction
step as well as the overall orchestration and examining the various
performance tradeoffs are important future research directions.

\section{Open Challenges and Future SDON Research Directions} \label{sec:open}
We have outlined open challenges and future
Software Defined Optical Network (SDON) research directions
for each sub-category of surveyed SDON studies in the Summary and
Discussion subsections in the preceding survey sections.  In this
section, we focus on the overall cross-cutting open challenges that
span across the preceding considered categories of SDON studies.
That is, we focus on open challenges and research directions
that span the vertical (inter-layer) and horizontal (inter-domain)
SDON aspects. The vertical SDON aspects encompass the
seamless integration of the various (vertical) layers of the
SDON architecture; especially the optical layer, which is not considered
in general SDN technology.
The horizontal SDON aspects include the integration of SDONs with
existing non-SDN optical networking elements, and the internetworking
with other domains, which may have  similar or different
SDN architectures.
A key challenge for SDON research is to enable the use of SDON concepts
in operational real-time network infrastructures.
Importantly, the SDON concepts need to demonstrate performance
gains and cost reductions to be considered by network and service providers.
Therefore, we cater some of the open challenges and
future directions towards enabling and demonstrating the successful use
of SDON in operational networks.

The SDON research and development effort to date
have resulted in insights for making the use of SDN in optical transport
networks feasible and have demonstrated advantages of SDN based optical network
management.
However, most network and service providers
depend on optical transport to integrate with multiple industries to
complete the network infrastructure. Often, network and service providers
struggle to integrate hardware components and to provide accessible software
management to customers. For example, companies that develop hardware
optical components do not always have a complete associated software stack
for the hardware components. Thus, network and service providers using the
hardware optical components often have to maintain a software development team
to integrate the various
hardware components through software based management into their network,
which is often a costly endeavor.
Thus, improving SDN technology so that it seamlessly integrates with
components of various industries
and helps the integration of components from various industries
is an essential underlying theme for future SDON research.

\subsection{Simplicity and Efficiency}   \label{simpl_fut:sec}
Optical network structures typically span heterogeneous devices ranging
from the end user nodes and local area networks via ONUs and OLTs in the
access networks to edge routers and metro network nodes and on
to backbone (core) network infrastructures.
These different devices often come from different vendors.
The heterogeneity of devices and their vendors often requires
manual configuration and maintenance of optical networks.
Moreover, different communication technologies typically
require the implementation of native functions that are specific
to the communication technology characteristics, e.g., the transmission and
propagation properties.
By centralizing the optical network control in an
SDN controller, the SDN networking paradigm creates a unified view
of the entire optical network.
The specific native functions for specific communication devices
can be migrated to the software layer and be implemented by a central node,
rather than through manual node-by-node configurations.
The central node would typically be readily accessible and could reduce
the required physical accesses to distributed devices at their
on-site locations.
This centralization can simplify the network management and
reduce operational expenditures.
An important challenge in this central management is the efficient
SDN control of components from multiple vendors.
Detailed vendor contract specifications of open-source middleware may be
needed to efficiently control components from different vendors.

The heterogeneity of devices may reduce the efficiency of
network infrastructures
due to the required multiple software and hardware modules for a
complete networking solution. Future research should
investigate efficient mechanisms for making complete networking solutions
available for specific use cases. For example, the use of
SDON for an access network provider may require multiple SDN controllers
co-located within the OLT to enable the control of the access network
infrastructure from one central location.  While the SDON studies reviewed in
this survey have led initial investigations of simple and dynamic
network management, future research needs to refine these management
strategies and optimize their operation across combinations of
network architecture structures and across various network
protocol layers.
Simplicity is an essential part of this
challenge, since overly complex solutions are generally not deployed
due to the risk of high expenditures.

\subsection{North Bound Interface} \label{nbi:fut}
The NorthBound Interface (NBI) comprises the communication from the
controller to the applications. This is an important area of
future research as applications and their needs are generally
the driving force for deploying SDON infrastructures.
Any application, such as video on demand, VoIP, file
transfer, or peer-to-peer networking, is applied from the NBI to the
SDN controller which consequently conducts the necessary actions to implement
the service behaviors on the physical network infrastructure.
Applications often require specific service behaviors that
need to be implemented on the overall network infrastructure.
For example,
applications requiring high data rates and reliability, such as Netflix,
depend on data centers and the availability
of data from servers with highly resilient failure protection mechanisms.
The associated management network needs to stack redundant devices as
to safeguard against outages. Services are provided as
policies through the NBI to the SDN controller, which in turn generates flow
rules for the switching devices. These flow rules can be
prioritized based on the customer use cases.
An important challenge for future NBI research is to
provide a simple interface for a wide variety of
service deployments without vendor lock-in, as vendor lock-in
generally drives costs up.
Also, new forms of communication to the controller, in
addition to current techniques, such as
REpresentational State Transfer (REST)~\cite{REST15} and HTTP, should
be researched.
Moreover, future research should develop an NBI framework that spans
horizontally across multiple controllers, so that service customers are not
restricted to using only a single controller.

Future research should examine control mechanisms that optimally
exploit the central SDN control to provide simple and efficient
mechanisms for automatic network management and dynamic service
deployment~\cite{ZhYZ14}.  The NBI of SDONs is a
challenging facet of research and development because of the multitude
of interfaces that need to be managed on the physical layer and
transport layer. Optical physical layer components and infrastructures
require high capital and
operational expenditures and their management is generally not associated with
network or service providers but rather with optical component/infrastructure
vendors. Future research should develop novel Application Program Interfaces
(APIs) for optical layer components and infrastructures that facilitate
SDN control and are amenable to efficient NBI communication.
Essentially, the challenge of efficient NBI communication with the
SDN controller should be
considered when designing the APIs that interface with the
physical optical layer components and infrastructures.

One specific strategy for simplifying network management and operation
could be to explore the grouping of control policies of similar
service applications, e.g., applications with similar QoS requirements.
The grouping can reduce the number of control policies at the
expense of slightly coarser granularity of the service offerings.
The emerging Intent-Based Networking (IBN) paradigm, which
drafts intents for services and policies,
can provide a specific avenue for simplifying
dynamic automatic configuration and virtualization~\cite{BlDM08,CoBR13}.
Currently network applications are deployed based on how the network
should behave for a specific action. For example, for inter domain routing,
the Border Gateway Protocol (BGP) is used, and the network gateways
are configured to communicate with the BGP protocol.
This complicates the provisioning of
services that typically require multiple protocols and limits the
flexibility of service provisioning. With IBN, the application gives an intent,
for example, transferring video across multiple domains.
This intent is then
associated with automated dynamic configurations of the network elements
to communicate data over the domains using appropriate protocols.
The grouping of service policies, such as intents, can facilitate
easy and dynamic service provisioning.
Intent groups can be described in a graph to simplify the compilation of
service policies and to resolve conflicts~\cite{PrLT15}.

\subsection{Reliability, Security, and Privacy} \label{futworksec:sec}
The SDN paradigm is based on a centrally managed network.
Faulty behaviors, security infringements, or failures of the
control would likely result in extensive disruptions and
performance losses that are exacerbated by the centralized nature of the
SDN control. Instances of extensive disruptions and losses due to
SDN control failures or infringements would likely reduce the trust in
SDN deployments. Therefore, it is very important to ensure reliable network
operation~\cite{rak2016inf}
and to provision for security and privacy of the communication.
Hence, reliability, security, and privacy are prominent SDON research
challenges. Security in SDON techniques is a fairly open research
area, with only few published findings.
As a few reviewed studies (see Section~\ref{app_fail_rec:sec}) have explored,
the central SDN control can facilitate reliable network service through speeding
up failure recovery.
The central SDN control can continuously scan the network and
the status messages from the network devices.
Or, the SDN control can
redirect the status messages to a monitoring service that analyzes
the data network.  Security breaches can be controlled by broadcasting
messages from the controller to all affected devices to block traffic
in a specific direction. Future research should refine these
reliability functions to optimize automated fault and performance
diagnostics and reconfigurations for quick failure recovery.

Network failures can either occur within the physical layer
infrastructure, or as errors within the higher protocol
layers, e.g., in the classical data link (L2), network (L3), of
transport (L4) layers.
In the context of SDONs, physical
layer failures present important future research opportunities.
Physical layer devices need to be carefully monitored by sending
feedback from the devices to the controller.
The research and development on communication between
the SDN controller and the network devices has mainly focused on sending
flow rules to the network devices while feedback communicated from the devices
to the controller has received relatively little attention.
For example, there are three types of OpenFlow messages, namely Packet-In,
Packet-Out, and Flow-Mod. The Packet-In messages are
sent from the OpenFlow switches to the controller,
the Packet-Out message is sent from
the controller to the device, and the Flow-Mod message is used to
modify and monitor
the flow rules in the flow table. Future research should examine extensions of
the Packet-In message to send specific status updates in support of
network and device failure monitoring to the controller.
These status messages could be monitored
by a dedicated failure monitoring service.
The status update messages could be broadly defined to
cover a wide range of network management aspects, including
system health monitoring and network failure protection.

A related future research direction is to secure configuration and
operation of SDONs through trusted encryption and key management
systems~\cite{ahm2015sec}. Moreover, mechanisms to
ensure the privacy of the communication should be explored.
The security and privacy mechanisms should strive to exploit the
natural immunity of optical transmission segments to electro-magnetic
interferences.

In summary, security and privacy of SDON communication are largely open
research areas.
The optical physical layer infrastructure has traditionally not
been controlled remotely, which in general reduces the occurrences
of security breaches.
However, centralized SDN management and control increase the risk of
security breaches, requiring extensive research on SDON security, so
as to reap the benefits of centralized SDN management and control in a
secure manner.

\subsection{Scalability} \label{scalability:fut}
Optical networks are expensive and used for high-bandwidth
services, such as long-distance network access and data center interconnections.
Optical network infrastructures either
span long distances between multiple geographically distributed locations,
or could be short-distance incremental additions (interconnects)
of computing devices. Scalability in multiple dimensions
is therefore an important aspect for future SDON research.
For example, a
myriad of tiny end devices need to be provided with network access in
the emerging Internet of Things (IoT) paradigm~\cite{wan2015nov}.
The IoT requires access network architectures and
protocols to scale vertically (across protocol layers and technologies)
and horizontally (across network domains).
At the same time, the ongoing growth of
multimedia services requires data centers to scale up optical network bandwidths
to maintain the quality of experience of the multimedia services.
Broadly speaking, scalability includes in the vertical
dimension the support for multiple network devices and technologies.
Scalability in
the horizontal direction includes the communication between
a large number of different domains as well as support for existing non-SDON
infrastructures.

A specific scalability challenge arising with SDN infrastructure is that the
scalability of the control plane (OpenFlow protocol signalling)
communication
and the scalability of the data plane communication which transports
the data plane flows
need to be jointly considered. For example, the OpenFlow protocol~1.4
currently supports 34 Flow-Mod messages~\cite{OF2016}, which can
communicate between the network devices and the controller. This
number limits the functionality of the SBI communication. Recent studies
have explored a protocol-agnostic
approach \cite{HuLi2015, bosshart2014p4}, which is a data plane
protocol that extends the use of multiple protocols for
communication between the control plane and data plane.
The protocol-agnostic approach resolves
the challenges faced by OpenFlow and, in general, any particular protocol.
Exploring this novel protocol-agnostic approach presents many new SDON
research directions.

Scalability would also require SDN technology to overlay and scale
over existing non-SDN infrastructures. Vendors provide support
for known non-SDN devices, but this area is still a challenge. There
are no known protocols that could modify the flow tables of existing
popularly described ``non-OpenFlow'' switches. In the case of optical
networks, as SDN is still being incrementally deployed, the overlaying
with non-SDN infrastructure still requires significant attention.
Ideally, the overlay mechanisms should ensure seamless integration and should
scale with the growing deployment of SDN technologies while incurring only
low costs.
Overall, scalability poses highly important future SDON research directions
that require economical solutions.

\subsection{Standardization}   \label{std:sec}
Networking protocols have traditionally followed a uniform
standard system for all the
communication across multiple domains. Standardization has helped vendors
to provide products that work in and across different network infrastructures.
In order to ensure the compatible inter-operation of SDON components (both hardware and software) from a various vendors,
key aspects of the inter-operation protocols need to be standardized.
Towards the standardization goal,
communities, such as Open Networking Foundation (ONF), have created boards
and committees to standardize protocols, such as OpenFlow.
Standardization should ensure that SDON infrastructures can be flexibly
configured and operated with components from various vendors.
The use of open-source
software can further facilitate the inter-operation.
Proprietary hardware and software components generally create vendor lock-in,
which restricts the flexibility of network operation and reduces
the innovation of network and service providers.

As groundwork for standardization, it may be necessary to develop
and optimize a common (or a small set) of SDON architectures
and network protocol configurations that can serve as a basis
for standardization efforts.
The standardization process may involve a common platform that
is built thorough the cooperation of multiple manufacturers.
Another thrust of standardization groundwork could be the development of
open-source software that supports SDON architectures.
For example, Openstack is a cloud based management framework that has been
adopted and supported by multiple networking vendors. Such efforts should
be extended to SDONs in future work.

\subsection{Multilayer Networking} \label{multilayer:fut}
As discussed in Section~\ref{orch:sec}, multilayer networking
involves vertical multilayer networking
across the vertical layers as well as horizontal multilayer (multidomain)
networking across multiple domains.
We proceed to outline open challenges and
future research directions for vertical multilayer networking in the context of
SDON, which includes an optical physical layer, in this subsection.
Horizontal multilayer (multidomain) networking is considered in
Section~\ref{multidomain:fut}.

For the vertical multilayer networking in a single domain,
the optical physical layer is the key distinguishing feature of
SDONs compared to conventional SDN architectures for general IP networks.
Most of the higher layers in SDONs have similar multilayer
networking challenges as general IP networks.
However, the optical physical layer
requires the provisioning of specific optical
transmission parameters, such as wavelengths and signal strengths.
These parameters are managed by optical devices, such as the OLT in PON
networks. For SDON networks, so-called \textit{optical orchestrators},
which are commercially available, e.g., from ADVA Optical Networking,
provide a single interface to provision the optical layer parameters.
We illustrate this optical orchestrator layer in the context of an
SDON multilayer network in the rightmost branch of Fig.~\ref{fig_control_orch}.
The optical orchestrator resides
above the optical devices and below the SDN controller.
The optical orchestrator uses common SDN SBI interface protocols,
such as OpenFlow,
to communicate with the optical devices in the south-bound direction
and with the controller in the north-bound direction.

The SDN controller in the control plane is responsible for the management of the
SDN-enabled switches, potentially via an optical orchestrator.
Communicating over the SBI using different
protocols can be challenging for the controller.
This challenge can
be addressed by using south-bound renderers. South-bound renderers are
APIs that reside within the
controller and provide a communication channel to any desired
SBI protocol.  Most SDN controllers currently have an
OpenFlow renderer to be able to communicate to OpenFlow
enabled network switches.
But there are also SNMP and NETCONF-based renderers, which
communicate with traditional non-OpenFlow switches. This enables the
existence of hybrid networks with already existing switches.
The effective support of such hybrid networks, in conjunction with
appropriate south-bound renderers and optical orchestrators, is an important
direction for future research.

\subsection{Multidomain Networks} \label{multidomain:fut}
A network domain usually belongs to a single organization that
owns (i.e., financially supports and uses) the network domain.
The management of multidomain networking involves the
important aspects of configuring the access control as well as the
authentication, authorization, and accounting. Efficient
SDN control mechanisms for configuring these multidomain networking
aspects is an important direction for future research and development.

Multidomain SDONs may also need novel routing
algorithm that enhance the capabilities of the currently used BGP
protocol. Multidomain research \cite{phe2014dis} has now taken
interest in the Intent-Based Networking (NBI) paradigm for
SDN control, where Intent-APIs can solve the
problems of spanning across multiple domains. For instance, the intent of an
application to transfer information across multiple domains is
translated into service instances that access configurations between
domains that have been pre-configured based on contracts.
Currently, costly manual configurations between
domains are required for such applications.
Future research needs to develop concrete models for
NBI based multidomain networking in SDONs.

\subsection{Fiber-Wireless (FiWi) Networking} \label{wireless:fut}
The optical (fiber) and wireless network domains have
many differences.
At the physical layer, wireless networks are characterized
by varying channel qualities, potentially high losses, and generally lower
transmission bit rates than optical fiber.
Wireless end nodes are typically mobile and may connect dynamically
to wireless network domains.
The mobile wireless nodes are generally the end-nodes in a
FiWi network and connect via intermediate optical nodes to
the Internet.
Due to these different characteristics, the management of
wireless networks with mobile end nodes is very different from the
management of optical network nodes.
For example, wireless access points should maintain their own
routing table to accommodate access to dynamically connected mobile
devices. Combining the control of both wireless and
optical networks in a single SDN controller requires
concrete APIs that handle the respective control functions of
wireless and optical networks.
Currently, service providers maintain separate physical management services
without a unified logical control and management plane for
FiWi networks.
Developing integrated controls for FiWi networks
can be viewed as a special case of multilayer networking and integration.

Developing specialized multilayer networking strategies for
FiWi networks is an important future research directions as many aspects of
wireless networks have dramatically advanced in recent
years. For instance, the cell structure of wireless cellular
networks~\cite{Ohlen2016} has advanced to femtocell
networks~\cite{cha2008fem} as well as heterogeneous and multitier
cellular structures~\cite{els2013sto,lou2011tow}. At the same time,
machine-to-machine communication~\cite{has2013ran,lay2014ran} and
energy savings~\cite{ana2015opt,has2011gre} have drawn research attention.

\subsection{QoS and Energy Efficiency}  \label{futworkene:sec}
Different types of applications have vastly different
traffic bit rate characteristics and QoS requirements.
For instance, streaming high-definition video requires high bit rates,
but can tolerate
some delays with appropriate playout buffering. On the other hand,
VoIP (packet voice) or video conference applications have
typically low to moderate bit rates, but require low latencies.
Achieving these application-dependent QoS levels in an energy-efficient
manner~\cite{has2011gre,ShZL14,wan2015ene}
is an important future research direction.
A related future research direction is to
exploit SDN control for QoS adaptations of real-time media and
broadcasting services.
Broadcasting services involve typically data rates ranging from
3--48~Gb/s to deliver video at various resolutions to
the users within a reasonable time limit.
In addition to managing the QoS, the network
has to manage the multicast groups for efficient
routing of traffic to the users.
Recent studies \cite{Butler2015, Ellerton2015} discuss
the potential of SDN, NFV, and optical technologies
to achieve the growing demands of broadcasters and media.
Moreover, automated
provisioning strategies of QoS and the incorporation of quality of
protection and security with traditional QoS are important direction for
future QoS research in SDONs.

\subsection{Performance Evaluation}
Comprehensive performance evaluation methodologies and metrics need to
be developed to assess the SDON designs addressing the
preceding future research directions ranging from simplicity and
efficiency (Section~\ref{simpl_fut:sec}) to optical-wireless networks
(Section~\ref{wireless:fut}).
The performance evaluations need to encompass the data plane,
the control plane, as well as the overall data and control plane interactions
with the SDN interfaces and need to take virtualization
and orchestration mechanisms into
consideration. In the case of the SDON infrastructure, the performance
evaluations will need to include the optical
physical layer~\cite{Azodolmolky2014}.
While there have been some efforts to develop evaluation frameworks for
general SDN switches~\cite{OFTest,rot2014ope}, such evaluation frameworks
need to be adapted to the specific characteristics of SDON architectures.
Similarly, some evaluation frameworks for general SDN controllers have
been explored~\cite{jar2012fle,jar2014ofc}; these need to be extended
to the specific SDON control mechanisms.

Generally, performance metrics obtained with SDN and virtualization
mechanisms should be benchmarked against the corresponding
conventional network without any SDN or virtualization components.
Thus, the performance tradeoffs and costs of the flexibility
gained through SDN and virtualization mechanism can be quantified.
This quantified data would then need to be assessed and compared
in the context of business needs. To identify some of the important aspects of
performance we analyze the sample architecture in Fig.~\ref{fig_app_i2rs}.
The SDN controller in the SDON architecture in Fig.~\ref{fig_app_i2rs}
spans across multiple elements, such as ONUs, OLTs,
routers/switches in the metro-section, as well as PCEs in the core section.
A meaningful performance evaluation of such a network
requires comprehensive analysis of data plane performance aspects and
related metrics,
including noise spectral analysis, bandwidth and link rate monitoring,
as well as evaluation of failure resilience.
Performance evaluation mechanisms need to be
developed to enable the SDON controller to obtain and analyze these
performance data. In addition, mechanisms for control layer
performance analysis are needed.
The control plane performance evaluation should, for instance
assess the controller efficiency and performance characteristics,
such as the OpenFlow message rates and the rates and delays of flow table
management actions.

\section{Conclusion}  \label{sec:conclusion}
We have presented a comprehensive survey of software defined
optical networking (SDON) studies to date.
We have mainly organized our survey according to the SDN
infrastructure, control, and application layer structure. In addition,
we have dedicated sections to SDON virtualization and orchestration
studies.
Our survey has found that SDON infrastructure studies
have examined optical (photonic) transmission and switching components
that are suitable for flexible SDN controlled operation.
Moreover, flexible SDN controlled switching paradigms and optical
performance monitoring frameworks have been investigated.

SDON control studies have developed and evaluated SDN control
frameworks for the wide range of optical network transmission
approaches and network structures. Virtualization allows for
flexible operation of multiple Virtual Optical Networks (VONs) over a given
installed physical optical network infrastructure.
The surveyed SDON virtualization studies have examined the provisioning
of VONs for access networks, exploiting the
specific physical and Medium Access Control (MAC) layer characteristics of
access networks. The virtualization studies have also examined
the provisioning of VONs in metro and
backbone networks, examining algorithms for embedding the VON topologies
on the physical network topology under consideration of the
optical transmission characteristics.

SDON application layer studies have developed mechanisms for achieving
Quality of Service (QoS), access control and security, as well as
energy efficiency and failure recovery.
SDON orchestration studies have examined coordination mechanisms
across multiple layers (in the vertical dimension of the network protocol
layer stack) as well as across multiple network domains (that may belong
to different organizations).

While the SDON studies to date have established basic principles
for incorporating and exploiting SDN control in optical networks,
there remain many open research challenges. We have
outlined open research challenges for each individual category of
studies as well as cross-cutting research challenges.

\section*{Acknowledgement}
A part of this article was written while Martin Reisslein visited the
Technische Universit\"at M\"unchen (TUM), Germany, in 2015. Support
from TUM and a Friedrich Wilhelm Bessel Research Award from the
Alexander von Humboldt Foundation are gratefully acknowledged.

\bibliographystyle{IEEEtran}


\begin{IEEEbiography}
  {Akhilesh S. Thyagaturu} is currently pursuing his Ph.D. degree
  in Electrical Engineering at Arizona State University, Tempe. He received the M.S. degree in Electrical Engineering from Arizona State University, Tempe, in 2013. He received B.E. degree from Visveswaraya Technological University (VTU), India, in 2010. He worked for Qualcomm Technologies Inc., San Diego, CA,
  as an Engineer between 2013 and 2015. 
\end{IEEEbiography}

\vspace{-1.65cm}

\begin{IEEEbiography}
{Anu Mercian} is currently a Sr. Systems Engineer, at Hewlett Packard
Enterprise, Networking R\&D, Palo Alto, CA. She earned her Bachelors in
Electronics Engineering from College of Engineering, Trivandrum, India,
and her Masters and Ph.D. in Electrical Engineering from Arizona State
University (ASU), Tempe. She is also a Research Affiliate with
ASU and Lawrence Berkeley
Labs (LBL) as well as a Contributor with the OpenDayLight (ODL) and
Open Networking Foundation (ONF) Communities. 
\end{IEEEbiography}

\vspace{-1.65cm}

\begin{IEEEbiography}
  {Michael P. McGarry} (M'98-SM'13) received the B.S. in Computer
  Engineering from the Polytechnic Institute of NYU in 1997. He
  received the M.S. and Ph.D. in Electrical Engineering from Arizona
  State University in 2004 and 2007, respectively.  He is currently an
  Assistant Professor in the Dept. of Electrical and Computer
  Engineering at the University of Texas at El Paso. In 2009, he was a
  co-recipient of the IEEE ComSoc. Best Tutorial Paper
  award. His research interests include optical and software defined networks,
  as well as cloud computing.
\end{IEEEbiography}

\vspace{-1.65cm}

\begin{IEEEbiography}{Martin Reisslein} (A'96-S'97-M'98-SM'03-F'14) 
is a Professor in the School of Electrical, Computer, and Energy
Engineering at Arizona State University (ASU), Tempe. He received the
Ph.D. in systems engineering from the University of Pennsylvania in
1998. He currently serves as Associate Editor for the
\textit{IEEE Transactions on Education} as well as \textit{Computer
  Networks} and \textit{Optical Switching and Networking}.
\end{IEEEbiography}

\vspace{-1.65cm}

\begin{IEEEbiography}{Wolfgang Kellerer} (M'96-SM'11) 
is full professor at the Technische Universit\"at M\"unchen (TUM),
heading the Chair of Communication Networks in the Department of
Electrical and Computer Engineering since 2012. Before, he has been
director and head of wireless technology and mobile network research
at NTT DOCOMO's European research laboratories, DOCOMO Euro-Labs, for
more than ten years. His research focuses on concepts for the dynamic
control of networks (Software Defined Networking), network
virtualization and network function virtualization, and 
application-aware traffic management. In the area of wireless networks the
emphasis is on Machine-to-Machine communication, Device-to-Device
communication and wireless sensor networks with a focus on resource
management towards a concept for 5th generation mobile
communications (5G). His research resulted in more than 200
publications and 29 granted patents in the areas of mobile networking
and service platforms. He is a member of the ACM and the VDE ITG, and a Senior
Member of the IEEE.
\end{IEEEbiography}

\end{document}